\documentclass[aps,prd,superscriptaddress,floatfix,nofootinbib,preprintnumbers,eqsecnum,twocolumn]{revtex4-1}
\pdfoutput=1  

\usepackage{amsmath,float,graphicx,placeins,amssymb,multirow,pgfplots,pgfplotstable}
\usepackage{tikz}
\usepackage{soul}
\setstcolor{red}
\usepackage{graphicx}

\usepackage{comment}
\usepackage{enumerate,slashed,cancel,comment,color,bbm,graphicx,rotating,placeins,mathtools,multirow,hhline}
\usepackage[normalem]{ulem}
\usepackage{array}
\usepackage{enumitem}
\usepackage{hyperref}
\usepackage{color}
\hypersetup
 { 
   colorlinks,
   linkcolor={blue!80!black},
   citecolor={blue!70!black},
   urlcolor={blue!70!black}
 }

\usepackage{lipsum}
\usepackage{diagbox}
\usetikzlibrary{shapes.geometric,arrows}
\graphicspath{}

\def\qb{\overline{q}}
\def\mfrac{\mbox{\small $ \left( \frac{M^2}{4\pi\calM^2} \right) $}}
\def\mfrac{ {\widetilde M^2} }
\def\mmfrac{ {\widetilde M^4} }
\def\mmmfrac{ {\widetilde M^6} }
\def\mmmmfrac{ {\widetilde M^8} }

\def\ampX{{\langle \calX \rangle }}

\def\muresc{\hat{\mu}}

\def\msmall#1 {\mbox{\small{$#1$}}}
\def\qbar{{\overline{q}}}

\def\beq{\begin{equation}}
\def\eeq{\end{equation}}
\def\beqn{\begin{eqnarray}}
\def\eeqn{\end{eqnarray}}
\def\apo{\mbox{\small ${\frac{\alpha'}{2}}$}}
\def\half{\mbox{\small ${\frac{1}{2}}$}}

\def\sqapo{\mbox{\tiny $\sqrt{\frac{\alpha'}{2}}$}}
\def\sqap{\mbox{\tiny $\sqrt{{\alpha'}}$}}
\def\sqapxtwo{\mbox{\tiny $\sqrt{2{\alpha'}}$}}
\def\aptwo{\mbox{\tiny ${\frac{\alpha'}{2}}$}}
\def\apofour{\mbox{\tiny ${\frac{\alpha'}{4}}$}}
\def\bosqtwo{\mbox{\tiny ${\frac{\beta}{\sqrt{2}}}$}}
\def\btosqtwo{\mbox{\tiny ${\frac{\tilde{\beta}}{\sqrt{2}}}$}}
\def\apofour{\mbox{\tiny ${\frac{\alpha'}{4}}$}}
\def\sqaptwo{\mbox{\tiny $\sqrt{\frac{\alpha'}{2}}$}  }
\def\apoeight{\mbox{\tiny ${\frac{\alpha'}{8}}$}}
\def\sapoeight{\mbox{\tiny ${\frac{\sqrt{\alpha'}}{8}}$}}

\newcommand{\newc}{\newcommand}
\def\calZ{{\cal Z}}
\def\calM{{\cal M}}
\def\calV{{\cal V}}
\def\calF{{\cal F}}
\def\calG{{\cal G}}
\def\calS{{\cal S}}
\def\calY{{\cal Y}}
\def\calX{{\cal X}}

\def\calK{{\cal K}}
\def\calA{{\cal A}}
\def\calB{{\cal B}}
\def\eulerg{{\gamma_E}}
\def\twotwo{{\mathbbm{T}^2}}

\def\mm{{\widetilde m}}
\def\nn{{\widetilde n}}
\def\RR{{\widetilde R}}

\def\mathbbX{{\mathbbm{X}}}

\def\mathbbZ{{\mathbbm{Z}}}
\def\mathbbT{{\mathbbm{T}}}
\def\mathbbA{{\mathbbm{A}}}
\def\mathbbB{{\mathbbm{B}}}
\def\bQ{{\bf Q}}
\def\bT{{\bf T}}

\def\Qs{{\bf q}}

\def\gx{g_{\calX}}
\def\calE{{\cal E}}
\def\calO{{\cal O}}

\def\ie{{\it i.e.}\/}
\def\eg{{\it e.g.}\/}
\def\etc{{\it etc}.\/}
\def\inbar{\,\vrule height1.5ex width.4pt depth0pt}
\def\IR{\relax{\rm I\kern-.18em R}}
 \font\cmss=cmss10 \font\cmsss=cmss10 at 7pt
\def\IQ{\relax{\rm I\kern-.18em Q}}
\def\IZ{\relax\ifmmode\mathchoice
 {\hbox{\cmss Z\kern-.4em Z}}{\hbox{\cmss Z\kern-.4em Z}}
 {\lower.9pt\hbox{\cmsss Z\kern-.4em Z}}
 {\lower1.2pt\hbox{\cmsss Z\kern-.4em Z}}\else{\cmss Z\kern-.4em Z}\fi}

\def\oneRes{{       
     \underset{s=1}{\rm Res}
  }}
\def\Str{{\rm Str}}
\def\zStr{{         
      \underset{M=0}{\rm Str}
       \,}}
\def\mStr{{         
      \underset{{\rm mass}=M}{\rm Str}
       \,}}

\def\effStr{{         
      \underset{\small 0<M\lesssim \mu}{\rm Str}
       \,}} 
\def\effStrprime{{         
      \underset{\small 0<M\lesssim \mu}{{\rm Str}'}
       \,}}
\def\effTr{{         
      \underset{\small 0<M\lesssim \mu}{\rm Tr}
       \,}}

\begin{document}

\title{A New Non-Renormalization Theorem from UV/IR Mixing}

\def\andname{\hspace*{-0.5em}} 
\author{Steven Abel}
\email[Email address: ]{s.a.abel@durham.ac.uk}
\affiliation{Institute for Particle Physics Phenomenology, Durham University, DH1 3LE, Durham, UK}
\author{Keith R. Dienes}
\email[Email address: ]{dienes@arizona.edu}
\affiliation{Department of Physics, University of Arizona, Tucson, AZ 85721 USA}
\affiliation{Department of Physics, University of Maryland, College Park, MD 20742 USA}
\author{Luca A. Nutricati}
\email[Email address: ]{l.nutricati@ucl.ac.uk}
\affiliation{Department of Mathematical Sciences, Durham University, Durham, DH1 3LE UK}
\affiliation{University College London, Gower St., London WC1E 6BT UK}

\begin{abstract}
In this paper, we prove a new non-renormalization theorem which arises from UV/IR mixing.  This theorem and its corollaries
are relevant for all four-dimensional tachyon-free closed string theories which can be realized from higher-dimensional theories via geometric compactifications.  As such, they therefore 
apply regardless of the presence or absence of spacetime supersymmetry and regardless of the gauge symmetries or matter content involved.  
This theorem resolves a hidden clash between modular invariance and the process of decompactification, 
and enables us to uncover a number of surprising phenomenological properties of these theories.
Chief among these is the fact that certain 
physical quantities within such theories
cannot exhibit logarithmic or power-law running and instead enter an effective fixed-point regime above the compactification scale. This cessation of running occurs as the result of the UV/IR mixing inherent in the theory. 
These effects apply not only for gauge couplings but also for the Higgs mass and other quantities of phenomenological interest, thereby eliminating the  logarithmic and/or power-law running that might have otherwise appeared for such quantities.  These results illustrate the power of UV/IR mixing to tame divergences --- even without supersymmetry --- and reinforce the notion that UV/IR mixing may play a vital role in resolving hierarchy problems without supersymmetry.
\end{abstract}
\maketitle
  \tableofcontents

\def\ie{{\it i.e.}\/}
\def\eg{{\it e.g.}\/}
\def\etc{{\it etc}.\/}
\def\taubar{{\overline{\tau}}}
\def\qbar{{\overline{q}}}
\def\kbar{{\overline{k}}}
\def\bQ{{\bf Q}}
\def\calT{{\cal T}}
\def\calN{{\cal N}}
\def\calF{{\cal F}}
\def\calM{{\cal M}}
\def\calZ{{\cal Z}}

\def\beq{\begin{equation}}
\def\eeq{\end{equation}}
\def\beqn{\begin{eqnarray}}
\def\eeqn{\end{eqnarray}}
\def\apo{\mbox{\small ${\frac{\alpha'}{2}}$}}
\def\half{\mbox{\small ${\frac{1}{2}}$}}
\def\sqapo{\mbox{\tiny $\sqrt{\frac{\alpha'}{2}}$}}
\def\sqap{\mbox{\tiny $\sqrt{{\alpha'}}$}}
\def\sqapxtwo{\mbox{\tiny $\sqrt{2{\alpha'}}$}}
\def\aptwo{\mbox{\tiny ${\frac{\alpha'}{2}}$}}
\def\apofour{\mbox{\tiny ${\frac{\alpha'}{4}}$}}
\def\bosqtwo{\mbox{\tiny ${\frac{\beta}{\sqrt{2}}}$}}
\def\btosqtwo{\mbox{\tiny ${\frac{\tilde{\beta}}{\sqrt{2}}}$}}
\def\apofour{\mbox{\tiny ${\frac{\alpha'}{4}}$}}
\def\sqaptwo{\mbox{\tiny $\sqrt{\frac{\alpha'}{2}}$}  }
\def\apoeight{\mbox{\tiny ${\frac{\alpha'}{8}}$}}
\def\sapoeight{\mbox{\tiny ${\frac{\sqrt{\alpha'}}{8}}$}}

\newc{\gsim}{\lower.7ex\hbox{$\;\stackrel{\textstyle>}{\sim}\;$}}
\newc{\lsim}{\lower.7ex\hbox{$\;\stackrel{\textstyle<}{\sim}\;$}}
\def\calM{{\cal M}}
\def\calV{{\cal V}}
\def\calF{{\cal F}}
\def\bQ{{\bf Q}}
\def\bT{{\bf T}}
\def\Qs{{\bf q}}

\def\ie{{\it i.e.}\/}
\def\eg{{\it e.g.}\/}
\def\etc{{\it etc}.\/}
\def\inbar{\,\vrule height1.5ex width.4pt depth0pt}
\def\IR{\relax{\rm I\kern-.18em R}}
 \font\cmss=cmss10 \font\cmsss=cmss10 at 7pt
\def\IQ{\relax{\rm I\kern-.18em Q}}
\def\IZ{\relax\ifmmode\mathchoice
 {\hbox{\cmss Z\kern-.4em Z}}{\hbox{\cmss Z\kern-.4em Z}}
 {\lower.9pt\hbox{\cmsss Z\kern-.4em Z}}
 {\lower1.2pt\hbox{\cmsss Z\kern-.4em Z}}\else{\cmss Z\kern-.4em Z}\fi}

\section{Introduction and overview of results}

Non-renormalization theorems are powerful tools for the study of quantum field theories.  Historically, the most famous non-renormalization theorems are those that arise within the context of theories with unbroken spacetime supersymmetry (SUSY).~  Such supersymmetric non-renormalization theorems can be understood as the result of relatively straightforward ``level-by-level'' pairwise cancellations between states
with similar masses, with the renormalization contributions from each particle  in the spectrum cancelling against the contributions of corresponding superpartners.    
However, given that unbroken supersymmetry does not appear to be a feature of the natural world, 
such SUSY-based non-renormalization theorems cannot be exact in any phenomenologically viable model.  Historically, another (somewhat related) motivation for SUSY was to address the hierarchy problems of the Standard Model, such as the gauge (Higgs) hierarchy and the cosmological-constant problems.   However, given the collider data that has been collected over the past decade,  it is also becoming increasingly unlikely that electroweak-scale SUSY plays a significant role in addressing these hierarchy problems.

Recently, increasing attention has focused on the extent to which hierarchy problems can be alternatively addressed through symmetries that involve UV/IR-mixing (see, \eg, Refs.~\cite{Dienes:1994np, Dienes:1995pm,Dienes:2001se, 
Cohen:1998zx,Abel:2005rh,
Abel:2006wj,
Bramante:2019uub,
Craig:2019zbn,
Blinov:2021fzl,
Charalambous:2021kcz,
Craig:2021ksw,Castellano:2021mmx,
Abel:2021tyt,Arkani-Hamed:2021xlp,
Freidel:2022ryr,
Berglund:2022qsb,
Abel:2023hkk};  for recent reviews see Refs.~\cite{
Craig:2022eqo,Berglund:2022qcc}).
Within such scenarios, UV physics is directly related to IR physics as well as to physics at intermediate scales, implying that 
potential solutions to hierarchy problems within such theories might emerge through what might initially appear to be conspiracies across many or all energy scales.    However, given the ongoing speculation about the possible existence of such UV/IR-mixed approaches to hierarchy problems, we are then led to ask the further question as to whether such UV/IR-mixed symmetries might also give rise to non-renormalization theorems.   Such non-renormalization theorems would then 
emerge not as the consequence of pairwise level-by-level cancellations (such as those that arise in theories with unbroken supersymmetry), but instead as the consequence of symmetries which operate across all scales simultaneously.

In this paper, we investigate this issue by focusing on one of the most important UV/IR-mixed symmetries in string theory, specifically {\it worldsheet modular invariance}\/.
Worldsheet modular invariance is an exact fundamental symmetry within closed string theories, and remnants of modular invariance even exist for open strings as well.
Within a given string model, modular invariance governs the string spectrum and its interactions regardless of the presence or absence of spacetime supersymmetry, regardless of the gauge group and particle content of the model, and even regardless of its assumed spacetime dimensionality. While modular invariance has numerous implications for the low-energy phenomenologies of such strings, this paper is devoted to demonstrating that worldsheet modular invariance has an additional effect which has not previously been noticed, namely that it also gives rise to a powerful non-renormalization theorem. 

As we shall find, this theorem emerges within the context of 
four-dimensional closed string theories which can be realized as  geometric compactifications of higher-dimensional string theories.
In other words, our theorem applies to all four-dimensional closed string theories which have self-consistent decompactification limits. This gives our theorem a broad and nearly universal applicability.   

Rigorously stating and proving this theorem is one of the primary goals of this paper.  However, in order to gain a very rough understanding of the content of our theorem, let us begin by recalling that as the spacetime dimensionality of an ordinary quantum field theory  increases, it tends to become more finite in the IR but more divergent in the UV.~   This is the direct result of the different numbers of momentum components associated with the states propagating in loops.
However, in theories with UV/IR-mixing, it turns out that this behavior is generally different.
For example, within closed string theories there is only one potential divergence for a given one-loop amplitude.  Indeed, by making use of the UV/IR-mixed symmetries of the theory, one can recast this divergence as either a UV divergence or an IR divergence~\cite{Abel:2021tyt}.   
Moreover, as the dimensionality of such a string theory increases, it turns out that the theory tends to become {\it more}\/ finite (or equivalently, its divergences tend to become less
severe).  
This feature arises because
additional internal cancellations or constraints come into play across the string spectrum as the spacetime dimensionality of our string theory increases --- constraints 
which soften or eliminate these divergences and which are thereby
ultimately responsible for these additional finiteness properties~\cite{Dienes:1994np,Dienes:1995pm}.

This situation becomes particularly interesting for string theories which have decompactification limits.
If a given string theory has a {\it bona fide}\/ decompactification limit, then its spectrum  must satisfy the extra constraints discussed above in the full decompactification limit at which our theory becomes higher-dimensional.  By contrast, there is no need for such extra constraints to apply to the compactified theory, since the compactified theory by definition is lower-dimensional.
What we find, however, is that these extra constraints apply not only to the higher-dimensional theory that emerges in the 
full decompactification limit, {\it but also to the original compactified theory, regardless of the compactification volume}\/! 
In particular, stated succinctly, in this paper we shall prove:
\begin{quote}
{\it \underbar{Theorem}:~
Any four-dimensional closed string theory which can be realized as a geometric compactification from a higher-dimensional string theory will inherit the precise stricter internal cancellations of the higher-dimensional theory from which it is obtained despite the compactification.    }
\end{quote}
Thus, as long as our four-dimensional theory has a decompactification limit, its spectrum must satisfy not only the constraints that would normally be associated with its existence in four dimensions, but also the additional constraints that would be required in higher dimensions.  Indeed, this remains true even if our four-dimensional theory is nowhere near the decompactification limit and is thus fully four-dimensional!
As we shall demonstrate,
this theorem and the cancellations it requires are realized 
across all energy scales at once, and the mechanism by which it operates has no field-theoretic analogue or approximation. 

This theorem leads to many surprising phenomenological consequences for our original four-dimensional theory.  One of these consequences is that there are new,
unexpected UV/IR-mixed supertrace constraints which operate across the entire four-dimensional string spectrum at all energy scales, similar to those which were originally obtained in Ref.~\cite{Dienes:1995pm} 
and more recently generalized in Refs.~\cite{Abel:2021tyt,Abel:2023hkk}.  Like these previous supertrace constraints,  our new supertrace constraints are the results of an underlying so-called ``misaligned supersymmetry''~\cite{Dienes:1994np, Dienes:1995pm} that governs the spectra of all tachyon-free modular-invariant theories --- even those that lack spacetime supersymmetry. We emphasize that these new constraints are completely unexpected from the perspective of our original four-dimensional theory. They  nevertheless secretly govern the spectra of such theories at all energy scales.

Another surprising conclusion of our theorem concerns the effective field theories (EFTs) that are derived from such string theories.  In particular, within such theories we find that the couplings can at most run {\it only until the compactification scale is reached.}\/   Beyond this point, our theorem asserts that {\it all running ceases}\/ --- even if this compactification scale is much lower than the string scale.  Indeed, the theory necessarily enters a ``fixed-point'' regime in which the beta-functions of the gauge couplings vanish.  This too is a result of the extra ``hidden'' constraints discussed above.   This gives us an important corollary to our main theorem:
\begin{quote}
    {\it \underbar{Non-renormalization corollary}:~ Within any modular-invariant theory which has $\delta\equiv D-4$ large extra dimensions opening up at a scale $1/R$, misaligned supersymmetry and UV/IR-mixing eliminate all running for $\mu \gtrsim R^{-1}$ regardless of the value of $\delta$. For $\mu < R^{-1}$, these same phenomena eliminate all running for $\delta>2$, and leave at most logarithmic running for $\delta =2$.  }
\end{quote}

The above results arise most naturally within the context of string  theories exhibiting
a single decompactification limit (along with a corresponding $T$-dual limit).
However, most string theories have {\it multiple}\/ decompactification limits, and the different higher-dimensional theories which emerge in these limits may even have different spacetime dimensionalities.
Such cases nevertheless continue to satisfy our theorem.
In particular, we shall find that four-dimensional theories with multiple decompactification limits will simultaneously satisfy {\it all}\/ of the different constraints that emerge from each individual  decompactification.   Moreover, while our discussion here will focus on the case of four-dimensional theories, there is nothing intrinsically special to four dimensions, and similar results  apply for theories in other spacetime dimensions as well, as long as such theories continue to have their own decompactification limits.

Along the way, we also prove another potentially important result. Specifically, we prove that the one-loop contributions to certain string amplitudes have a universal behavior in the limit of large compactification volume.
In particular, we define a modular-invariant compactification volume $\widetilde V_T$, and then demonstrate that all such amplitudes necessarily scale as
$(\pi/3)\widetilde V_T $ as $\widetilde V_T\to\infty$.

In this Introduction we have merely sought to describe 
our theorem and how it operates in a rough intuitive sense. Needless to say, there are many subtle details which we are omitting here.
These details will be discussed in subsequent sections. 
Moreover, as might be imagined, this paper is somewhat technical and relies on a number of results which were established in previous papers by the present authors and others.   We have therefore attempted to keep this paper entirely self-contained by including an initial section  in which we summarize those aspects of previous work which will ultimately be relevant for proving our theorem. 

Accordingly, this paper is organized as follows. 
First, in Sect.~\ref{sec:ingredients}, we assemble all of the conceptual and calculational ingredients that will ultimately be needed for proving our theorem.
Then, in Sect.~\ref{sec:theorem}, we discuss the fundamental clash that emerges between modular invariance and the process of decompactification, and explain how our theorem automatically resolves this clash.   Thus Sect.~\ref{sec:theorem} may be regarded as the central crux of this paper in which we present our theorem and discuss how it fits into (and emerges from) the larger theoretical framework.     
In Sect.~\ref{sec:implications}, we then proceed to discuss two of the most important implications of our theorem.   These include not only new supertrace constraints to which our theorem gives rise, but also tight restrictions on the running of couplings in these UV/IR-mixed theories.
Finally, for the experts, in Sect.~\ref{KDL} we provide an explicit example in which all of these results are illustrated through direct calculation.
We then conclude in Sect.~\ref{sec:conclusions}
with a discussion of some of the additional consequences of our theorem for low-energy string phenomenology.

 \section{Assembling the ingredients}
 \label{sec:string-comp}
\label{sec:ingredients}

In this section we collect the central ingredients that will be required in order to formulate and interpret our non-renormalization theorem.

\subsection{Operator insertions\label{subsec:operator_insertions}}

In general, a given closed string theory formulated in $D$ uncompactified spacetime dimensions will have a partition function $Z^{(D)}$ which is a function of a modular parameter $\tau\equiv \tau_1+i\tau_2$ with $\tau_i\in \IR$, 
and which can be written as a double power-series expansion in $q\equiv e^{2\pi i\tau}$ and $\overline{q}\equiv e^{-2\pi i \overline{\tau}}$ of 
 the form
\begin{equation}
Z^{(D)}(\tau) ~=~ \tau_2^k \, \sum_{m,n} \, a_{mn} \, 
\qb^m q^n~.
\label{eq:Z4}
\end{equation}
Here the summation is over all left-moving and right-moving worldsheet energy levels of the string, respectively denoted $(m,n)$, and $a_{mn}$ is the net (bosonic minus fermionic) number of degrees of freedom in the string spectrum  with worldsheet energies $(m,n)$.   Physical consistency of the partition function requires that it be modular invariant, \ie, that $Z(\tau+1)=Z(-1/\tau)=Z(\tau)$.  It is the latter invariance under $\tau\to -1/\tau$ which ties together the degeneracies of states $a_{mn}$ at all energies $(m,n)$ across the string spectrum, thereby connecting UV and IR physics in a highly non-trivial way.
The quantity $k$ is the {\it modular weight}\/  of the partition function, and for a string theory formulated in $D$ uncompactified spacetime dimensions generically has the value
\beq
 k~=~ 1-D/2~.
 \label{kD}
\eeq
We thus have $k= -1$ for $D=4$.
In general, the space-time  mass $M$ of any state with worldsheet energies $(m,n)$ is given by 
\begin{equation}
    M^2 ~=~ \frac{1}{2} (M_R^2 +M_L^2) ~=~ \frac{2}{\alpha'} (m+n)~,
    \label{eq:M2}
\end{equation}
where $m=\alpha' M_R^2/4$, 
$n=\alpha' M_L^2/4$,
and $\alpha'\equiv 1/M_s^2$,
where $M_s$ is the string scale.
The states with $m=n$ are considered ``on-shell'' or ``physical'' and can serve as in- and out-states,
while the ``off-shell'' states with $m\not=n$ are intrinsically stringy or ``unphysical'' and contribute only in loop amplitudes.   We shall assume  throughout this paper that we are dealing with {\it tachyon-free}\/ string theories (\ie, theories for which $a_{nn}=0$ for all $n<0$).  However, we shall not make any assumption that our theory exhibits spacetime supersymmetry.   Thus we will not assume that $a_{nn}=0$ for all non-negative values of $n$, or make any other similar assumptions regarding the vanishing of the $a_{mn}$ coefficients beyond our tachyon-free requirement that $a_{nn}=0$ for all $n<0$.
In this connection,
we note that no phenomenologically viable model can remain exactly supersymmetric.  By contrast, all string models must maintain an exact modular invariance as part of their internal self-consistency constraints.

In this paper, we consider the one-loop amplitudes associated with various physical quantities in four dimensions.   In particular, we focus on amplitudes in which there are no (or small) external momenta, or alternatively amplitudes in which such external momenta can be factored out from the one-loop integration.   This is a large and crucial class of amplitudes, and we shall see explicit examples below.
We shall let $\zeta$ generally denote the phenomenological quantities for which such amplitudes provide the one-loop contributions.   

In general we begin the calculation of such one-loop amplitudes by calculating the  modular-invariant tally of the contributions to $\zeta$ coming from each string state.   With the assumptions described above, this tally will take the form
\begin{equation}
Z_\calX^{(4)} ~=~ \tau_2^{-1} \sum_{n,m}  \, a_{mn} \, \calX_{mn} \,
\qb^m q^n~.
\end{equation}
This clearly resembles the partition function $Z^{(4)}$ but also includes a factor $\calX_{mn}$ which denotes the contribution to $\zeta$ from each $(m,n)$ state. The resulting one-loop contribution $\zeta$ is then given by 
\beq
 \zeta ~\equiv~ \langle \calX\rangle
\label{zetadef}
\eeq
where
\beq 
   \langle \calX \rangle~\equiv~ 
 \int_\calF \frac{d^2\tau}{\tau_2^2}\,Z^{(4)}_\calX~.
 \label{amplitudeA}
\eeq
Here $d^2\tau/\tau_2^2$ is the standard modular-invariant measure and 
$\calF$ is the fundamental domain of the modular group
\beq
\calF ~\equiv~ 
 \lbrace \tau: ~~
  |\tau_1|\leq 1/2\, , ~|\tau| \geq 1\,, \tau_2 >0\rbrace~
\label{Fdef}
\eeq
with $\tau_1\equiv {\rm Re}\,\tau$ and $\tau_2\equiv {\rm Im}\,\tau$ respectively.

In general,
these factors $\calX_{mn}$ are the eigenvalues of an {\it operator insertion}\/ $\calX$ into the partition function.
In general, there is a different operator insertion $\calX$ for each physical quantity $\zeta$.  In this paper we concentrate on operator insertions $\calX$ which take the form
\beq
   \calX~=~ \mathbbX_0 + \tau_2 \mathbbX_1 + \tau_2^2 \mathbbX_2
\label{X1X2presplit}
\eeq
where each $\mathbbX_\ell$ is $\tau$-independent.  
However it turns out that
the operator insertions for any physical quantity in four dimensions either take the form in Eq.~(\ref{X1X2presplit}) directly or can be reduced to it.   Thus, we may consider the operator-insertion form in Eq.~(\ref{X1X2presplit}) to be completely general.

In general, just like the partition function $Z^{(4)}$ itself, the resulting $\calX$-weighted spectral tally $Z_\calX^{(4)}$ must also be modular invariant.
This in turn implies that $\calX$ must be a modular-invariant operator insertion, which further implies that the $\mathbbX_\ell$-coefficients in Eq.~(\ref{X1X2presplit}) are tightly linked together by modular invariance.   Thus, knowledge of any one of these $\calX_\ell$-insertions permits the determination of the others through a process of  modular completion~\cite{Abel:2021tyt}.
In all cases, however, 
the requirements of modular invariance ensure that $\mathbbX_0$ can be at most proportional to the identity operator ${\bf 1}$\/. Thus the one-loop contribution to $\zeta$ from $\mathbbX_0$ will be proportional to the four-dimensional one-loop cosmological constant 
\beq
\Lambda ~\equiv ~ 
- \frac{\calM^4}{2}
\,\langle {\bf 1}\rangle ~=~
- \frac{\calM^4}{2}
\int_\calF \frac{d^2 \tau}{\tau_2^2} \,
Z^{(4)} ~
\label{lambdadef}
\eeq
where $\calM$ is the reduced string scale $M_s/(2\pi)$.

Later in Sect.~\ref{sec:implications} 
and Appendix~\ref{ap:entwined}
we will extend our analysis to certain cases in which the $\mathbbX_\ell$ can carry a holomorphic dependence on $\tau$.   We shall find, however, that such cases do not disturb our main results.

As we shall see, our theorem and its proof will not require any further details regarding these operator insertions $\mathbbX_\ell$.
However, it may be useful to recall what these insertions look like in various phenomenologically relevant cases.
As a first example, 
we may consider $\zeta$ to be the one-loop contribution to the mass of a scalar Higgs field in an arbitrary heterotic string model.
This calculation is
discussed in detail in
Ref.~\cite{Abel:2021tyt}, where it shown that the 
corresponding operator-insertion coefficients $\mathbbX_\ell$ turn out to be given by
\begin{align}
  \mathbbX_0 ~&\equiv ~ \, - \frac{\xi \calM^2 }{8\pi^2 }  \nonumber \\
  \mathbbX_1 ~&\equiv ~ \, \frac{\calM^2}{8\pi } \left. \partial_\phi^2 M^2\right|_{\phi=0}\nonumber \\
    \mathbbX_2 ~&\equiv ~ \, - \frac{1}{32\pi^2 } \left. (\partial_\phi M^2)^2\right|_{\phi=0}~
\label{XforHiggs}
\end{align}
where $M(\phi)$ describes the mass of a given  string state as a function of the fluctuation $\phi$
in the particular Higgs field in question. Thus, the functions $M(\phi)$ --- and whether they are zero or not --- essentially encode which Higgs field $\phi$ is under discussion. The parameter $\xi$ is a function of the shifts induced on the mass spectrum by the Higgs field.

To make these expressions for $\mathbbX_\ell$  more explicit, we may re-express them in terms of charge operators ${\bf Q}$ which fill out the 
so-called ``charge lattice'' associated with the string spectrum~\cite{Abel:2021tyt}.
In general, these charge operators take the form
${\bf Q} =({\bf Q}_L,{\bf Q}_R)$ where the `L' and `R' components correspond to the charges associated with our left-moving and right-moving worldsheet degrees of freedom.  For perturbative heterotic strings in four spacetime dimensions, these lattices have maximal dimensionalities $(22,10)$;
the left-moving charge components are generally associated with the corresponding gauge group while the 
right-moving components generally correspond to additional factors such as spacetime helicities.  
In general, these charge lattices are also {\it Lorentzian}\/, meaning that the scalar dot-product between two  charge operators is defined as 
$\bQ\cdot \bQ' \equiv \bQ_L \bQ'_L- \bQ_R \bQ'_R $.   In terms of these lattice operators, the action of a non-zero Higgs VEV  $\phi$ is to induce a shift in  the lattice of $\bQ$ charges~\cite{Abel:2021tyt}:
\begin{align}
     \bQ ~\to ~ \bQ + \frac{\phi}{M_s} \calT \cdot \bQ +
     \calO(\phi^2)~ 
\end{align}
where the ``response'' matrix $\calT$ can be decomposed in a  left-right block-diagonal fashion as
\begin{align}
    \calT ~&=~ \left(
\begin{array}{cc}
\calT_{LL} & \calT_{LR} \\
-\calT^t_{LR} & \calT_{RR} 
\end{array}
    \right)~.
\end{align}
This then yields alternative expressions for the $\mathbbX_1$ and $\mathbbX_2$ operators in Eq.~(\ref{XforHiggs}) as sums over charges:
\begin{align}
\mathbbX_1 &~=~ \frac{\calM^2}{2\pi} \left( \bQ\cdot \calS \cdot \bQ\right)  ~\nonumber \\
\mathbbX_2 &~=~ - \calM^2 \left( \bQ \cdot \calT \cdot \bQ\right)^2  ~
\end{align}
where we have defined
\begin{align}
    \calS ~&=~ \left(
\begin{array}{cc}
\calT_{LR}\calT^t_{LR} & -\calT_{LL}\calT_{LR} \\
\calT_{RR}\calT^t_{LR} & \calT^t_{LR}\calT_{LR} 
\end{array}
    \right)~.
\end{align}
Meanwhile $\mathbbX_0$ is given by~\cite{Abel:2021tyt}
\beq
\xi ~\equiv~ \frac{1}{2} {\rm Tr} \,\calS~,
\label{eq:xi}
\eeq
whereupon
\beq
  \mathbbX_0 ~\equiv ~  - \frac{ \calM^2 }{16\pi^2 }  \,
  {\rm Tr}\, {\cal S}~.
\eeq

In a similar fashion, we can also consider the case in which $\zeta$ is related to the one-loop contribution to the gauge coupling $g_G$ for any group factor $G$ in a given string model.  This case is discussed in detail in Ref.~\cite{Abel:2023hkk}.
To perform this calculation, we evaluate these couplings $g_G$ to one-loop order and then separate out the tree-level contributions.   In general, these  quantities are related through
\beq 
    \left. \frac{16\pi^2}{g_G^2} \right|_{\substack{{\rm total~thru}\\ {\rm one\hbox{-}loop}\\ {\rm order}}}
    ~=~
    \left. \frac{16\pi^2}{g_G^2} \right|_{\rm tree} ~+~\Delta_G
\label{oneloopcontribution}
\eeq
where in string theory we have
$g_G|_{\rm tree}\sim e^{-\langle \phi\rangle}$ with $\langle \phi\rangle$ denoting the VEV of the dilaton $\phi$ 
and where $\Delta_G$ denotes the one-loop contribution to $16\pi^2/g_G^2$.
We may thus now take $\zeta$ to be the one-loop contribution $\Delta_G$,
whereupon the corresponding operator insertions are given by~\cite{Abel:2023hkk}
\beqn
\mathbbX_0 ~&=&~ 0 \nonumber\\
\mathbbX_1 ~&\equiv&~ \frac{\xi}{2\pi} 
 \left( \overline{Q}_H^2  -\frac{\overline{E}_2}{12} \right) \nonumber\\
\mathbbX_2 ~&\equiv&\, -2 \left( \overline{Q}_H^2  -\frac{\overline{E}_2}{12} \right) Q_G^2~
\label{eq:Xs}
\eeqn 
where $\overline{Q}_H$ is the (right-moving) spacetime helicity operator (a specific component of ${\bf Q}_R$) and where $Q_G^2$ is the quadratic Casimir of $G$ (comprised out of components of ${\bf Q}_L$). 
In Eq.~(\ref{eq:Xs}), the quantity $\overline{E}_2$ is the anti-holomorphic Eisenstein function which is defined in Eq.~\eqref{defG2here}.   As discussed above, this case therefore provides an example in which the $\mathbbX_\ell$ are
not $\tau$-independent but rather carry a holomorphic $\tau$-dependence.  Such cases will be considered in Sect.~\ref{sec:implications}, but we shall find that they will not induce significant departures from the main results we shall be presenting.

\subsection{Divergences and regulator function\label{sec:regulator}}

In general, with operator insertions $\calX$ of the form in Eq.~(\ref{X1X2presplit}), it is possible that the four-dimensional amplitude in Eq.~(\ref{amplitudeA})
experiences a logarithmic divergence. 
Indeed, such a divergence will arise in four-dimensional theories if
\beq
  \zStr\,\mathbbX_2 ~\not=~0~
\label{eq:finite4D}
\eeq
where $\zStr$ denotes a supertrace over only the massless states.
Indeed, given that our operator insertions generally take the form in Eq.~(\ref{X1X2presplit}),
this is the most severe divergence that can arise.

In such cases, we must regulate the amplitude without disturbing its modular invariance. 
Following Refs.~\cite{Abel:2021tyt,Abel:2023hkk} we will carry out this procedure by multiplying the
 integrand of Eq.~\eqref{amplitudeA} by 
 a suitable modular-invariant regulator function $  \calG(a,\tau)$
 where $\tau$ is the one-loop modular parameter and $a$ schematically represents other possible parameters within this function.
 In order to serve its purpose as a regular, such a function must vanish more rapidly than logarithmically  as $\tau\to i\infty$.
We also demand that
$ \calG\approx 1$ elsewhere within the fundamental domain, so that this regulator does not significantly disturb our theory within regions of integration that do not lead to divergences. 

An explicit regulator function satisfying these criteria was given in  Ref.~\cite{Abel:2021tyt}. 
However, given that the specific form of this function will not be needed for any of our main arguments, 
we shall defer discussion of this function  until Sect.~\ref{sec:implications}, when we work out a specific example of our results.

Thus, our procedure for regulating a divergent one-loop string amplitude amounts to deforming the amplitude according to our regulator function $\calG$:
\beq
\langle \calX\rangle ~\to~
 \langle \calX\rangle_{\calG}
 ~\equiv~ 
   \langle 
   \calX 
   \,\calG\rangle~.
\label{eq:Ctilde}
\eeq
As we shall see, there can also be other reasons for introducing such a regulator function, even for amplitudes that are in principle finite.  

\subsection{Rankin-Selberg transform and supertraces over
physical string states}
\label{subsec:supertraces}

In general, for arbitrary insertion $\calX$ and arbitrary dimension $D$, the one-loop amplitude $\langle \calX\rangle$ is given by the $D$-dimensional version of 
Eq.~\eqref{amplitudeA}, namely
\beq
\langle \calX \rangle~\equiv~
\int_\calF \frac{d^2 \tau}{\tau_2^2} \,
Z^{(D)}_\calX ~.
\label{amplitudeAD}
\eeq
We thus see that string states with
all allowed combinations of worldsheet energies $(m,n)$ contribute.   This is true not only for the on-shell states with $m=n$ but also the off-shell states with $m\not=n$;  indeed, while the former contribute at all values of $\tau_2$ within the fundamental domain in Eq.~(\ref{Fdef}), the latter also contribute, albeit within only the $\tau_2<1$ region.
These contributions can nevertheless be sizable.

It turns out that this amplitude may be expressed in another form which depends only on the {\it on-shell}\/ states with $m=n$.
This alternate form will be critical for our eventual theorem, and exists for all dimensions $D$ and
for all situations in which the amplitude $\langle \calX\rangle$ is finite. Specifically, as long as the amplitude $\langle \calX\rangle$ in Eq.~\eqref{amplitudeAD} is finite and modular invariant, 
a powerful result in modular-function theory  due to Rankin~\cite{rankin1a,rankin1b} and Selberg~\cite{selberg1} allows us to re-express this amplitude as
\beq
\langle \calX \rangle ~=~ \frac{\pi}{3}\,\, \oneRes\,\, \int_0^\infty 
    d\tau_2 \,\, \tau_2^{s-2}\,\, g(\tau_2)
\label{RSdeff2}
\eeq 
where
\beqn
 g(\tau_2) ~&\equiv&~ 
 \int_{-1/2}^{1/2}  d\tau_1\,
 Z_\calX^{(D)} (\tau)\nonumber\\
 ~&\equiv&~ \tau_2^{k} \,\int_{-1/2}^{1/2} d\tau_1 \, \sum_{m,n}  \,a_{mn} \,\calX_{mn} \, \qbar^m q^n \nonumber\\
 &=&~  \tau_2^{k} \,\sum_{n} \, a_{nn}\, \calX_{nn} \,
         e^{-\pi \alpha' M_n^2 \tau_2}~
 \label{RSdeff3}
\eeqn 
with $\alpha' M_n^2=4 n$, as in Eq.~\eqref{eq:M2}.  
This tells us that the original string amplitude $\langle\calX\rangle$ is nothing but the Mellin transform
of $g(\tau_2)/\tau_2$.  This in turn allows us to write $g(\tau_2)$ directly as the inverse Mellin transform of the amplitude, which yields the alternative result~\cite{zag,Kutasov:1990sv}
\beq
\langle \calX \rangle ~=~ \frac{\pi}{3}\, \lim_{\tau_2\to 0}
 \, g(\tau_2)~,
\label{RSdeff2alt}
\eeq   
where $g(\tau_2)$ continues to be given by Eq.~(\ref{RSdeff3}). This result 
is equivalent to that in Eq.~(\ref{RSdeff2}), but has the primary advantage that we can now evaluate $\langle\calX\rangle$ simply by taking the $\tau_2\to 0$ limit of $g(\tau_2)$ rather than by evaluating the residue of the $\tau_2$-integral of $g(\tau_2)$.  Indeed, inserting Eq.~(\ref{RSdeff3}) into 
Eq.~(\ref{RSdeff2alt}) yields
\beq
\langle \calX \rangle ~=~ \frac{\pi}{3} \lim_{\tau_2\to 0} 
 \tau_2^{k} 
 \sum_n a_{nn} \calX_{nn} \,e^{-\pi \alpha' M_n^2 \tau_2}~.
 \label{RSdeff2alt2}
\eeq
This, then, expresses the amplitude $\langle \calX\rangle$ in terms of the degeneracies $a_{nn}$ of only the {\it physical}\/ string states.

The issue then boils down to the evaluation of the right side of Eq.~(\ref{RSdeff2alt2}).
Following Ref.~\cite{Abel:2023hkk}, we shall do this by first defining the sum 
\beq
S^{(D)}(\tau_2)  ~\equiv~
\tau_2^{-k} g(\tau_2) ~=~
\sum_{n} a_{nn} \,\calX_{nn}\,
      e^{-\pi \alpha' M_n^2 \tau_2}~.
\label{SSdef}
\eeq
Thus $S^{(D)}$ encapsulates only that part of the $\tau_2$-dependence of $g(\tau_2)$ that comes from the $\calX$-weighted sums over the string states.
We can then expand $S^{(D)}(\tau_2)$ as a power series in $\tau_2$, \ie, 
\beq
S^{(D)}(\tau_2) ~\sim~ \sum_{j} \, C_j \tau_2^j ~~~{\rm as}~~\tau_2\to 0~.
\label{assumedtau2dep}
\eeq
Note that for complete generality we will not assume that only integer values of $j$ contribute in Eq.~(\ref{assumedtau2dep});  indeed  subleading contributions can generically also have fractional $j$. Inserting Eq.~(\ref{assumedtau2dep}) into
Eq.~(\ref{RSdeff2alt}),
we thus have
\beq
 \langle \calX\rangle~=~ \frac{\pi}{3}\, 
  \lim_{\tau_2\to 0} \, \sum_j \,
 C_j \, \tau_2^{j+k}  ~.
 \label{intresult}
\eeq

It is not difficult to determine the values of the $C$-coefficients for integer $j$.
Indeed, following Ref.~\cite{Dienes:1995pm},
we may ``invert''
Eq.~(\ref{assumedtau2dep})
by taking 
$\tau_2$-derivatives of both sides. In this way  we find 
\beq
C_0 ~=~ \lim_{\tau_2\to 0} \left[  \sum_{n} a_{nn} \calX_{nn} \,
      e^{-\pi \alpha' M_n^2 \tau_2} \right] 
\eeq
and
\beqn
C_1 &=&
\lim_{\tau_2\to 0} \left\lbrace 
\sum_n a_{nn} \left\lbrack \frac{d}{d\tau_2}
 \calX_{nn} \right\rbrack
      \,e^{-\pi \alpha' M_n^2 \tau_2}\right\rbrace \nonumber\\
&& ~~~~~
- \lim_{\tau_2\to 0} \left[  \sum_{n} a_{nn} \calX_{nn} \,(\pi \alpha' M_n^2)\,
      e^{-\pi \alpha' M_n^2 \tau_2}
      \right] \nonumber\\
&=& \, 
\lim_{\tau_2\to 0} \left\lbrack
\sum_n a_{nn} \left( \frac{d}{d\tau_2} -\pi \alpha' M_n^2\right)
 \calX_{nn}
      \,e^{-\pi \alpha' M_n^2 \tau_2}\right\rbrack \nonumber\\
&=& \, 
\lim_{\tau_2\to 0} \left\lbrack
\sum_n a_{nn} \left( D_{\tau_2} 
 \calX_{nn}\right)
      \,e^{-\pi \alpha' M_n^2 \tau_2}\right\rbrack
\label{C1expression}
\eeqn
where $D_{\tau_2}$ is the modular-covariant derivative
\beqn 
D_{\tau_2}~&\equiv&~ \frac{d}{d\tau_2} 
       -  \pi \alpha'M^2 ~\nonumber\\
&=&~ \frac{d}{d\tau_2} 
       -  \frac{M^2}{4\pi \calM^2} ~.
\label{RSderivative}
\eeqn
Coefficients $C_j$ with integer $j\geq 2$ can be calculated in a similar fashion by taking additional $\tau_2$-derivatives, yielding the general result
\beq 
    C_j~=~ 
\frac{1}{j!} \lim_{\tau_2\to 0} \left[  \sum_{n} a_{nn} \left( D_{\tau_2}^j \calX_{nn}\right)  e^{-\pi \alpha' M_n^2 \tau_2} \right] ~.
\eeq

These results may be further simplified by defining the regulated supertrace~\cite{Dienes:1995pm}
\beq
  \Str \, \calA ~\equiv~   
       \lim_{\tau_2\to 0} \, \sum_n a_{nn} \,\calA_{nn} \,e^{- \pi \alpha' M_n^2 \tau_2}~.
\label{supertrace_regulated}
\eeq
Given that the $a_{nn}$ coefficients tally the net number of bosonic minus fermionic degrees of freedom with left- and right-moving worldsheet energies $n$, such supertraces are completely analogous to the standard spectral supertraces $\Str\,A\equiv \sum_n (-1)^F A_n$ that we would have in ordinary quantum field theory except that they yield finite results even for  UV/IR-mixed theories which contain infinite towers of states~\cite{Dienes:1995pm}. 
Indeed, even in such cases the summation in Eq.~(\ref{supertrace_regulated}) is finite thanks to the exponential damping factor $e^{-\pi \alpha' M_n^2 \tau_2}$, and remains finite even as $\tau_2\to 0$ and this damping factor is removed.
We shall therefore adopt this supertrace definition in what follows. Expressed in terms of these supertraces,
we then find that our $C_j$-coefficients with integer $j$ all take the relatively simple form
\beq
   C_j ~=~ \frac{1}{j!}\, \Str\, D_{\tau_2}^j \calX~~~~ {\rm for~all}~~ j\geq 0~.
\label{Cs-as-supertraces}
\eeq

Before proceeding further, we emphasize that the above derivation leading to the
supertrace expression for the $C$-coefficients in Eq.~(\ref{Cs-as-supertraces}) implicitly rested on an understanding that 
$\Str\,A=0$ if $A$ itself contains an uncancelled positive power of $\tau_2$.
This follows formally from the fact that the definition of the supertrace  in Eq.~(\ref{supertrace_regulated}) includes a limit taking $\tau_2\to 0$.  It may indeed seem somewhat unorthodox to have $\tau_2$ appear not only within the argument of the supertrace but also as the supertrace regulator, but expressions such as that in Eq.~\ref{Cs-as-supertraces}) have a clear operational definition and will cause no difficulty.  Thus, for example, if $\calX$ takes the form in Eq.~(\ref{X1X2presplit}) with $\tau_2$-independent coefficients $\mathbbX_\ell$, then
\beqn
  \Str\, \calX ~&=&~
        \Str\, \mathbbX_0~\nonumber\\
  \Str\, \frac{d}{d\tau_2} 
        \calX ~&=&~ \Str\, \mathbbX_1 \nonumber\\
  \Str \, \frac{d^2}{d\tau_2^2} \calX ~&=&~
     2\,\Str \,\mathbbX_2~.
\label{tobereferredto}
\eeqn 

The result in Eq.~(\ref{intresult}) enables us to express our string amplitude $\langle \calX\rangle$ in terms of the $C$-coefficients.  For example, taking 
$k= -1$ (as appropriate for string theories in four dimensions) and recalling that $\langle \calX\rangle$ is finite, 
we find 
\beq
 \langle \calX\rangle~=~ \frac{\pi}{3} \, C_1~.
 \label{amplitudeC1}
 \eeq
 However, this result assumes that we have also imposed the auxiliary conditions 
 \beq
     C_j ~=~0 ~~~~{\rm for~all}~~j <1~.
\eeq
In particular, from this we learn that
\beq  
     C_0~=~0~.
\label{eq:C0=0}
\eeq

The result in Eq.~(\ref{amplitudeC1}) allows us to express our string amplitude $\langle \calX\rangle$ in terms of supertraces over only the physical string states.
Indeed, combining Eqs.~(\ref{C1expression}) and (\ref{amplitudeC1}) we have
\beqn
 \langle \calX \rangle ~&=&~ \frac{\pi}{3} \,
\Str \left(
 \frac{d}{d\tau_2} \calX \right)
 -  \frac{\pi}{3} \,
 \Str \left[ \calX \,(\pi \alpha' M^2) \right] ~\nonumber\\
&=&~ \frac{\pi}{3} \,
\Str \left( D_{\tau_2} \calX \right)~.
\label{finalRSresult}
\eeqn
Likewise, our auxiliary condition
in Eq.~(\ref{eq:C0=0})
now takes the form
\beq
       \Str \, \calX~=~0~.
\label{singlesupertrace}
\eeq
Note that these results apply for any modular-invariant operator insertion $\calX$ in four dimensions so long as this insertion results in a finite amplitude $\langle \calX\rangle$.

Finally, we note that we may occasionally be required to evaluate supertraces of quantities --- such as the
 $\mathbbX_\ell$ in Eq.~(\ref{eq:Xs}) --- which involve the Eisenstein function $E_2(\tau)$ defined in
 Eq.~\eqref{defG2here}.   The presence of such a function introduces a number of subtleties into the procedure for evaluating supertraces.    These subtleties are fully described in Ref.~\cite{Abel:2023hkk} and summarized in Appendix~\ref{ap:entwined}, with the end result that the usual notion of supertrace is generalized in a straightforward manner.  

The results in Eqs.~(\ref{amplitudeC1}) and (\ref{eq:C0=0}) were derived for $k= -1$, as appropriate for four-dimensional theories.   However, these results can be easily generalized to higher dimensions $D$.   Indeed, given 
the relation in Eq.~(\ref{kD}), we obtain
\beqn 
&& \phantom{1}D=6:~~ \begin{cases}         
               &\,\,\,C_0 ~=~ C_1~=~0 \\
           &\langle \calX\rangle ~=~ \displaystyle{\frac{\pi}{3}} \,C_2~
          \end{cases}\nonumber\\
&& \phantom{1}D=8:~~ \begin{cases}         
               &\,\,\,C_0 ~=~ C_1~=~C_2~=~0 ~~~~~~~\\  
           &\langle \calX\rangle ~=~ \displaystyle{\frac{\pi}{3}} \,C_3~
          \end{cases}\nonumber\\
&& D=10:~~ \begin{cases}         
               &\,\,\,C_0 ~=~ C_1~=~C_2~=~C_3~=~0~~~~ \\
           &\langle \calX\rangle ~=~ \displaystyle{\frac{\pi}{3}} \,C_4~.
          \end{cases}~\nonumber\\
\label{genRSresults}
\eeqn
We thus observe, as discussed in the Introduction,  that theories in higher dimensions exhibit more internal cancellation constraints than do theories in lower dimensions.  

 Note that for convenience we shall restrict our attention in this paper to spacetime dimensionalities which are even, with $D\in 2\mathbbZ$.  These are the dimensionalities for which the modular weight $k$ is an integer, and in which chiral theories can exist.  Similar results also arise for odd $D$, but with additional complications that obscure the underlying physics.   We shall therefore focus on theories with even $D$ in what follows. 

\section{The theorem}
\label{sixd}
\label{sec:theorem}

Having assembled the ingredients that will be needed for our theorem, we now turn our attention to the theorem itself.   Our theorem ultimately rests on modular invariance and misaligned supersymmetry, as do most of the results quoted in Sect.~\ref{sec:ingredients}.~
As we shall see,  there is a deep clash between
 modular invariance and the process of decompactification.  This clash is intrinsic to the UV/IR mixing inherent in string theory, and does not exist in ordinary quantum field theory.    Our theorem emerges as the result of this clash, and ultimately provides the means by which these two features
can be reconciled.

\subsection{The fundamental clash between decompactification and modular invariance}

\label{subsec:clash}

Let us begin by examining the properties of modular-invariant four-dimensional theories in the  presence of a large-volume $\delta$-dimensional compactification.  
In other words, we shall consider a $(4+\delta)$-dimensional modular-invariant theory compactified on a manifold of the form $\calM_4 \times \calK_\delta$ where $\calM_4$ is ordinary uncompactified Minkowski space and where $\calK_\delta$
is our compactified $\delta$-dimensional space whose characteristic dimensions we shall consider to be large, with a corresponding $\delta$-dimensional volume $V_\delta \gg M_s^{-\delta}$ where $M_s$ is the string scale.
Our goal is to study how the resulting theory evolves as we take $V_\delta\to \infty$. 

For simplicity, we shall start by considering {\it untwisted}\/ compactifications, temporarily deferring our analysis of situations with twisted compactifications (such as arise in orbifold compactifications) to Sect.~\ref{twisted}.~
We also remark that in the case of four-dimensional closed strings, we would generally be compactifying from ten dimensions to four dimensions.  There are therefore a total of six compactified dimensions, and we choose $\delta$ to represent the number of such dimensions whose characteristic sizes we wish to consider growing increasingly large.
Thus $0\leq \delta\leq 6$.  Moreover, in keeping with the observation at the end of the previous section, we shall focus on the cases with $\delta \in 2\mathbbZ>0$.  

Within such theories, we shall concentrate on physical quantities $\zeta$ for which the corresponding one-loop contributions $\langle \calX\rangle$ are finite for all $V_\delta$.
In four spacetime dimensions, $\langle \calX\rangle$ can generally have at most a logarithmic divergence. From direct inspection of Eq.~(\ref{amplitudeA}), and as noted in Eq.~(\ref{eq:finite4D}), 
we see that such a logarithmic divergence is proportional to $\zStr \mathbbX_2$ where the supertrace is restricted to the massless 
$\mathbbX_2$-charged states in the string spectrum.  Thus, we shall concentrate on quantities $\zeta$ for which  
\beq
\zStr\,\mathbbX_2 ~=~ 0~.
\label{finite}
\eeq
Of course, for physical quantities $\zeta$ for which this condition is not satisfied, it would be necessary to introduce a regulator function $  \calG$, as described in Sect.~\ref{sec:regulator}.~   This would introduce several subtleties into the proof of our main theorem but will not alter the main result.

Given that we are temporarily focusing on untwisted compactifications, the $\calX$-inserted partition function  $Z^{(4)}_\calX$ of our compactified four-dimensional theory can be factorized, 
\ie,
\beq
 Z^{(4)}_\calX ~=~  Z^{\rm (base)}_\calX \,\cdot\, Z_{\rm KK/winding}~,
 \label{Zorig}
\eeq    
where $Z_{\rm KK/winding}$ is the trace over all of the Kaluza-Klein (KK) and winding modes associated with the $\delta$ large compactified extra dimensions and where the ``base'' partition function 
$Z^{\rm (base)}_\calX$ can be written as 
\beq
      Z^{\rm (base)}_\calX   ~=~ \tau_2^{-1}
      \sideset{}{'}\sum_{mn}
      a_{mn} \,\calX_{mn} \,\qbar^m q^n~
\label{Zint}
\eeq
where $\sum'_{mn}$ indicates a sum over the states {\it excluding}\/ the KK and winding modes associated with the $\delta$ large dimensions.
In other words, our original theory can be viewed 
as a “base” theory tensored with a cloud of KK/winding states stemming from the compactification, with each state in the base theory accruing the same set of KK/winding excitations.

In general, $Z^{\rm (base)}_\calX$ contains the information regarding our specific theory independently of the compactification.   This is thus the portion of the original $\calX$-inserted partition function that depends on the particular operator insertion $\calX$ but which is independent of the compactification volume $V_\delta$.   By contrast, $Z_{\rm KK/winding}$ contains the information regarding the specific geometry associated with the large compactified dimensions.   As such, $Z_{\rm KK/winding}$ is independent of $\calX$ but depends on $V_\delta$.
   For example, if we specialize to $\delta=1$ and define the dimensionless radius $\RR\equiv M_s R=R/\sqrt{\alpha'}$, we have
\beqn
 Z_{\rm KK/winding}~&=&~
         \sum_{\mm,\nn} \,  \qbar^{ (\mm/\RR - \nn \RR)^2/4}
   \,  q^{ (\mm/\RR + \nn \RR)^2/4}~\nonumber\\
   ~&=&~
   \sum_{\mm,\nn} \,e^{ -\pi \tau_2 \left( 
   \mm^2/\RR^2 +\nn^2 \RR^2\right) } \,
    e^{2\pi i \tau_1 \mm\nn}~\nonumber\\
  &=&~ Z_{\rm circ}/\sqrt{\tau_2}
\label{KKwindingdef}
\eeqn
where $\mm$ and $\nn$ respectively index the KK (momentum) and winding modes associated with this large extra dimension and where $Z_{\rm circ}$ is the modular-invariant circle partition function
\beq
     Z_{\rm circ}(\tilde R,\tau)~\equiv~  
    \sqrt{ \tau_2}\, \sum_{\mm,\nn \in\mathbb{Z}} \,
   \overline{q}^{(\tilde m /\tilde R-\tilde n \tilde R)^2/4}  \,q^{(\tilde m \tilde R+\tilde n /\tilde R)^2/4}
   ~.
\label{Zcircdef}
\eeq  
In this case we thus see from the middle line of Eq.~(\ref{KKwindingdef})
that $Z_{\rm KK/winding}$ traces over KK/winding states with masses
\beq
        M_{\mm,\nn}^2 ~=~ 
         \frac{\mm^2}{R^2} + 
            \nn^2 M_s^4 R^2~,
\label{KKwmasses}
\eeq
as required. 
Likewise, for a $\delta$-dimensional (square) toroidal compactification we may take $Z_{\rm KK/winding}=  (Z_{\rm circ}/\sqrt{\tau_2})^\delta$.

Even though the masses of the KK states in Eq.~(\ref{KKwmasses}) are the same as we would expect for a five-dimensional field theory compactified on a circle, 
this summation also includes the contributions from winding modes and is thereby modular invariant, so that the full $\calX$-inserted partition function $Z_\calX^{(4)}$ in Eq.~(\ref{Zorig}) is modular invariant.
Indeed, even though the individual factors in Eq.~(\ref{Zorig}) are not separately modular invariant, we may reshuffle factors of $\tau_2$ in order to write
\beq
 Z^{(4)}_\calX ~=~  \left(  \tau_2^{-\delta/2}\, Z^{\rm (base)}_\calX \right) \,\cdot\,  \left( \tau_2^{\delta/2} \,Z_{\rm KK/winding}\right)~,
\label{Zorigrefactored}
\eeq    
where now each factor is individually modular invariant.
For compactifications of the sort we are discussing,
a modular-invariant reshuffling such as that in Eq.~(\ref{Zorigrefactored}) is completely general, independent of the spacetime geometry. 
Indeed, for a square $\delta$-dimensional toroidal compactification, the final factor is nothing but the modular-invariant sum $\left(Z_{\rm circ}\right)^\delta$.  

Let us now ask what happens
as $V_\delta\to \infty$.
It is once again easiest to focus on the case of square toroidal compactifications as a guide 
and ask what happens when the radius
associated with these $\delta$ dimensions becomes large, with $R^{-1}\ll M_s$ or equivalently $\RR\gg 1$.   In the $M_s R\to\infty$ limit we can disregard all excited winding-mode states with $\nn \not=0$, as
the masses of these states become infinitely great. Likewise, the KK masses become essentially continuous.  In this limit we can then evaluate $Z_{\rm KK/winding}$,
obtaining
\beqn
Z_{\rm KK/winding} ~&\approx& ~
   \sum_\mm \, e^{-\pi \tau_2 \mm^2/\RR^2}
   \nonumber \\ 
   ~& = &~\frac{M_sR}{\sqrt{\tau_2}}\, \sum_{\ell\in\mathbbZ} e^{-\pi \ell^2 (M_s R)^2 /\tau_2} ~\nonumber\\
   ~&\approx& ~
\frac{M_s R}{\sqrt{\tau_2}}~.
\label{KKlimit}
\eeqn
Note that in passing from the first line of Eq.~(\ref{KKlimit}) we have employed an exact Poisson resummation, while the passage to the third line
then follows by taking the $R\to \infty$ limit.   Of course, we could have obtained the same results by approximating the sum in the first line as an integral, which would have led to third line directly. 

Likewise, for $\delta$ orthogonal dimensions of radius $R$, we obtain
\beq
Z_{\rm KK/winding}  ~\approx~ \frac{M_s^\delta R^\delta}{\tau_2^{\delta/2} }
  ~=~ 
  \frac{\calM^\delta V_\delta}{\tau_2^{\delta/2} }
  \label{MVexpression}
\eeq
where $\calM\equiv M_s/(2\pi)$ is the reduced string scale and where $V_\delta = (2\pi R)^\delta$ is the compactification volume.  
Once again, we note that the final expression in Eq.~(\ref{MVexpression}) is completely general, holding independently of the (factorized) compactification geometry.

However, partition functions of compactified string theories in different spacetime dimensions are generally related via
\beq
Z^{(D+\delta)} ~\equiv~
 \lim_{V_\delta\to \infty} 
  \left[ \frac{1}{\calM^\delta V_\delta} \, Z^{(D)} \right]
\label{smoothness}
\eeq
where the theory corresponding to $Z^{(D)}$ has $\delta$ large compactification radii with compactification volume $V_\delta$.
Indeed, as $V_\delta\to\infty$, the partition function $Z^{(D)}$ develops a divergence which scales as the volume $V_\delta$ of compactification;  dividing out by this volume as in Eq.~(\ref{smoothness}) then yields the finite higher-dimensional partition function $Z^{(D+\delta)}$.
Putting the pieces together, we therefore find that 
\beq
Z_\calX^{(4+\delta)} ~=~
   \lim_{V_\delta\to \infty} 
   \left[ \frac{1}{\calM^\delta V_\delta} \, Z_\calX^{(4)} \right]
~=~ \tau_2^{-\delta/2}\, Z^{\rm (base)}_\calX~
\label{Zintnew}
\eeq    
where $Z^{\rm (base)}_\calX$ is given in Eq.~(\ref{Zint}). 
This observation allows us to identify the ``base'' factor within our four-dimensional theory in terms of the higher-dimensional theory:
\beq
   Z_\calX^{\rm (base)} ~=~ 
      \tau_2^{\delta/2}\, Z_\calX^{(4+\delta)}~.
\label{baseequalshigherD}
\eeq
Note that both sides of this relation are indeed $V_\delta$-independent.

As we have seen, 
Eq.~(\ref{smoothness}) relates modular-invariant theories in different dimensions.
We shall refer to this equation as a ``smoothness'' constraint because it ensures that the four-dimensional partition function smoothly becomes a $(4+\delta)$-dimensional partition function in the $V_\delta\to\infty$ limit. In this context, we note 
that Eq.~(\ref{smoothness}) directly implies a similar smoothness relation for the corresponding one-loop {\it amplitudes}\/.  Indeed, defining
\beq
 \langle \calX\rangle^{(4+\delta)} ~\equiv~
 \int_\calF \frac{d^2\tau}{\tau_2^2}\,
  Z_\calX^{(4+\delta)}~,
\label{higherdimamp}
\eeq
we have
\beq
   \langle \calX \rangle^{(4+\delta)} ~=~ 
      \lim_{V_\delta\to \infty}\, \frac{1}{\calM^\delta V_\delta} \,\langle\calX\rangle^{(4)}~.
\label{amprelation0}
\eeq
However, while Eq.~(\ref{smoothness}) implies Eq.~(\ref{amprelation0}), the converse is not true.  Indeed, while two equal partition functions lead to identical amplitudes, identical amplitudes only imply equality of the corresponding partition functions modulo functions $f(\tau,\taubar)$ whose $\calF$-integrals vanish.  Indeed, many such non-zero functions $f(\tau,\taubar)$ with vanishing $\calF$-integrals are known to exist~\cite{Moore:1987ue,Dienes:1990ij,Dienes:1990qh}.   
We also note that $\langle \calX\rangle^{(4)}$ scales as the compactification volume $V_\delta$ as $V_\delta\to\infty$;  dividing out by $V_\delta$ as in Eq.~(\ref{amprelation0}) then leads to a finite higher-dimensional amplitude $\langle \calX\rangle^{(4+\delta)}$.
In this connection we note that this divergence of $\langle \calX\rangle^{(4)}$ as $V_\delta\to\infty$ is associated with a mere overall multiplicative factor.   In particular,  it
is {\it not}\/ associated with the modular integration of $Z^{(4)}$ over the fundamental domain $\calF$ (such as might arise  due to certain massless or tachyonic states).  

The (re-)emergence of a higher-dimensional theory in the large-volume limit is certainly not a surprise.
Indeed, geometric decompactification is an intrinsically smooth and continuous process.
{\it However, the extra factor $\tau_2^{-\delta/2}$ which appears in Eq.~(\ref{Zintnew})
is of critical importance}\/.
This extra factor indicates that if $Z_\calX^{\rm (base)}$ has a $\tau_2$-dependent prefactor $\tau_2^{-1}$, as appropriate for a four-dimensional theory, then $Z_\calX^{(4+\delta)}$ has a $\tau_2$-dependent prefactor  $\tau_2^k$ where $k=1-D/2$ with $D=4+\delta$.
The appearance of the new factor $\tau_2^{-\delta/2}$ in Eq.~(\ref{Zintnew})
thus reflects the change in dimensionality of any modular-invariant theory when an extra uncompactified spacetime dimension comes into existence.

{\it It is here that we witness
the fundamental clash between the smoothness of the decompactification process and the discrete integer nature of the number of uncompactified spacetime dimensions}\/
(or equivalently the half-integer nature of the modular weight $k\in \IZ/2$).
Indeed, while the $V_\delta\to \infty$ limit is essentially a smooth one as far as the resulting physics is concerned, the powers of $\tau_2$ change in this limit in what is ultimately a discontinuous way according to Eq.~(\ref{kD}).

It is important to understand the nature of this discontinuity.
Toward this end, let us revisit Eq.~(\ref{eq:Z4}).   We may regard the form of this expression as the``canonical'' form for a partition function.   Indeed, this form consists of a discrete double power series in $q$ and $\qbar$, where $q\equiv e^{2\pi i\tau}$, along with an overall factor of $\tau_2$ raised to a certain power $k$.   The canonical form 
of the partition function is of utmost importance because it is only in this form that one can read off a value of $k$ which can be interpreted as a modular weight --- indeed, the same modular weight $k$ which appears throughout the Rankin-Selberg procedure.
In other words, it is only when we cast our partition function into the canonical form that we expose the true modular weight $k$ of our theory.

Such a partition function can also depend on a compactification radius $R$, which is a continuous variable.   As we have stated above, the underlying physics of our theory must have a smooth $R\to \infty$ limit.  Indeed, for every value of $R$ (including infinity), it is possible to recast our partition function into the canonical form in Eq.~(\ref{eq:Z4});  moreover, for every {\it finite}\/ value of $R$, the value of $k$ that appears in the canonical form remains the same
(equalling $-1$ for four-dimensional theories).   {\it However, in the $R\to\infty$ limit, the value of $k$ 
that appears in the canonical form jumps to a new value}.
For example, in the case of a four-dimensional theory with a single extra dimension, we now have $k= -3/2$ in the $R\to \infty$ limit, consistent with Eq.~(\ref{kD}).   This is the ``discontinuity'' to which we are referring.  

We stress that the underlying physics is not discontinuous in this limit;  {\it it is merely the passage to the canonical form comprising a discrete power double power series that becomes discontinuous}\/.
Indeed, this discontinuity arises from the fact that
in the infinite-radius limit 
$Z_{\rm KK/winding}$ 
can no longer be written in the same canonical form as for finite radius.  Instead, what happens at infinite radius is that our discrete spectrum becomes continuous.  As a result, in this limit, $Z_{\rm KK/winding}$ takes the form of a divergent volume factor multiplying a volume-independent expression.  However,  this volume-independent expression is in the canonical form, but now with a different value of $k$.   The passage to the higher-dimensional theory as in Eq.~(\ref{smoothness}) then eliminates this volume factor, but leaves us with a new canonical form with a new value of $k$.

This, then, is the fundamental clash between modular invariance and the process of decompactification.  We know that the process of decompactification must ultimately be smooth, even in the decompactification limit.  On the other hand, the value of $k$ within the canonical form changes in a discontinuous way in the decompactification limit --- with an extra factor of $\tau_2^{-\delta/2}$ appearing in Eq.~(\ref{Zintnew}) ---
and we know that $k$ is a quantity which is absolutely fundamental in describing the modular properties of our theory.  
How then can these two features be reconciled?

 Before proceeding further, we note that this is not the first time such clashes have arisen within string theory, or even simply within conformal field theory.   For example, let us consider the case of a boson compactified on a circle of radius $R$, as discussed in Ref.~\cite{Kani:1989im}.   If $R$ is rational, it can be expressed in lowest form as $p/q$ for some integers $(p,q)$, and the resulting decomposition of the partition function into left- and right-moving CFT characters depends critically on the values of $p$ and $q$.
Thus, as we sweep through rational values of $R$, it would seem that the corresponding partition functions --- and therefore the properties of the resulting CFTs --- will vary hugely and discontinuously.   The existence of irrationals amongst the rationals only introduces further potential discontinuities into the mix.   Yet we know that the physics must ultimately be smooth as we vary $R$.

How can this clash be resolved?  In Ref.~\cite{Kani:1989im}, it was shown that modular invariance --- specifically the relevant GSO projections between the left- and right-moving sectors of the theory ---  must always connect the different CFTs in a way that is responsible for restoring continuity to all physical amplitudes as a function of $R$.  In our case, by contrast, we are dealing with a clash between modular invariance (specifically a discontinuous change in the modular weight) and the process of decompactification.  What then is the analogous resolution to this puzzle?

\subsection{
 Resolving the clash
\label{subsect:resolving}} 

To answer this question, let us proceed by examining the consequences of this extra factor of $\tau_2^{-\delta/2}$ in Eq.~(\ref{Zintnew}).
Although there are several ways in which we might incorporate this factor into our analysis, the most straightforward way is to bundle it with the leading prefactor $\tau_2^{-1}$ in Eq.~(\ref{Zintnew}).   As noted above, this then produces the net prefactor $\tau_2^{1-D/2}$ that we expect of a fully $D$-dimensional theory, with $D=4+\delta$.   

However, what is perhaps less obvious is how this new factor of $\tau_2^{-\delta/2}$ affects our results for the amplitude $\langle \calX\rangle$.   To see this, let us now proceed to apply the Rankin-Selberg procedure in order to analyze the one-loop amplitudes $\langle \calX\rangle^{(4)}$ and $\langle\calX\rangle^{(4+\delta)}$
that correspond to $Z_\calX^{(4)}$ and $Z_\calX^{(4+\delta)}$
in Eqs.~(\ref{Zorig}) and (\ref{Zintnew}) 
respectively. 
The corresponding $g$-functions $g(\tau_2)\equiv \int_{-1/2}^{1/2} d\tau_1 Z(\tau,\taubar)$ can then be written as
 $g^{(4)}=\tau_2^{-1}S^{(4)}$
and $g^{(4+\delta)} = \tau_2^{-1-\delta/2} S^{(4+\delta)}$ where
\beqn
S^{(4)}~&\equiv&~
\int_{-1/2}^{1/2}d\tau_1 
  \Bigl\lbrack \Bigl( \sideset{}{'}\sum_{mn} a_{mn} \calX_{mn}
         \qbar^m q^n\Bigr) \nonumber\\
    && ~~~~~~~~~~~~~~~~~~~~~~~~~~ \times~  Z_{\rm KK/winding} \Bigr\rbrack~~\nonumber\\
S^{(4+\delta)}~&\equiv&~
  \sideset{}{'} \sum_n a_{nn} \calX_{nn}
         (\qbar q)^n~.
\label{Sdef}
\eeqn
where we remind the reader that the primes on the summation  $\sum'_{mn}$, just as in Eq.~(\ref{Zint}),
indicate sums over the states {\it excluding}\/ the KK and winding modes associated with the $\delta$ large dimensions. Indeed, we shall generally use primes to indicate quantities uniquely associated with the ``base'' theory in Eq.~(\ref{Zint}) rather than the full theory which also includes the compactification factor $Z_{\rm KK/winding}$.

Following Eq.~(\ref{assumedtau2dep}),
we can then expand $S^{(4)}$ and $S^{(4+\delta)}$ in powers of $\tau_2$ as $\tau_2\to 0$, \ie,
\beqn
 S^{(4)}  ~&\sim& ~ \sum_{j}   C_j \,
   \tau_2^j\nonumber\\
 S^{(4+\delta)}  ~&\sim& ~ \sum_j  C'_j \,
   \tau_2^j~,
\label{Cexpansions}
\eeqn 
where the $C$ and $C'$ coefficients correspond to $S^{(4)}$ and $S^{(4+\delta)}$ respectively.
Indeed, given the expression for $S^{(4+\delta)}$ in Eq.~(\ref{Sdef}),
we immediately have
\beqn
    C'_0 ~&=&~ \Str'\, \calX \nonumber\\
    C'_1 ~&=&~ \Str'\,  \frac{d\calX}{d\tau_2}
                - \pi \alpha'\,\Str'(\calX M^2)\nonumber\\
    C'_2 ~&=&~ \frac{1}{2} \,\Str'\,  \frac{d^2\calX}{d\tau_2^2}
                - \pi \alpha'\,\Str'\left(\frac{d\calX}{d\tau_2} M^2\right)~~~~~~~~~~~~
                \nonumber\\
        && ~~~~~~~~~~~~~~~~~~
                + \frac{1}{2} \pi^2 (\alpha')^2 \, \Str' \left(
                 \calX M^4 \right)~\nonumber\\
    ~&\vdots &~  \nonumber\\
C'_n ~&=&~\frac{1}{n!} \, \Str'\, D_{\tau_2}^n \calX
\label{Cresults}
\eeqn
where $D_{\tau_2}$ is defined in Eq.~(\ref{RSderivative}) and where the primes on the supertraces in Eq.~(\ref{Cresults}) continue to indicate that the KK and winding states associated with the $\delta$ large dimensions are excluded.

Given the $C$- and $C'$-coefficients, the Rankin-Selberg procedure outlined in Sect.~\ref{subsec:supertraces} then tells 
us that
\beqn
\langle \calX\rangle^{(4)} ~&=&~ \frac{\pi}{3} 
 \lim_{\tau_2\to 0}  \tau_2^{-1} \, S^{(4)}
 \nonumber\\
\langle \calX\rangle^{(4+\delta)} ~&=&~ \frac{\pi}{3} 
 \lim_{\tau_2\to 0}  \tau_2^{-1-\delta/2} \, S^{(4+\delta)}~.
 \label{RSrelationss}
\eeqn
The presumed finiteness of $\langle \calX\rangle^{(4)}$
then allows us to conclude that the $C$-coefficients satisfy
\beq   
 \begin{cases}         
               &\,\,\,\,\,\,~ \,C_0 ~=~ 0~~~~~~~ \\
           &\ampX^{(4)} ~=~ \displaystyle{\frac{\pi}{3}} \,C_1~,
          \end{cases}
\label{usualRS}
\eeq   
as expected for any four-dimensional theory.
Likewise, for the $C'$-coefficients, the corresponding finiteness of $\langle\calX\rangle^{(4+\delta)}$ allows us to obtain
results which depend critically on $\delta$:
\beqn 
&& \delta=2:~~ \begin{cases}         
               & ~~\, \,\,\,\,C'_0 ~=~ C'_1~=~0 \\
           &\ampX^{(6)} ~=~ \displaystyle{\frac{\pi}{3}} \,C'_2~
          \end{cases}\nonumber\\
&& \delta=4:~~ \begin{cases}         
               &~~\, \,\,\,\,C'_0 ~=~ C'_1~=~C'_2~=~0 ~~~~~~~\\  
           &\ampX^{(8)} ~=~ \displaystyle{\frac{\pi}{3}} \,C'_3~
          \end{cases}\nonumber\\
&&  \delta=6:~~ \begin{cases}         
               &~~\, \,\,\,\,\,\,C'_0 ~=~ C'_1~=~C'_2~=~C'_3~=~0~~~~~\\
           &\ampX^{(10)} ~=~ \displaystyle{\frac{\pi}{3}} \,C'_4~.
          \end{cases}~\nonumber\\
\label{genRSresults}
\eeqn
 Indeed, these results are all manifestations of the misaligned supersymmetry 
that governs the spectra of modular-invariant string theories in different dimensions, even without spacetime supersymmetry.  
Moreover, these results provide a direct illustration of our  earlier assertion that the spectra of modular-invariant string theories exhibit increasingly many internal cancellations as the spacetime dimension increases.  

These results allow us to rephrase and
sharpen the ``discontinuity'' that occurs for our compactified string theory as $V_\delta \to \infty$.  
To see this, let us recall, as in Eq.~(\ref{Zorig}), that our compactified string theory consists of two components tensored together:  a ``base'' theory and  a ``cloud'' of KK/winding-mode excitations.
We have also seen in Eq.~(\ref{baseequalshigherD}) that the base theory is nothing but the higher-dimensional theory prior to compactification. Finally, we have been assuming that the overall physical amplitude $\langle \calX\rangle$ associated with our theory remains finite for all $V_\delta$.    Given these assumptions, we can ask what constraints must be satisfied by our theory as a function of $V_\delta$.   In general, there are two classes of constraints:  
\begin{itemize}
    \item $C$-constraints that govern the entire spectrum of the four-dimensional string model;   and
    \item $C'$-constraints that govern that portion of the spectrum associated with the ``base'' theory alone.
\end{itemize}

For any finite $V_\delta$, the finiteness of our overall amplitude $\langle \calX\rangle$ (\ie, the finiteness of $\langle \calX\rangle^{(4)}$)
requires --- at a bare minimum --- that
\beq
\begin{cases}
  & \bullet  ~~C_0 ~=~ 0 \\
  & \bullet  ~~C'_j~{\rm arbitrary}~.
\label{eins}
\end{cases}
\eeq
However, let us now consider what happens as
$V_\delta\to\infty$.  In this limit, 
the overall amplitude $\langle \calX\rangle^{(4)}$ technically accrues a ``spurious'' 
divergence due to the infinite volume factor in 
Eq.~(\ref{amprelation0}).
However, as $V_\delta\to\infty$, our theory is now in higher dimensions.  This means, according to Eq.~(\ref{amprelation0}), that we should divide out by this volume in order to continue to obtain the corresponding amplitude $\langle \calX\rangle$.  Indeed, the resulting amplitude is now nothing but $\langle \calX\rangle^{(4+\delta)}$.
Thus, as $V_\delta\to\infty$,  the continued finitenss of our overall amplitude $\langle \calX\rangle$ now translates to the  finiteness of $\langle \calX\rangle^{(4+\delta)}$,
which in turn requires 
\beq
\begin{cases}
  & \bullet  ~~C'_0 ~=~ C'_1 ~=~ ... ~=~ C'_{\delta/2}~=~0~. 
\end{cases}
\label{zwei}
\eeq
This sudden shift in the constraints on our string model as $V_\delta\to\infty$ is the manifestation of  the clash between the decompactification limit and the requirements of modular invariance.

Before proceeding further, we emphasize how and why these different sets of 
constraints arise.   In the discussion above, we have let $\langle \calX\rangle$ represent the physical amplitude of our theory.   When $V_\delta$ is finite, this quantity is nothing other than $\langle \calX\rangle^{(4)}$.    However, when $V_\delta$ is infinite, this quantity is nothing other than 
$\langle \calX\rangle^{(4+\delta)}$.
What we are demanding is simply that this transition as $V_\delta\to\infty$ be a {\it smooth}\/ one, with no discontinuity in the physical amplitude $\langle \calX\rangle$.
If $\langle \calX\rangle$ is finite for {\it all}\/ $V_\delta$
[where we have already compensated for the spurious volume divergence via Eq.~(\ref{amprelation0})], then we are saying that our theory has no choice but to satisfy the constraints in Eq.~(\ref{eins}) for all finite $V_\delta$ and to satisfy the constraints in Eq.~(\ref{zwei}) for infinite $V_\delta$.
This sudden shift in the constraints on our string model as $V_\delta\to \infty$ is the manifestation of the apparent discontinuity we are seeking to resolve.

Ultimately, there is only one way in which these two sets of constraints can be reconciled for all $V_\delta$:   our string theory must actually satisfy the more stringent constraints
\beq
\begin{cases}
  & \bullet  ~~C_0 ~=~ 0 \\
  & \bullet  ~~C'_0 ~=~ C'_1 ~=~ ... ~=~ C'_{\delta/2}~=~0~
\label{drei}
\end{cases}
\eeq
{\it for all compactification volumes $V_\delta$}.
Indeed, this is tantamount to demanding that the extra constraints
$C'_j=0$ for all $0\leq j\leq \delta/2$ 
apply not only in the $V_\delta\to\infty$ limit, but rather {\it for all}\/ $V_\delta$. Note that we are not {\it introducing}\/ a new set of constraints for string models with decompactification limits.   What we are instead asserting is that such string models must {\it already}\/ have been satisfying these constraints, even if these constraints had not been explicitly noticed before.  Indeed, {\it it is these properties that allow the decompactification limits to exist}\/.

This assertion represents the content of our theorem. Specifically, we have
\begin{quote}
{\it \underbar{Theorem}:
Any four-dimensional closed string theory which can be realized as a geometric compactification from a higher-dimensional string theory will inherit the precise stricter internal cancellations of the higher-dimensional theory despite the compactification.}
\end{quote}
We shall prove this theorem in Sect.~\ref{sec:theorem_proof}.~
In this connection, we remind the reader that we have been limiting our discussion here to theories whose partition functions can be factored as in Eq.~(\ref{Zorig}) --- \ie, theories whose compactifications are untwisted.   However, as we shall soon discuss, 
the above theorem can actually be trivially generalized to apply to  {\it any}\/ compactification, twisted or untwisted.   

In Sect.~\ref{sec:implications},
we shall see why we may regard this as a non-renormalization theorem.
For now, however, we simply note that this theorem may also conversely be viewed as providing an important constraint on the construction of compactified string models.
Indeed, as already noted, our compactified string theory consists of a ``base'' theory tensored with a cloud of KK/winding states.   We might then ask 
whether we can tensor such a cloud of KK/winding states to {\it any}\/ base theory.  In a field-theoretic context, the answer is yes.   However, in string theory, the requirements of modular invariance imply that we cannot do this  unless certain (primed) supertraces in the base theory vanish.   These are the primed supertraces associated with the $C'_j$-coefficients for $0\leq j\leq \delta/2$.  Indeed this is the only way in which we can smoothly and self-consistently absorb the extra powers of $\tau_2$ which arise in the decompactification limit.  

We conclude this discussion of our theorem with one final comment. In general, while
the $C_j$-coefficients are are messy functions of compactification radius and geometry, the $C'_j$-coefficients are by definition independent of any details of  compactification.
For example, a six-dimensional theory compactified on a two-torus and the same six-dimensional theory compactified on a two-sphere will give rise to {\it distinct}\/ four-dimensional theories.   However, these four-dimensional theories will nevertheless share the same $C_j'$-coefficients because they flow to the same six-dimensional theory as the volume $V_\delta$ becomes large. 
Thus, the space of compactified four-dimensional theories can be separated into different {\it equivalence classes}\/ based on their internal $C'$-constraints 
--- \ie, equivalence classes which depend on the higher-dimensional theories to which they flow at large volume.

\subsection{Proving the theorem\label{sec:theorem_proof}}

We shall begin by proving that any four-dimensional closed string theory which can be realized as a geometric compactification from a $(4+\delta)$-dimensional string theory with arbitrary compactification volume $V_\delta$ satisfies the constraints given in Eq.~(\ref{drei}) rather than those given in Eq.~(\ref{eins}).  
To do this, let us study the relationship between the $C_j$-coefficients and the $C'_j$-coefficients.

Given that these coefficients respectively describe our original four-dimensional theory and the ``base'' of that theory, and given that this base is nothing but the original $(4+\delta)$-dimensional theory, any relationship between these two sets of coefficients must stem from a relationship between the compactified and uncompactified theories.   However, we have already seen such a relationship: 
  this is our ``smoothness'' constraint in Eqs.~(\ref{amprelation0}).   Performing a $\tau_1$-integration of both sides of this relation over $-1/2<\tau_1\leq 1/2$,
inserting the expansions in Eq.~(\ref{Cexpansions}), and matching terms with equal powers of $\tau_2$ then yields the relation
\beq
  C'_j ~=~ \lim_{V_\delta\to \infty} 
  \frac{1}{\calM^\delta V_\delta} \, C_{j-\delta/2}~~~{\rm for~all}~~j~.
\label{C-to-C_relation}
\eeq
As discussed above, the shifting of the $j$-index reflects the extra powers of $\tau_2$ that emerge upon the decompactification of the large spacetime dimensions. Indeed, this index shifting is required by modular invariance and our smoothness requirement.  

We have seen that our smoothness constraint on the partition functions in Eq.~(\ref{smoothness}) leads directly to the smoothness constraint in Eq.~(\ref{amprelation0})  on the corresponding {\it amplitudes}\/ $\langle \calX\rangle^{(4)}$ and $\langle \calX\rangle^{(4+\delta)}$.
Indeed, it is the finiteness of
$\ampX^{(4)}$ that implies the auxiliary condition that $C_j=0$ for all $j< 1$, which includes the constraint $C_0=0$.
From the relation in Eq.~(\ref{C-to-C_relation}) we then find
\beq
    C'_j~=~0~~~~{\rm for~all}~~ j< 1+ \delta/2~.  
\label{firstCprimeconstraint}
\eeq
 This result is consistent with the results quoted in Eq.~(\ref{genRSresults}), and is tantamount to the assertion that if $\ampX^{(4)}$ is finite (dividing out, of course, the overall volume factor which will diverge in the $V_\delta\to\infty$ limit), then 
 $\ampX^{(4+\delta)}$
is also finite.  In other words, the $\calX$-amplitude of our four-dimensional compactified theory cannot suddenly grow a new divergence in the $V_\delta\to\infty$ limit.
This then completes the proof of our theorem.

In fact, we can even push things one step further.
Thus far, we have seen that we have two kinds of constraints:  our $C$-constraints which govern the entire four-dimensional theory and our $C'$-constraints which govern the ``base'' portion of that theory (or equivalently which govern its higher-dimensional decompactification limit).
However, we shall now demonstrate that there is in fact a universal relation between these two groups of constraints.  Indeed, this relation will apply to any theory that has a decompactification limit regardless of the degree to which its partition function factorizes. 

To proceed let us begin with two fundamental observations:
\begin{itemize}
\item   The result in Eq.~(\ref{C-to-C_relation}) does not depend on the compactification geometry.   All that is assumed is that the partition function of any string theory in $D$ spacetime dimensions has a leading prefactor of $\tau_2^k$ where $k=1-D/2= -1-\delta/2$ where $\delta=D-4$.  This is a general result for any compactification.
This also does not assume an {\it untwisted}\/ compactification (\ie, it does not assume that the four-dimensional partition function $Z^{(4)}$ factorizes), for the same reason.  Indeed, the $C$-expansion in Eq.~(\ref{Cexpansions}) is completely general regardless of the precise form of the quantity $S^{(4)}$ in
Eq.~(\ref{Sdef})
as long as $S^{(4)}$ corresponds to a four-dimensional theory,  so that $g^{(4)}(\tau_2)=\tau_2^{-1} S^{(4)}$.
\item The results in Eqs.~(\ref{usualRS}) and (\ref{genRSresults}) are also completely general, following from the same feature as described above.
\end{itemize}
Given these observations, our task is to now to relate the $C_j$-constraints from the $C'_j$-constraints --- not just at infinite volume but even at finite volume. 

The easiest way to proceed is to consider the {\it difference}\/ between our partition functions
\beq
 \Delta Z_{\calX} ~\equiv~ Z_\calX^{(4+\delta)} -\frac{1}{\calM ^\delta V_\delta} Z_\calX^{(4)}~.
 \label{eq:DZX}
 \eeq
Note that in constructing this difference we are {\it not}\/ taking the $V_\delta\to \infty$ limit;  thus this difference is a function of $V_\delta$.
By considering only the difference in this way we avoid making any assumptions about the behavior of $\langle \calX\rangle^{(4)}$ at finite $V_\delta$.
In this connection we note
that the difference of two partition functions is not necessarily itself the partition function of any self-consistent string model (see, \eg, Refs.~\cite{Dienes:1990qh, Dienes:1990ij}).   
However, such a property is not required for our proof.   

Given this definition for $\Delta Z_\calX$, we can then define the corresponding amplitude
\beq 
\langle \Delta \calX\rangle ~=~ \int \frac{d^2\tau}{\tau_2^2} \,\Delta Z_{\calX} ~,
\label{stevenew} 
\eeq
the corresponding $g$-function
\beq
g_{\Delta \calX} ~\equiv ~ \int d\tau_1 \,Z_{\Delta \calX} ~,
\label{gdeltadef}
\eeq
and the corresponding sum
\beq
   S_{\Delta\calX} ~\equiv~ \tau_2 \, g_{\Delta \calX} ~.
\eeq
We can also consider expanding $S_{\Delta \calX}$ in powers of $\tau_2$ in the $\tau_2\to 0$ limit, \ie,
\beq 
S_{\Delta\calX} ~=~ \tau_2^{-\delta/2} \sum_j C''_j \, \tau_2^j 
~~~~{\rm as}~~\tau_2\to 0~,
\label{eq:gDX2}
\eeq
thereby defining a new set of $C''_j$-coefficients.

Let us now discuss the finiteness of $\Delta Z$.   ~Of course,  we learn from Eq.~(\ref{smoothness})  that $\Delta Z_{\calX}\to 0$  as $V_\delta\to\infty$.
However, in order for this limit to exist, we also learn that $\Delta Z_{\calX}$ must be {\it finite} for large $V_\delta$ (\ie, for $\calM^\delta V_\delta\gg 1$).
Thus, 
we see that $g_{\Delta\calX}\to 0$  as $V_\delta\to\infty$ and that
$g_{\Delta \calX}$ remains finite for
$\calM^\delta V_\delta\gg 1$.   Indeed, these latter assertions follow because the $\tau_1$-integration in Eq.~(\ref{gdeltadef}) is incapable of producing a new divergence, given that this integration merely selects the zero-mode of the partition-function Fourier series.
Likewise, we find that
 $S_{\Delta\calX}\to 0$  as $V_\delta\to\infty$ and that
$S_{\Delta \calX}$ remains finite for
$\calM^\delta V_\delta\gg 1$.

Because these quantities are all finite, the expression for $\Delta Z$ in terms of the difference between
$Z^{(4+\delta)}$ and $Z^{(4)}$ allows us to write $g_{\Delta \calX}$ and 
$S_{\Delta \calX}$ as analogous differences, and thereby ultimately express $C''_j$ in terms of $C_j$ and $C'_j$.    Following this chain of steps, we thus have
\beq
C''_j ~=~ C'_j -  \frac{1}{\calM ^\delta V_\delta} C_{j-\delta/2}~.
\label{eq:Cpp}
\eeq
Once again we stress that the $C''_j$ coefficients are generally functions of $V_\delta$, with these relations holding for all
$\calM^\delta V_\delta\gg 1$.  Likewise, these relations hold as functions of $\tau_2$ for all $\tau_2$.

The final step of our analysis rests on the properties of $\langle \Delta \calX\rangle$.
As discussed below Eq.~(\ref{zwei}), the smoothness of the $V_\delta\to\infty$ limit requires that $\langle \Delta \calX\rangle$ be finite for large $V_\delta$. Thus, we can take the Rankin-Selberg  transform of the finite amplitude $\langle\Delta \calX\rangle$, \ie,
\beq
    \langle \Delta \calX\rangle ~=~
       \frac{\pi}{3} \lim_{\tau_2\to 0}
    \, \left[ \tau_2^{-1} S_{\Delta \calX}(\tau_2)\right]~
\eeq
to find that 
\beq
  C''_j~=~0 ~~~~{\rm for~all}~~ j< 1+\delta/2~.
\eeq
Indeed, it is the fact that our relations hold for all $\tau_2$ which enables us to take the $\tau_2\to 0$ limit without difficulty.
It then follows that
\beq
C'_j ~=~ \frac{1}{\calM^\delta V_\delta} \,C_{j-\delta/2}~~~~ {\rm for~all}~~j<1+\delta/2~.
\label{C-to-C_relation2}
\eeq

This is the result we have been seeking.  It provides a direct relationship between the $C_j$ and $C'_j$ coefficients for $j< 1+\delta/2$ and thereby relates our different sets of constraints to each other.
It is important to note that Eq.~(\ref{C-to-C_relation2}) is different from Eq.~(\ref{C-to-C_relation}) because it holds {\it regardless}\/ of the compactification volume $V_\delta$.  On the other hand, it holds only for $j< 1+\delta/2$.

Given the result in Eq.~(\ref{C-to-C_relation2}),
we see that
\beq
C'_{\delta/2}~=~0 ~~\Longrightarrow~~
    C_0~=~0~,
\label{proofresult}
\eeq
with $C_j$ for all $j<0$ vanishing as well.
This also implies that
\beq
 \langle \Delta \calX \rangle ~=~
   \langle \calX\rangle^{(4+\delta)} 
   - \frac{1}{\calM^\delta V_\delta} 
   \langle \calX\rangle^{(4)}~,
\label{amprelation}
\eeq
Indeed, although the left side of this equation is finite, any divergences that appear within the expressions on the right side must be identical so that they cancel in the difference.   

Note that Eq.~(\ref{proofresult}) holds for all volumes $V_\delta$.
Indeed, there is only one possible exception to this conclusion. In particular, as $V_\delta$ becomes smaller, it is possible that a physical, on-shell tachyon might appear.    However, the appearance of such a tachyon would signify a breakdown of our compactified theory and automatically result in divergent one-loop amplitudes in any case.   This would therefore correspond to taking our theory to a point at which it becomes ill-defined.
Thus, we conclude that
these results hold for all volumes $V_\delta$ which correspond to tachyon-free compactifications.

These results provide an additional perspective on our theorem. As we have seen, our theorem states that any four-dimensional string model with a {\it bona fide}\/ decompactification limit satisfies
not only  
a  $C$-constraint $C_0=0$ but also a set of additional new $C'$-constraints.
However, we now see that this $C$-constraint can be {\it replaced}\/ by the $C'$-constraints without any loss of generality.  Thus the $C'$ constraints are not only {\it necessary}\/
(as implied by our theorem)
but also {\it sufficient}\/.  Indeed, any model which satisfies our new $C'$-constraints will already satisfy our $C_0=0$ constraint.   

Of course, the $C'$-constraints that we have discovered here go beyond the $C$-constraint that was already known~\cite{Dienes:1995pm}).
Indeed, as originally discussed in Ref.~\cite{Dienes:1995pm}), all four-dimensional closed string theories must satisfy the $C_0=0$ constraint of Eq.~\eqref{usualRS} as long as they are finite (free of on-shell physical tachyons).
However, what we are now learning from our theorem is that if we additionally demand that our four-dimensional theory also have a self-consistent decompactication limit, then this theory must additionally satisfy the $C'$-constraints which not only are more powerful than the original $C_0=0$ constraint 
but even subsume it.

\subsection{The $T$-volume scaling rule}
\label{subsec:T-volume} 

We shall now present another result which we call the {\it $T$-volume scaling rule}.    This result follows from our previous results but now focuses on the first {\it non-zero}\/ coefficients $C_1$ and $C'_{1+\delta/2}$.
From Eq.~\eqref{amprelation} we find that our original four-dimensional amplitude $\langle \calX\rangle^{(4)}$ is given by
\beq
\langle \calX\rangle^{(4)} ~=~ 
\calM^\delta V_\delta \left[
\langle \calX\rangle^{(4+\delta)} - ~ \langle \Delta \calX \rangle \right]  ~.
\label{eq:collapse}
\eeq
In principle, this represents a complicated volume dependence for $\langle \calX\rangle^{(4)}$ because $\langle \Delta \calX\rangle$ is itself $V_\delta$-dependent even though $\langle \calX\rangle^{(4+\delta)}$ is not.
However we know that $\langle \Delta \calX \rangle\to 0$ at large volume. We therefore expect that 
\beq
\langle \calX\rangle^{(4)} ~\approx ~ 
\calM^\delta V_\delta \,
\langle \calX\rangle^{(4+\delta)}  ~~~~{\rm for}~~ \calM^\delta V_\delta\gg 1~.
\label{eq:collapse2}
\eeq
In other words, for large volumes, we expect that our amplitude $\langle \calX\rangle^{(4)}$ scales as the volume itself.   Indeed, this is nothing but the volume divergence discussed earlier.   We also see that the coefficient of this scaling is given by the full amplitude of the original higher-dimensional theory.

As expressed above, however, this result is not consistent with $T$-duality.
Indeed, from $T$-duality considerations  we know that our four-dimensional amplitude should scale as the  compactification volume
not only for very {\it large}\/ compactification volumes but also for very {\it small}\/ ones.
Towards this end, we seek to define a new kind of (dimensionless) volume --- a so-called {\it $T$-volume} $\widetilde V_T$ --- which is consistent not only with modular invariance but also with $T$-duality.  One natural proposal for such a quantity would be 
\begin{equation}
\widetilde V_T ~
\stackrel{?}{\equiv}~
\frac{3}{\pi}\, \int _\calF \frac{d^2\tau}{\tau_2^2} \, \tau_2^{\delta/2 } \, Z_{\rm KK/winding}  ~.
\label{eq:lamd}
\end{equation}
Indeed, this definition has the advantage that it results from a modular-invariant integral and also depends directly on the $T$-duality-invariant $Z_{\rm KK/winding}$ partition-function factor.
Indeed, if we were to na\"ively apply the Rankin-Selberg procedure 
to this integral, we would find
\beq
\widetilde V_T ~\equiv ~
  \frac{3}{\pi}\,
  \lim_{\tau_2\to 0} \,
  g_{\rm KK/winding}
\label{eq:newdef}
\eeq 
where
\beq
\label{eq:gKK}
g_{\rm KK/winding} ~\equiv~
\int_{-1/2}^{1/2}  d\tau_1\, 
 \tau_2^{\delta/2} \,Z_{\rm KK/winding}~.~~
 \eeq
However, we can immediately see that 
the expression in Eq.~(\ref{eq:lamd}) is actually {\it divergent}\/  for all $\delta\geq 2$.   Given that $Z_{\rm KK/winding}$ has a $(q,\qbar)$ expansion that necessarily begins with a non-zero constant term,
this divergence arises in the $\tau_2\to\infty$ region due to the extra power $\tau_2^{\delta/2}$ that was needed for  modular invariance.
This divergence invalidates the Rankin-Selberg procedure that leads to Eq.~(\ref{eq:newdef}).  Indeed, this failure of the Rankin-Selberg procedure can be seen from the fact that Eq.~(\ref{eq:lamd}) is
finite only for $\delta<2$ while Eq.~(\ref{eq:newdef}) is finite for all $\delta$.

Given that Eq.~(\ref{eq:newdef}) is finite for all $\delta$, 
we shall therefore {\it define}\/
$\widetilde V_T$ to be given by Eq.~(\ref{eq:newdef}) rather than by Eq.~(\ref{eq:lamd}).
As we shall see below, this ensures a finite value of $\widetilde V_T$ for all $\delta$.  Moreover, we shall find that it is this definition that leads to meaningful results, and indeed this is all we shall ever need. 

This definition for $\widetilde V_T$  in Eq.~(\ref{eq:newdef}) provides us with a dimensionless compactification volume which respects $T$-duality for the factorized compactifications which have been our focus thus far.  The quantity $\widetilde V_T$ thereby substitutes for the quantity $\calM^\delta V_\delta$ that we have been writing until now.
Furthermore, the overall  normalization factor $3/\pi$ in Eqs.~(\ref{eq:lamd}) and (\ref{eq:newdef}) ensures that $\widetilde V_T=1$ for the trivial $\delta=0$ case in which $\tau_2^{\delta/2} Z_{\rm KK/winding}\to 1$.
Of course, in the special case with $\delta=1$, we find that $\widetilde V_T$ is also given by Eq.~(\ref{eq:lamd}).
Indeed, for the simple case of compactification on a circle of (dimensionless) radius $\widetilde R\equiv M_s R=R/\sqrt{\alpha'}$, as in Eq.~(\ref{KKwindingdef}), 
we find 
\beq
   \widetilde V_T ~= ~  \widetilde R + \frac{1}{\widetilde R}~.
\eeq
We thus see that $\widetilde V_T\to\infty$ both at large radius and at small radius, and thereby subsumes both cases  in a $T$-duality-invariant manner.

Proceeding with this definition of $\widetilde V_T$,
we will now show that the 
coefficient $C_1$ is indeed given in terms of $\widetilde V_T$ by
\beq
 C_1 ~\approx ~ 
     \widetilde V_T \,
    C'_{1+\delta/2} ~~~~
    {\rm for}~ \widetilde V_T \gg 1~, 
\eeq
or equivalently
\beq
 \langle \calX\rangle^{(4)}  ~\approx ~  \widetilde V_T \,
 \langle \calX\rangle^{(4+\delta )}~~~~ 
 {\rm for}~ \widetilde V_T \gg 1~.
 \label{QueenCamilla}
 \eeq
We thus have:
\begin{quote}
    \label{cor:32}
    {\it \underbar{$T$-volume scaling rule}:~\break
    Within any four-dimensional closed string theory
which can be realized as a geometric compactification from a higher-dimensional string
theory, the one-loop amplitude $\langle \calX\rangle^{(4)}$ in the large-volume limit is given
by the product of the dimensionless $T$-volume of compactification and the corresponding amplitude of the original higher-dimensional theory.}
\end{quote}
While our proof of this result will hold for 
untwisted compactifications, we shall see that it can be easily generalized in order to hold for twisted compactifications as well.

To prove this result, let us recall from Eq.~(\ref{RSrelationss}) 
that 
$\langle \calX\rangle^{(4)}$ 
is given as
\beq
\langle \calX \rangle^{(4)} ~=~ 
  \frac{\pi}{3} \lim_{\tau_2\to 0} g^{(4)}(\tau_2)
\label{KingCharles}
\eeq
where
\beq
  g^{(4)}(\tau_2)~=~ 
\int_{-1/2}^{1/2}d\tau_1 \,
   Z_\calX^{\rm (base)} \, Z_{\rm KK/winding}~.
\label{vscalingproof1}
\eeq
In general, both $Z_\calX^{\rm (base)}$ and  $Z_{\rm KK/winding}$ are double power series in $q$ and $\qbar$.   Indeed, the latter power series depends on the particular compactification geometry, with an example given in Eq.~(\ref{KKwindingdef}) for the case of a one-dimensional compactification on a circle.  It is for this reason that $\langle \calX\rangle^{(4)}$ generally depends in a highly non-trivial way on the compactification geometry. 
Indeed, according to Eq.~(\ref{vscalingproof1}) we would need to multiply these two power series together, thereby producing a new power series for the product,
whereupon the $\tau_1$-integration would project us down to terms with equal coefficients of $q$ and $\qbar$ in the product.
However, because of the fact that our integrand is a {\it product}\/ of two independent power series, the terms that have equal powers of $q$ and $\qbar$ in the product need not themselves have equal powers of $q$ and $\qbar$ for each factor individually.  Phrased somewhat differently, if we follow 
Eq.~(\ref{Zorigrefactored}) and define
\beq
g_{\rm base} 
~\equiv~\int_{-1/2}^{1/2} d\tau_1\,  \tau_2^{-\delta/2}\, Z^{\rm (base)}_\calX~
\eeq
along with the definition of $g_{\rm KK/winding}$ in Eq.~(\ref{eq:gKK}), we see that
\beq
    g^{(4)} ~\not= ~
       g_{\rm base} \cdot
        g_{\rm KK/winding}.
\eeq
Indeed,
$g_{\rm base}$ and $g_{\rm KK/winding}$ are non-trivially {\it entwined}\/ in forming $g^{(4)}$.
This phenomenon was discussed in detail in Ref.~\cite{Abel:2023hkk}.

To proceed, let us therefore write
\beq
    g^{(4)} ~= ~
       g_{\rm base} \cdot
        g_{\rm KK/winding} + g_{\rm entwined}~
\eeq
where $g_{\rm entwined}$ represents the ``error'' term that prevents us from performing a full factorization of $g^{(4)}$.
Recalling the definition of $\widetilde V_T$ in Eq.~(\ref{eq:newdef}),
we then find from Eq.~(\ref{KingCharles})
that
\beqn
\langle \calX\rangle^{(4)} ~&=&~
 \frac{\pi}{3} \lim_{\tau_2\to 0} 
  \left[ g_{\rm base} \cdot
        g_{\rm KK/winding} + g_{\rm entwined}\right] \nonumber\\
&=&~  \frac{\pi}{3}\,
   \widetilde V_T \lim_{\tau_2\to 0} 
     g_{\rm base}  +  \frac{\pi}{3}
     \lim_{\tau_2\to 0}   g_{\rm entwined} ~\nonumber\\
&=&~ \widetilde V_T \langle \calX\rangle^{(4+\delta)}  + 
   \frac{\pi}{3}   \lim_{\tau_2\to 0}   g_{\rm entwined} ~,    
\label{PrinceWilliam}
\eeqn
where we have used Eqs.~(\ref{baseequalshigherD}) and (\ref{higherdimamp})
in passing to the final line.
Moreover, as promised earlier, we see from the final line of Eq.~(\ref{PrinceWilliam}) that $\widetilde V_T$ --- as defined in Eq.~(\ref{eq:newdef}) --- is indeed finite because it serves as the proportionality constant between the finite quantities 
$\langle \calX\rangle^{(4)}$ and
$\langle \calX\rangle^{(4+\delta)}$.
Comparison with Eq.~\eqref{eq:collapse} and replacing $\calM^\delta V_\delta\to \widetilde V_T$ then allows us to identify
\beq
\langle \Delta \calX \rangle ~=~  - \frac{\pi }{3} \, \frac{1}{\widetilde V_T }  \lim_{\tau_2\to 0}  g_{\rm entwined}~.
\eeq
Thus we see that $\langle \Delta \calX \rangle$ encapsulates the entwinement between $g_{\rm base}$ and $g_{\rm KK/winding}$ in the contribution to $\ampX^{(4)}$.    Indeed, contributions from such entwined terms are generally exponentially suppressed relative to those that are unentwined.   

We have therefore proven the $T$-volume scaling rule, as expressed in  Eq.~(\ref{QueenCamilla}), with ``error'' terms that become increasingly small (indeed, exponentially suppressed) as $\widetilde V_T\to\infty$.   
 
\subsection{General applicability:  Twisted compactifications and multiple constraints
\label{twisted}} 

As we have repeatedly stressed,  our  theorem in Sect.~\ref{subsect:resolving} has been  derived within the context of factorized theories [\ie, theories with factorized partition functions, as in Eq.~(\ref{Zorig})]
for which one factor $Z_{\rm KK/winding}$ completely describes the compactification geometry.   This generally corresponds to untwisted compactifications.  

However, there also exist {\it twisted}\/ compactifications for which this sort of factorization does not arise.
These include compactifications on orbifolds; 
coordinate-dependent Scherk-Schwarz compactifications of the kind discussed in Ref.~\cite{Abel:2015oxa};  and  also compactifications involving Wilson-line breaking of gauge symmetries.   Likewise, there exist theories (such as Type~I strings, or non-perturbative closed strings involving $D$-branes) which have some sectors which are modular invariant as well as other sectors which are not modular invariant. 
It therefore remains to determine the extent to which our theorem applies to such theories as well.

As we shall demonstrate, our theorem
applies to such theories as well.   In particular, our theorem will apply to any  modular-invariant portion of any four-dimensional theory which itself becomes a $(4+\delta)$-dimensional theory as a corresponding compactification modulus becomes large.   

The issue as to whether or not the partition function factorizes is not merely an algebraic distinction.  Instead, it reflects the manner in which the compactification deforms the theory.
For an untwisted compactification, the partition function factorizes because the precise KK/winding spectra are the same for each state in the underlying ``base'' theory.
These spectra are thus independent of the quantum numbers associated with the states in the base theory.   However, for a twisted theory this is no longer true:   the KK and winding numbers of entire towers of states in the $\delta$-dimensional compact space are shifted by amounts that depend on their four-dimensional quantum numbers.   It is this feature that breaks that factorizability of the partition functions of such theories.

As a result of these observations, 
it follows that the algebraic structure of the partition function depends critically on the numbers and types of twists involved in the compactification.
Indeed, one generally obtains a partition function which can be written schematically as the sum of contributions from different {\it sectors}\/, \ie, 
\beq
Z_\calX^{(4)} ~=~ \sum_{{\rm sectors}~s} Z^{\rm (base)}_s   \,\cdot\, Z^{(s)}_{\rm KK/winding} 
\label{eq:schem}
\eeq
where each sector $s$ is associated with its own $\calX$-dependent ``base function'' $Z^{\rm (base)}_s$ and its own set of KK/winding states associated with
$Z^{(s)}_{\rm KK/winding}$.

In order to understand how our theorem can apply in such situations, it will prove simplest to analyze a particular example.
Accordingly, for simplicity, we shall consider the case in which our four-dimensional theory is realized as  a one-dimensional compactification of a five-dimensional theory, taking our compactification geometry to be that of a circle modded out by a single $\IZ_2$ twist.   In this case, we find that the resulting four-dimensional theory has a partition function of the form in Eq.~(\ref{eq:schem}) with only four sectors, \ie, $s=1,...,4$.  

For this scenario, it is not difficult to identify the resulting $Z^{\rm (base)}_s$ and $Z_{\rm KK/winding}^{(s)}$ factors.
Following Ref.~\cite{Rohm:1983aq} while adopting the conventions in Ref.~\cite{Abel:2015oxa}, we may take the $Z_{\rm KK/winding}^{(s)}$ functions to be
none other than $\calE_{0,1/2}/\sqrt{\tau_2}$ and $\calO_{0,1/2}/\sqrt{\tau_2}$, where these functions are defined to be the same as
$Z_{\rm circ}$ in Eq.~(\ref{Zcircdef}) except that their summation
variables are restricted and shifted as follows:
\beqn
       \calE_0 &=& \lbrace  \tilde m\in\IZ,~\tilde n~{\rm even}\rbrace\nonumber\\
       \calE_{1/2} &=& \lbrace  \tilde m\in\IZ+\half ,~\tilde n~{\rm even}\rbrace\nonumber\\
       \calO_0 &=& \lbrace  \tilde m\in\IZ,~\tilde n~{\rm odd}\rbrace\nonumber\\
       \calO_{1/2} &=& \lbrace  \tilde m\in\IZ+\half ,~\tilde  n~{\rm odd}\rbrace~.
\label{EOfunctions}
\eeqn
The half-integer modings for $\tilde m$ and the even/odd sensitivity for $\tilde n$ are both related to the $\IZ_2$ nature of the orbifold twist.

Likewise, the corresponding base functions $Z_s^{\rm (base)}$ in each sector consist of those parts of the original base theory which are either even (untwisted) or odd (twisted)
under the action of the orbifold.
Specifically, using standard notation, we may identify $Z^{+}_+$
as the partition function of the original base theory prior to compactification, $Z^+_-$ as that of its projection sector, $Z^{-}_+$ as that of the corresponding twisted sector, and $Z^-_-$ as that of the projection sector of the twisted sector. 
Note that according to the standard conventions for such orbifold-sector partition functions $Z^\pm_\pm$ in four dimensions (see, \eg, Ref.~\cite{Abel:2015oxa}) such factors 
already include factors of $\tau_2^{-3/2}$.
We can then identify
\beqn
        Z^{\rm (base)}_1 &=& \half \sqrt{\tau_2} \left 
         ( Z^+_+ + Z^-_+ \right)\nonumber\\
        Z^{\rm (base)}_2 &=& \half\sqrt{\tau_2} \left( Z^+_- + Z^-_- \right)\nonumber\\
        Z^{\rm (base)}_3 &=& \half \sqrt{\tau_2} \left( Z^+_+ - Z^-_+ \right)\nonumber\\
        Z^{\rm (base)}_4 &=& \half\sqrt{\tau_2} 
 \left( Z^+_- - Z^-_- \right)~.
\label{Zifactors}
\eeqn

Given these identifications, 
our final orbifolded theory then has a partition function of the form
\beqn
 Z_\calX^{(4)} ~&=&~
\frac{1}{\sqrt{\tau_2}} \,
\biggl[ Z^{\rm (base)}_{1} \calE_0 +
     Z^{\rm (base)}_{2} \calE_{1/2}\nonumber\\
     && ~~~~~~~~+
     Z^{\rm (base)}_3 \calO_0 +
     Z^{\rm (base)}_4 \calO_{1/2}\biggr]~~~~~~~~
\label{eq:Z4_fourterms}
\eeqn
where the $Z^{\rm (base)}_{1,...,4}$ functions continue to have the $\calX$-insertions.  In writing Eq.~(\ref{eq:Z4_fourterms}) we recall that the $\calE,\calO$ functions have leading $\sqrt{\tau_2}$ factors while the $Z_i^{\rm (base)}$ functions have leading factors $\tau_2^{-1}$ in four dimensions.  Our final result for $Z_\calX^{(4)}$ thus has a leading $\tau_2^{-1}$ factor, as expected.

What will be important for us are the limits of these geometric functions $\calE_{0,1/2}$ and $\calO_{0,1/2}$ as their radii $\widetilde R\equiv M_s R= R/\sqrt{\alpha'}$
are taken to be extremely large or small.  These can be determined by explicit calculation, yielding
\beqn
\widetilde R\to\infty:&
  ~~~& 
     \calE_0, \, 
      \calE_{1/2} \to
            \widetilde R~,
    ~~~
    \calO_0,\, \calO_{1/2}
    \to 0
    \nonumber\\
   \widetilde R\to 0:&
  ~~~& 
     \calE_0, \,\calO_0 \to \frac{1}{2\widetilde R}~,~~~
     \calE_{1/2},\,
    \calO_{1/2} \to 0~.~~~~\nonumber\\
\label{eq:EOlimits}
\eeqn
From Eq.~(\ref{eq:Z4_fourterms}) 
it therefore follows that
\beqn
\widetilde R\to\infty:&
  ~~~
     Z_\calX^{(4)}\to 
     \frac{\widetilde R}{\sqrt{\tau_2}} & \!\!\left(Z^{\rm (base)}_1+Z^{\rm (base)}_2\right)\nonumber\\
\widetilde R \to 0:& 
~~~
    Z_\calX^{(4)}\to  
  \frac{1}{2 \sqrt{\tau_2} \widetilde R}  &\!\! \left(Z^{\rm (base)}_1+Z^{\rm (base)}_3\right) ~.~~~\nonumber\\
\eeqn
We thus see that our original four-dimensional theory with partition function $Z_\calX^{(4)}$ flows to different theories in the $\widetilde R\to \infty$ and $\widetilde R\to 0$ limits! Indeed, from Eq.~(\ref{Zintnew}) and identifying 
$\calM^{1/2} V_{1/2}$ as $\widetilde R/2$,
we find
\beqn
\widetilde R\to\infty:&
  ~~~& 
     Z_\calX^{(5)} ~=~
     \frac{2}{\sqrt{\tau_2}} \left(
     Z_1^{\rm (base)}
    +
    Z_2^{\rm (base)}\right)
    \nonumber\\
    && 
     \phantom{Z_\calX^{(5)} ~} =~
    Z^+_+ + Z^+_- + Z^-_+ + Z^-_-\nonumber\\
\widetilde R\to 0 :&
  ~~~& 
     Z_\calX^{(5)} ~=~
     \frac{2}{\sqrt{\tau_2}} \left(
     Z_1^{\rm (base)}
    +
    Z_3^{\rm (base)}\right)
    ~~~\nonumber\\
    && 
     \phantom{Z_\calX^{(5)} ~} =~
    Z^+_+~.
\label{5Dconsistent}
\eeqn
We thus see that 
 $Z_\calX^{(4)}$ flows to the original five-dimensional {\it untwisted}\/ theory in the $\widetilde R\to\infty$ limit, while it flows to the five-dimensional {\it twisted}\/ theory in the $\widetilde R\to 0$ limit. This kind of interpolation between different decompactified theories is completely standard, and  the breaking of $T$-duality in this case is the effect of the twist in the compactification.  

Our discussion thus far has centered around theories with one large extra dimension compactified on $S^1/\IZ_2$.
However, similar treatments will also apply to more complicated compactifications from higher dimensions.
For example, it is possible to consider the compactification of a {\it six}\/-dimensional theory on a {\it two}\/-dimensional compactification geometry.  In order to exploit the above results, we can consider this compactification geometry to be $(S^1/\IZ_2)_{\widetilde R_5} 
\otimes (S^1/\IZ_2)_{\widetilde R_6}$ where $\widetilde R_{5,6}$ are the dimensionless radii associated with the fifth and sixth dimensions respectively.
Our four-dimensional partition function will then have {\it sixteen}\/ sectors and
takes the form
\beq 
     Z_\calX^{(4)} ~=~ 
     \sum_{p=1}^4
      \sum_{q=1}^4
      ~ Z_{pq}^{\rm (base)}
     \,  Z_{\rm KK/winding}^{(pq)}~
\label{tensor_prod_theory}
\eeq
where $p$ and $q$ respectively correspond to the fifth and sixth dimensions  
and where
\beqn
  Z_{\rm KK/winding}^{(1,1)} ~&=&~
   \tau_2^{-1}\,  \calE_0 \cdot \calE_0~\nonumber\\
     Z_{\rm KK/winding}^{(1,2)} ~&=&~  \tau_2^{-1}\,
    \calE_0 \cdot \calE_{1/2}~\nonumber\\
 &\vdots &  \nonumber\\
 Z_{\rm KK/winding}^{(4,4)} ~&=&~  \tau_2^{-1}\,
     \calO_{1/2} \cdot 
     \calO_{1/2} ~.
\eeqn
Defining
\beq 
g'_{pq}(\tau_2)~\equiv~\int_{-1/2}^{1/2} d\tau_1 \, 
        Z_{pq}^{\rm (base)} ~
\eeq
and further defining $S'_{pq}(\tau_2) = \tau_2\, g_{pq}(\tau_2)$, we may expand
\beq
   S'_{pq}(\tau_2) ~\sim~ \sum_j C^{\prime (pq)}_j \tau_2^j~~~~~~{\rm as}~~ \tau_2\to 0~.
\label{Cprimabdefs}
\eeq
These $C^{\prime (pq)}_j$-coefficients
thus correspond to the $C'_j$ coefficients of the simpler untwisted compactification, except that we now have a different set of $C^{\prime (pq)}_j$-coefficients for each base theory in Eq.~(\ref{tensor_prod_theory}), \ie, for each value of $p$ and $q$.

For such a four-dimensional theory, there will now be {\it four}\/ ways of decompactifying in order to produce a six-dimensional theory.  These  correspond to taking $\widetilde R_5\to 0,\infty$ and $\widetilde R_6\to 0,\infty$, each yielding a {\it different}\/ six-dimensional theory.
The partition functions of these six-dimensional theories will be different combinations of our sixteen underlying $Z_{pq}^{\rm (base)}$ functions in Eq.~(\ref{tensor_prod_theory}). However, we observe (just as in the five-dimensional case) that no single base function $Z_i^{\rm (base)}$ by itself corresponds to a decompactified theory.  Indeed, this only happens when there is a single sector --- \ie, an untwisted compactification.

We shall assume, as stated above, that each decompactification limit leads to a finite one-loop amplitude.   Following our previous discussions for the untwisted case, this means that each limit must {\it independently}\/ satisfy the same smoothness constraint that we imposed in the case of an untwisted compactification.
Thus, for the six-dimensional twisted compactification we have been considering here,
there are now {\it four independent}\/ smoothness constraints that must hold.    These limits represent the four different ways in which we might obtain a six-dimensional theory.

To formulate these constraints, we can follow 
our previous analysis in Eq.~(\ref{Sdef}) and
establish four distinct sums corresponding to these different decompactification limits:
\beq
S^{(6)}_Q ~=~  \tau_2^{2} \int_0^\infty  d\tau_1 
     \left[ \lim_Q  \, Z_{\calX}^{(4)}\right]
\eeq
where $Z_\calX^{(4)}$ is given in Eq.~(\ref{tensor_prod_theory}), where
$Q=1,...,4$ ranges over the different decompactification limits
$(\widetilde R_5,\widetilde R_6)\to 
 (\infty,\infty)$, 
 $(\infty,0)$, 
 $(0,\infty)$, and
$(0,0)$
respectively.
Each limit will have its own $\tau_2$-expansion.   To avoid confusion 
(assuming the reader is not already sufficiently confused), 
we shall let $D'$ denote the coefficients of such an expansion:
\beq
    S_Q^{(6)} ~\sim~ \sum_j D^{\prime (Q)}_j \tau_2^j
    ~~~~~{\rm as}~~ \tau_2\to 0~.
\eeq 
In general, these four sets of $D'_j$ coefficients (one for each $Q$) will be distinct from each other, with each corresponding to a distinct fully modular-invariant six-dimensional theory.

Given these coefficients, and given our previous discussion, 
there will be new constraints on each set of coefficients
that corresponds to a decompactification limit yielding a finite higher-dimensional amplitude.
For example, if the 
$\widetilde R_5\to \infty, \widetilde R_6\to \infty$ limit produces a finite string amplitude, then we learn that 
\beq
   D_0^{\prime(1)}=0~,~~~
   D_1^{\prime(1)}=0~.
\label{eq:example_cancellation}
\eeq
Likewise, if the $\widetilde R_5\to \infty, \widetilde R_6\to 0 $ limit also produces a finite string amplitude, then we also have
\beq
   D_0^{\prime(2)}=0~,~~~
   D_1^{\prime(2)}=0~,
\label{eq:example_cancellation2}
\eeq
and so forth.
Such results are the twisted analogues of our theorem in Sect.~\ref{subsect:resolving}, and the proof of these assertions follows directly from the Rankin-Selberg procedure.

Our goal, of course, is to express these $D^{\prime (Q)}$-constraints in terms of the $C^{\prime (pq)}_j$-coefficients  corresponding to our original four-dimensional partition function in Eq.~(\ref{tensor_prod_theory}).
These $C^{\prime (pq)}_j$-coefficients are defined in Eq.~(\ref{Cprimabdefs}).
However, using Eq.~(\ref{eq:EOlimits}), we may immediately relate these two sets of coefficients.   For example, we find
\beqn 
   D_j^{\prime (1)} ~&=&~   
   C_j^{\prime (1,1)}+C_j^{\prime (1,2)}+C_j^{\prime (2,1)}+C_j^{\prime (2,2)}~\nonumber \\
   D_j^{\prime (2)} ~&=&~   
   C_j^{\prime (1,1)}+C_j^{\prime (1,3)}+C_j^{\prime (3,1)}+C_j^{\prime (3,3)}~,~~~~
\eeqn
and so forth.
We thus find that our complete set of constraints becomes
\beqn
C_j^{\prime (1,1)}+C_j^{\prime (1,2)}+C_j^{\prime (2,1)}+C_j^{\prime (2,2)}~&=&~ 0\nonumber \\
C_j^{\prime (1,1)}+C_j^{\prime (1,3)}+C_j^{\prime (2,1)}+C_j^{\prime (2,3)}~&=&~ 0\nonumber \\
C_j^{\prime (1,1)}+C_j^{\prime (1,2)}+C_j^{\prime (3,1)}+C_j^{\prime (3,2)}~&=&~ 0\nonumber \\
C_j^{\prime (1,1)}+C_j^{\prime (1,3)}+C_j^{\prime (3,1)}+C_j^{\prime (3,3)}~&=&~0~~~~~~
\label{twistedCprimeconstraints}
\eeqn
for all  $j\leq 1$.

In the analogous case of an untwisted compactification, we obtained constraints on the $C'_j$-coefficients corresponding to the base theory.   By contrast, for a twisted compactification, we  see our theorem now yields {\it  multiple} constraint equations.
However, each of these constrains only a {\it linear combination}\/ of the coefficients corresponding to {\it different}\/ base theories.   Moreover, as indicated above, each of these constraint equations holds not only for $j=0$ but also for $j=1$.  As discussed in Sect.~\ref{subsect:resolving},
the latter reflects the emergence of the extra dimensions and is required for the consistency with our lower-dimensional theory upon decompactification. 

Of course, the constraints in Eq.~(\ref{twistedCprimeconstraints}) allow us to obtain results such as
\beq
C_j^{\prime (1,2)}+C_j^{\prime (2,2)}~=~ 
C_j^{\prime (1,3)}+C_j^{\prime (2,3)}~
\eeq
which do not correspond to any single decompactification limit.  
Moreover, our four-dimensional theory may also have other internal symmetries that are reflected in constraints on these $C_j^{\prime (pq)}$-coefficients.
For example, if $Z_{pq}^{\rm (base)}=
Z_{qp}^{\rm (base)}$ (as might occur for theories with a permutation symmetry between the fifth and sixth dimensions), it then follows that 
$C_j^{\prime (pq)}=C_j^{\prime (qp)}$ for all $j$.
In such cases, there are effectively fewer base partition functions and potentially fewer decompactifications as well.

In general, there can also be decompactification
limits which are tachyonic.
For example, a four-dimensional theory might be finite (tachyon-free) over a certain range of compactification volumes, yet encounter a tachyon as this volume increases towards infinity or decreases towards zero.   A well-known example of this occurs for the thermal analogue of a one-dimensional compactification, where we identify the compactification radius as an inverse temperature.  Such theories become tachyonic once the temperature exceeds a critical value, leading to the so-called Hagedorn transition~\cite{Hagedorn:1965st,Atick:1988si,Dienes:2012dc}.   Such transitions clearly lead to divergences in the one-loop amplitude.   As a result, the $C'$-constraints that emerge from this decompactification are valid only within the range of radii in which such tachyons do not appear.
There can also be situations in which no tachyon appears at any compactification radius, but in which certain states in the string spectrum become massless at specific compactification radii before becoming massive again (see, \eg, Ref.~\cite{Dienes:2012dc}).   The sudden appearance of such massless states will generally induce {\it higher-order}\/ Hagedorn-like phase transitions~\cite{Dienes:2012dc} which 
represent discontinuities that also violate our ``smoothness'' assumptions.   However, even though our theorem will not apply at or beyond such radii, the constraints emerging from our theorem will continue to apply before these states are reached.

Finally, it is interesting to note that
our compactification functions $\calE_{0,1/2}$ and $\calO_{0,1/2}$ --- like
{\it any}\/ compactification functions ---
have certain properties which guarantee that we can continue to use the Rankin-Selberg mapping.
In particular, {\it a priori}\/, one might have worried that additional $C$-constraints could appear upon compactification.
It is easy to see how such extra constraints might have arisen.
For this purpose, it is perhaps easiest to start with the compactified theory with the partition function given in Eq.~(\ref{eq:Z4_fourterms}) and ask what happens for large but finite $R$.  In this regime the terms involving $\calE$-functions dominate --- terms which we can rewrite in the form 
\beqn
Z_\calX^{(4)} &\approx &  \frac{1}{2} \,\tau_2^{-1/2} \,\biggl[ \left(  Z_1^{\rm (base)} + Z_2^{\rm (base)} \right)
    \left( \calE_0 + \calE_{1/2} \right) \nonumber\\
   &&~+
  \left(  Z_1^{\rm (base)} - Z_2^{\rm (base)} \right)
    \left( \calE_0 - \calE_{1/2} \right) \biggr] \,.~~
\label{rewritten}
\eeqn
The top line, of course, is entirely expected and does not yield any constraints beyond those we have already considered.  Indeed, $Z_1^{\rm (base)} + Z_2^{\rm (base)}$ is the $R\to\infty$ limit that we have already considered, and whose properties as $\tau_2\to 0$ have allowed us to formulate our existing constraints.  However, in principle, there is the possibility that the {\it second}\/ line of Eq.~(\ref{rewritten}) might to lead to additional constraints.
Indeed, such additional constraints could have arisen
if
$\calE_0 - \calE_{1/2}$ were for example to behave as $\tau_2/R$ as $\tau_2\to 0$ at finite $R$.    In such cases, one could take $R\to\infty$ prior to taking $\tau_2$ and thereby conclude that no new constraint comes from such a difference.  However, we could alternatively take the $\tau_2\to 0$ limit first, leaving us with a dangerous $1/R$ dependence whose cancellation would require an additional constraint.
 
 However, it is straightforward to verify that this does not happen.
 Indeed,
 direct calculation for our specific $\calE$-functions
 tells us that 
\beq
\calE _0(\widetilde R) - \calE _{1/2}(\widetilde R) ~\approx ~ (\widetilde R ^4 /\tau_2)\, e^{-\pi \widetilde R^2 /\tau_2} ~.
\label{fixedpoint} 
\eeq
Thus the {\it difference} between the $\calE$ functions decreases faster than any power of $\tau_2$ as $\tau_2 \to 0$.  In other words, the difference $\calE_0-\calE_{1/2}$ has an essential singularity at $\tau_2=0$. 
As a matter of principle this will be true for any compactification functions whose large-volume limits produce valid higher-dimensional theories.   Thus our theorem remains intact.

The discussion in this section has been somewhat technical.   However, the main idea is simple and can be summarized as follows.  For an untwisted compactification, there is only one decompactification limit. 
This then leads to a single extra $C'$-constraint, namely 
$C'_j=0$ for $j < 1+\delta/2$, 
which must be adjoined onto our original $C$-constraint $C_0=0$.
However, for a twisted compactification, there are generally multiple decompactification limits, each involving a different subset of the sectors in our theory.   Our theorem nevertheless applies exactly as before, with each decompactification leading to its own new $C'$-constraint, as illustrated above.  As a result, our original four-dimensional theory not only must satisfy the original constraint $C_0=0$, but also all of the individual $C'$-constraints that emerge 
from each different decompactification limit. Indeed, all of these extra $C'$-constraints must be satisfied {\it simultaneously}\/ within the original four-dimensional theory because this theory simultaneously contains all of these different possibilities for a self-consistent decompactification.

We close with a final comment.  As discussed in Sect.~\ref{sec:theorem_proof} for the case of an untwisted compactification, we have seen that the single extra constraint $C'_j=0$ for $j < 1+\delta/2$ actually {\it implied}\/ our original $C$-constraint $C_0=0$.   The same will be true for twisted compactifications.    In particular, even though each different decompactification leads to an independent $C'$-constraint, these $C'$-constraints collectively imply that $C_0=0$ as well.

\FloatBarrier
\section{Implications of the theorem}
\label{sec:implications}

As discussed in  Sect.~\ref{sec:theorem}, our theorem is completely general, providing us with new constraints on the $C'_j$-coefficients for any string model meeting the conditions outlined at the beginning of Sect.~\ref{sec:theorem} and for any operator insertion $\calX$ for which the corresponding 
physical quantity $\zeta\sim \langle \calX\rangle^{(4)}$ within that model is finite.  In this section we now proceed to consider two important implications of these new $C'$-constraints.
For concreteness and simplicity, we shall focus in this section on the case of untwisted compactifications.  As discussed in Sect.~\ref{sec:theorem}, the situation with twisted compactifications is similar and proceeds in an analogous way.

\subsection{New supertrace identities}
 
Just as the $C$-coefficients can be expressed as supertraces over the states within the full partition function $Z_\calX^{(4)}$, as in Eq.~(\ref{Cs-as-supertraces}), the $C'$-coefficients can likewise be expressed as supertraces of the states that contribute to $Z_\calX^{(4)}$  {\it without}\/ including the KK and winding states associated with the compactification under study.  Indeed, these are the states that reside within the base theory and contribute to 
$Z^{\rm (base)}_\calX$ alone.  

Given this observation, our new constraints on the $C'_j$-coefficients immediately yield new constraints for the supertraces over the states contributing solely to $Z^{\rm (base)}_\calX$.
In particular, the vanishing of the $C'_j$-coefficients for all $j< 1+ \delta/2$ --- as required by our theorem --- thus gives rise to the new
supertrace identities
\beq
  \Str'\,D_{\tau_2}^j \calX ~=~0 ~~~{\rm for~all}~~ 0\leq j \leq  \delta/2~
\label{newtrace1}
\eeq
where the prime on the supertrace indicates that only the states contributing to $Z^{\rm (base)}_\calX$ are included and where $D_{\rm \tau_2}$ is the derivative defined in Eq.~(\ref{RSderivative}). Likewise, as the
compactification $T$-volume $\widetilde V_T$ becomes large,
a similar supertrace formulation applies to the amplitude $\langle \calX \rangle^{(4)}$.
We know, of course, that 
$\langle \calX\rangle^{(4)}$ is given by $(\pi/3) C_1$ for all compactification volumes.
Indeed, $C_1$ is generally a complicated function of the compactification volume.   However, for $\widetilde V_T\gg 1$, we have seen that our $T$-volume scaling rule 
allows us to pull out the compactification volume as a single overall factor, leaving us with
\beq 
  \langle \calX \rangle^{(4)}~\approx~ \frac{\pi}{3}\, \widetilde V_T\, C'_{1+\delta/2}~.
\label{newtrace1.5}
\eeq
Thus, for $\widetilde V_T\gg 1$, we have
\beq
 \langle \calX\rangle^{(4)} 
 ~\approx~
\frac{\pi}{3}\, \frac{1}{(1+\delta/2)!} 
\,\widetilde V_T\, 
\Str'\, D_{\tau_2}^{1+\delta/2}\, \calX ~,
\label{newtrace2}
\eeq 
thereby once again yielding results depending on a primed supertrace.

The general results in Eqs.~(\ref{newtrace1}) and (\ref{newtrace2}) yield a host of new supertrace constraints on the spectrum of the base theory.  In general, the supertrace relations that emerge depend on the number $\delta$ of extra spacetime dimensions which are associated with our decompactification limits. Indeed, taking
$\calX$ of the form 
in Eq.~(\ref{X1X2presplit}),
we may extract these new identities directly from Eq.~(\ref{Cresults}).  

Our
 results are as follows.
Defining
\beq
   \widetilde M^2 ~\equiv~
   \frac{M^2}{4\pi \calM^2}~
\eeq
and recalling Eq.~(\ref{tobereferredto}), 
we see that our usual four-dimensional constraints $C_0=0$ and $\langle \calX\rangle^{(4)}= (\pi/3) C_1$ now take the form
\beqn 
\begin{cases}                &\Str\, \mathbbX_0  ~=~ 0~  \\
 &\langle \calX \rangle^{(4)} ~=~ 
            \displaystyle{\frac{\pi }{3}} \left( \Str \,\mathbbX_1 - \, \Str\,  \mathbbX_0 \mfrac  
            \right) ~
\end{cases} 
\label{eq:X_delta=0}
\eeqn
However, our theorem now tells us that for each decompactification limit we have the additional constraints
 \begin{widetext}
\beqn 
 && \delta=2:~~~ \begin{cases}  
 &\Str' \, \mathbbX_0  ~=~ 0~  \\
  &\Str'\, \mathbbX_1 -  \Str' \left(\mathbbX_0 \mfrac\right)   ~=~ 0~  \\
 &\langle \calX \rangle^{(4)} ~\approx ~          \displaystyle{\frac{\pi }{3}} \widetilde V_T \Bigg[\Str' \, \mathbbX_2 - \Str'\left( \mathbbX_1 
 \mfrac\right)
            + \frac{1}{2} 
            \Str' \left( \mathbbX_0 
 \mmfrac\right)  \Bigg] ~ \label{eq:X_delta=2} \\
 \end{cases} \nonumber\\
&& \delta=4:~~~ \begin{cases}      
&\Str' \, \mathbbX_0  ~=~ 0~  \\
&\Str' \, \mathbbX_1 -
 \Str'  \left( \mathbbX_0 \mfrac\right)  ~=~ 0~ \label{eq:X_delta=4} \\ 
&\Str' \, \mathbbX_2 - \Str'  \left( \mathbbX_1 \mfrac\right)  +  \frac{1}{2} \Str'  
\left(\mathbbX_0 \mmfrac \right)  ~=~ 0~  \\
&\langle \calX \rangle^{(4)} ~\approx~  \displaystyle{\frac{\pi }{3}}  \widetilde V_T \Bigg[  - \Str' \left(\mathbbX_2 \mfrac \right) + 
\frac{1}{2} \Str'  \left( \mathbbX_1 \mmfrac \right)  - 
\frac{1}{6}   \Str' \left( \mathbbX_0 \mmmfrac \right) \Bigg] ~   \\
          \end{cases}\nonumber\\
&& \delta=6:~~~ \begin{cases}  
&\Str' \, \mathbbX_0  ~=~ 0~  \\
&\Str' \, \mathbbX_1 -
 \Str' \left(  \mathbbX_0 \mfrac \right)   ~=~ 0~ \label{eq:X_delta=4} \\ 
&\Str' \, \mathbbX_2 - \Str' \left( \mathbbX_1 \mfrac\right)  +  \frac{1}{2} \Str'\left(  
\mathbbX_0 \mmfrac\right)  ~=~ 0~  \\
& \Str' \left( \mathbbX_2 \mfrac\right)  -
\frac{1}{2} \Str' \left( \mathbbX_1 \mmfrac\right)  +
\frac{1}{6}  \, \Str' \left( \mathbbX_0 \mmmfrac \right)  ~=~ 0~   \\
&\langle \calX \rangle^{(4)} ~\approx  ~\displaystyle{\frac{\pi }{6}} \widetilde V_T \bigg[ \Str' \left( \mathbbX_2 \mmmfrac \right) - \frac{1}{3}  \Str'\left( \mathbbX_1 \mmmfrac\right) 
+ \frac{1}{12}   \Str' \left( \mathbbX_0 \mmmmfrac\right)     \bigg]
~  , \\
          \end{cases}
\label{genXresults_2}
\eeqn
\end{widetext}
where our approximate expressions for $\langle \calX\rangle^{(4)}$ in each case are appropriate for $\widetilde V_T\gg 1$.
Indeed, given these results, we may regard 
the results in Eq.~(\ref{eq:X_delta=0}) as ``$\delta=0$'' constraints, with the understanding that $\Str'\to \Str$ in this case.  We shall continue to refer to our original four-dimensional results as $\delta=0$ results in the following.

Note that the exact value for $\langle \calX\rangle^{(4)}$ for {\it any}\/ compactification volume is given in Eq.~(\ref{eq:X_delta=0}).  The correctness of this result does not change even as the compactification volume becomes large.   However, the behavior of this amplitude for large compactification volume is not readily apparent from the expression in Eq.~(\ref{eq:X_delta=0}).   By contrast, in Eq.~(\ref{genXresults_2}), we have provided expressions for $\langle\calX\rangle^{(4)}$ which approximate this true value for $\widetilde V_T\gg 1$ and in which the compactification $T$-volume $\widetilde V_T$ emerges simply as an overall multiplicative factor.

 These results are completely general, written in terms of the arbitrary $\mathbbX_\ell$ insertions. 
 However, using Eqs.~(\ref{XforHiggs}) and (\ref{eq:Xs}), we can write these expressions directly in terms of the operators relevant for the Higgs mass, as in Ref.~\cite{Abel:2021tyt}, or the one-loop gauge coupling, as in Ref.~\cite{Abel:2023hkk}. 
 
 In the case of the Higgs mass, there is actually an important short-cut that we may exploit.   As shown in Ref.~\cite{Abel:2021tyt}, one can write the one-loop Higgs mass as
 \beq
    m_\phi^2 ~=~\left. \left( \partial_\phi^2 + \frac{\xi}{ 4\pi^2} \right) \Lambda(\phi) \,\right|_{\phi=0} ~
\label{CWrelation}
\eeq
where $\phi$ denotes a fluctuation of the Higgs field relative to its VEV within the Higgsed phase of any string model, where $\xi$ is a model-dependent numerical parameter defined in Eq.~(\ref{eq:xi}), and
where $\Lambda(\phi)$ 
is the amplitude corresponding to the trivial insertions 
\beq
 \mathbbX_0= -\half \calM^4~,~~~ \mathbbX_1=\mathbbX_2=0~
\label{LambdaXinsertions}
\eeq
where the masses of the states in the string spectrum are generally treated as functions of $\phi$.   It is the choice of such functions which specifies the particular scalar field whose mass is being calculated.
Indeed, thinking of $\Lambda$ as a kind of $\phi$-dependent Coleman-Weinberg potential, we thus see from
Eq.~(\ref{CWrelation}) that the Higgs mass is essentially given by the curvature of $\Lambda(\phi)$ around its minimum, augmented~\cite{Abel:2021tyt} by a stringy gravitational backreaction parametrized by $\xi$.
Indeed, $\Lambda\equiv \Lambda(\phi)|_{\phi=0}$ is nothing but the one-loop cosmological constant.

Thus, first performing our analysis for $\Lambda(\phi)$, we find that 
the constraints in Eq.~(\ref{newtrace1}) 
now take the simple form~\cite{Dienes:1995pm}
\beq
      \Str\,M^{2k}(\phi) ~=~0
~~~~{\rm for~all}\/~~~k\leq \delta/2~.
\label{Lambdafirst}
\eeq
Likewise  Eq.~(\ref{newtrace1.5}) yields
\beq 
\Lambda(\phi) ~ \approx ~ \calM^{-\delta}  \, \widetilde V _T \, \Lambda ^{(4+\delta)}(\phi) ~
\eeq
where the higher-dimensional cosmological constant
$\Lambda^{(4+\delta)}$
follows from Eq.~(\ref{newtrace2}) and is given by~\cite{Dienes:1995pm}
\beq
\Lambda ^{(4+\delta) }(\phi) ~=~
  \frac{\pi}{3} \,
  \frac{\calM^{4+\delta} }{2} \,
  \frac{(-1)^{\delta/2} }{\left(1+\delta/2 \right)! } 
\, 
\Str'
\left[ \left( \mfrac \right)^{1+\delta/2} \right] ~
\label{exactLam}
\eeq
where we also regard $\widetilde M^2$ as a $\phi$-dependent quantity.  We observe, in this context, that the $k=0$ equation 
within Eq.~(\ref{Lambdafirst}) is nothing but the constraint~\cite{Dienes:1995pm}
\beq
    \Str \, {\bf 1} ~=~ 0~.
\label{Strbone}
\eeq
This result, which actually applies to {\it all}\/ finite four-dimensional theories regardless of whether they have decompactification limits, implies that all such theories have equal numbers of bosonic and fermionic degrees of freedom {\it when summed across the entire string spectrum}\/.
Indeed, this observation holds even through there is no boson/fermion pairing, and is a non-trivial result of the UV/IR mixing inherent in such theories in which any surpluses of bosonic or fermionic degrees of freedom of a given mass are balanced against opposite surpluses at other masses throughout the string spectrum.     
Likewise, the exact result for the cosmological constant in four dimensions, specifically~\cite{Dienes:1995pm}
\beq
   \Lambda~=~ \frac{1}{24}\,\calM^2\, \Str\,M^2~,
\eeq
is equally surprising, telling us that the full one-loop zero-point amplitude in such theories is given simply as a supertrace of the squared masses of the string states across the string spectrum, with the regulator within the definition of the supertrace in Eq.~(\ref{supertrace_regulated}) ensuring a finite result.

Given these results for $\Lambda(\phi)$, we can now apply the result in Eq.~(\ref{CWrelation}) in order to obtain our corresponding results for the Higgs mass $m_\phi^2$.  In particular, for $\delta=0$ and $\delta=2$ we obtain
\begin{widetext}
\beqn 
&& \delta=0:~~~  
\begin{cases}
& \Str\,{\bf 1} ~=~0\\
& m^2_\phi ~= ~  
\left. \frac{\calM^2}{24}
\bigg[  
\Str\, (\partial_\phi^2 M^2) + \frac{\xi}{4\pi^2\calM^2 } \Str \,M^2  \bigg] \right|_{\phi=0} 
\end{cases}
\label{eq:mphidelta=0}  \nonumber\\
&& \delta=2:~~~ \begin{cases}               
& \Str'\,{\bf 1} ~=~0\\
&
 \Str' \, \partial_\phi^2 M^2
~=~ 0  \\
 & m_\phi^2  ~\approx ~ 
       \left.    -\frac{1}{48\pi  } \widetilde V_T  
           \bigg[
           \Str'\, (\partial^2_\phi M^4 )  +
      \frac{\xi }{4\pi^2 \calM^2 } 
\Str' \,M^4
  \bigg]\right|_{\phi=0} ~, \label{eq:mphidelta=2} \\
 \end{cases}
 \label{Higgsresults}
\eeqn
\end{widetext}
and so forth for higher $\delta$. 
In this connection, we note that the second constraint equation for $\delta=2$ actually takes the full form
\beq
\Str'\, (\partial_\phi^2 M^2) + \frac{\xi}{4\pi^2} \,\Str'M^2 ~=~0~.
\eeq
However, we see from Eq.~(\ref{Lambdafirst}) that the second term actually vanishes, thus leaving us with the simpler constraint equation in Eq.~(\ref{Higgsresults}).  Similar cancellations would likewise occur for $\delta>2$ involving $\Str\,M^{2k}$ for $k>1$.  
Likewise, we emphasize that the {\it constraint}\/ equations in Eq.~(\ref{Higgsresults}) actually hold for all $\phi$, whereas our expressions for $m_\phi^2$ hold by definition only for $\phi=0$.
Finally, we note that
the results in Eq.~(\ref{Higgsresults}) --- or more specifically, their truncations to $\phi=0$ --- could have been  obtained directly using the operator insertions in Eq.~(\ref{XforHiggs}).  
This provides an important cross-check on our calculations.

Before proceeding, we note that the constraint equations in Eq.~(\ref{Lambdafirst}) are highly non-trivial.   When truncated to $\phi=0$, these results provide tight constraints on the masses of all of the states in our theory.  Indeed,  for $\delta=6$, we learn that the states throughout the string spectrum must, through UV/IR mixing, arrange themselves so as to simultaneously cancel not only $\Str\,{\bf 1}$, as in Eq.~(\ref{Strbone}), but also $\Str'\, M^2$, $\Str'\,M^4$, and $\Str' \,M^6$.  However, Eq.~(\ref{Lambdafirst}) actually holds {\it as a function of $\phi$}\/ for all $\phi$. Therefore these equations also constrain how $\phi$ couples to all of the states in the string spectrum (and not just the massless states):  $\phi$ can only couple in a way that
 maintains these cancellations as $\phi$ varies.  This thereby restricts which kinds of Higgs fields are ultimately allowed in the theory.  Likewise, for any Higgs field, 
we may also view this   
as providing a significant constraint on the 
kinds of fluctuations which are ultimately permitted by the modular invariance of the underlying theory.

 Let us now turn to the case of the one-loop contributions to the gauge couplings.   As in Ref.~\cite{Abel:2023hkk}, these one-loop contributions to the quantity $16\pi^2/g_G^2$  will be 
 denoted $\Delta_G$ where $G$ is the corresponding gauge group.  Note in this context that $\Delta_G$ is the full one-loop contribution to $16\pi^2/g_G^2$, and not merely the contribution from the infinite towers of massive states.
 
Once again, just as with the Higgs mass, we may consider the cases with $\delta=0$, $\delta=2$, and $\delta=4$.
Recalling that $\mathbbX_0=0$ for this calculation, we find
\begin{widetext}
\beqn 
&& \delta = 0:~~~ \qquad\Delta_G ~=~ \frac{\xi}{6}  \bigg[ \Str\, \overline{Q}_H^2 - \frac{1}{12}\Str_E\, {\bf 1}\bigg]~\nonumber \\
&& \delta=2:~~~ \begin{cases}                &\Str'\,\overline{Q}_H^2 - \frac{1}{12}\Str_E' \,{\bf 1} ~=~ 0  \nonumber\\
 &\Delta_G ~\approx ~ 
            \displaystyle{\frac{\pi }{3}} \widetilde V_T \bigg[-2\, \Str' \,(Q_G^2  \overline{Q}_H^2) + \frac{1}{6}\Str_E' \,Q_G^2                 -\frac{\xi}{2\pi }   \Str' \left( \overline{Q}_H^2 \mfrac\right)  + \frac{\xi }{24\pi }  \Str_E' \,\mfrac   \bigg] \label{eq:DeltaGdelta=2} \\
 \end{cases} \nonumber\\
&& \delta=4:~~~ \begin{cases}                     &\Str'\, \overline{Q}_H^2 - \frac{1}{12}\Str_E' \,{\bf 1} ~=~ 0\label{eq:DeltaGdelta=4} \\ 
&
-2 \,\Str' \,(Q_G^2  \overline{Q}_H^2) + \frac{1}{6}\Str_E' \,Q_G^2 -\frac{\xi}{2\pi }   \Str' \,  \left(\overline{Q}_H^2 \mfrac \right) + \frac{\xi }{24\pi }  \Str_E'\, \mfrac 
 ~=~ 0  \\
&\Delta_G ~\approx \displaystyle{\frac{\pi }{3}}  \widetilde V_T \bigg[ 2 \,\Str' \,   \left(\overline{Q}_H^2 Q_G^2  \mfrac\right)   -\frac{1}{6} \Str'_E \,\left(Q_G^2  \mfrac \right) + \frac{\xi}{4\pi } \Str'\left(\overline{Q}_H^2  \mmfrac\right)  - \frac{\xi}{48\pi }  \Str'_E \,   \mmfrac   \bigg] ~.  ~~~~~ ~~~
          \end{cases}
\label{genDeltaresults}
\eeqn
\end{widetext}

We again emphasize that these supertrace identities involve the full infinite towers of states in our theories.   In general, these identities are not satisfied level-by-level, but rather represent conspiracies across the infinite string spectrum at all energy scales.   Indeed, these identities are satisfied through the hidden so-called ``misaligned supersymmetry'' that remains in the string spectrum even after spacetime supersymmetry is broken in any modular-invariant tachyon-free string theory~\cite{Dienes:1994np,Dienes:1995pm}. 

We note that we have now collected several combinations of supertraces that have to vanish exactly. One might imagine that these particular linear sums of supertraces could cancel between themselves in unique modular-invariant combinations. However {\it any modular-invariant integral that we can construct which we know to be finite provides its own vanishing supertrace constraint.} Thus we are perfectly at liberty to consider constraints from additional integrals that we also know to be finite even if they do not correspond to any physical process. In Ref.~\cite{toappear_offshell} we play this game and derive even stronger sets of supertrace constraints.

\subsection{Implications for one-loop running:  IR/UV limits, scale duality, and the absence of power-law running
\label{sec:implications_running}}

Our theorem also has 
 ramifications for the 
 effective field theories (EFTs) that might be 
associated with our string theories, and in particular the manner in which these EFTs evolve as we change the relevant energy scale $\mu$ at which they are probed.
 
To study this, let us first briefly recall how one can extract an EFT from a given string theory in a manner that is consistent with modular invariance and which naturally takes the full spectrum of the string theory into account.
Certain aspects of this procedure were discussed previously in Refs.~\cite{Abel:2021tyt, Abel:2023hkk}, where more details can be found.

First, for any four-dimensional amplitude $\langle \calX\rangle^{(4)}$, we introduce a regulator. Then, just as for  ordinary EFTs, we
identify the regulator parameter with a running physical scale $\mu$.

To carry out this procedure we will adopt the same regulator function $ \calG\equiv  \calG_\rho(a,\tau)$ (previously denoted $  \widehat\calG_\rho(a,\tau)$)
  utilized in Refs.~\cite{Abel:2021tyt,Abel:2023hkk}, namely 
\beq
\calG_\rho(a,\tau) ~=~
\frac{a^2}{1+\rho a^2} 
\frac{\rho}{\rho-1} 
\frac{\partial}{\partial a} 
\biggl[   Z_{\rm circ}(\rho a) - 
Z_{\rm circ} (a) \biggr]
\label{operator}
\eeq
where $Z_{\rm circ}$ is the circle-compactification partition function 
in Eq.~(\ref{Zcircdef}) and where $\rho$ is a constant which in effect plays the role of an RG scheme. 
It then follows that our regulated amplitude, which we shall denote $\langle\calX\rangle_\calG^{(4)}$, is given by
\beq
\langle \calX \rangle_\calG^{(4)}~=~
\frac{a^2}{1+\rho a^2} 
\frac{\rho}{\rho-1} 
\frac{\partial}{\partial a} \,
          \biggl[   P(\rho a) - P(a) \biggr]~
\label{operator2}
\eeq
where the ``reduced'' amplitude $P(a)$ is given by
\beq
       P(a) ~= ~ \Bigl\langle  \calX
      \, Z_{\rm circ}(a,\tau)\Bigr\rangle ~.
\label{mainint}
\eeq
Having adopted this convention, we shall then make the identification~\cite{Abel:2021tyt,Abel:2023hkk}
\beq
  \alpha' \mu^2 ~\equiv~   \rho a^2~.
\label{mudef}
\eeq
It then follows that our regulated physical quantities such as $\langle \calX\rangle_{  \calG}^{(4)}$
are functions of $\mu$.

In general, this choice of regulator function $\calG_\rho(a,\tau)$ vanishes exponentially quickly as $\tau_2\to \infty$ (thereby ensuring the effectiveness of this function as an IR regulator) but otherwise asymptotes to $1$ as $\tau_2\to 1$ (thereby preserving the original theory in this regime).
Indeed, a rough measure of the transition between these two behaviors occurs at $\tau_2\approx (\rho a^2)^{-1}$.  Thus $\rho a^2$ sets the ``scale'' at which a given state with mass $M$ is either included amongst or excluded from (or``integrated out'' from) the dynamical degrees of freedom in our analysis, thereby allowing us to make the identification in Eq.~(\ref{mudef}).
We also point out that while this identification procedure for $\mu$ is, strictly speaking, valid only for $\mu\ll M_s$, we shall treat this  as a  {\it definition}\/ of $\mu$ within the entire range $0\leq \mu\leq M_s$.

In the following, without loss of generality, we shall adopt the choice $\rho=2$.
Given these conventions, our goal is to understand how our theorem affects the {\it running}\/ of 
$\langle \calX\rangle_{  \calG}^{(4)}(\mu)$ for arbitrary insertion $\calX$
 --- \ie, affects the manner in which $\langle \calX\rangle_{  \calG}^{(4)}$
varies with $\mu$.
Our goal is also to understand the corresponding beta-function
\beq
\label{eq:beta_def}
\beta_\calX (\mu) ~\equiv~ 
\frac{\partial \langle \calX\rangle_\calG^{(4)} }{\partial \log\,\mu}~.
\eeq

In general, the result for $\langle \calX\rangle_\calG^{(4)}(\mu)$ for arbitrary $\calX$ and arbitrary $\mu$
was derived in Refs.~\cite{Abel:2021tyt, Abel:2023hkk}. This result will also be quoted in Sect.~\ref{KDL}, where explicit examples will be considered.
However, schematically, in cases for which 
$\zStr\,\mathbbX_2=0$ this result takes the form~\cite{Abel:2023hkk}
\beq
\langle \calX \rangle_\calG^{(4)}(\mu)
 ~=~ \langle \calX \rangle^{(4)}
  +  \sum_{i} c_i \,\calK(M_i/\mu)~
\label{eq:schematic}
\eeq
where $\sum_i$ denotes a sum over the on-shell states in the theory and where $\calK(x)$ schematically denotes a generic combination of Bessel functions.  Although a slight generalization of Eq.~(\ref{eq:schematic}) applies when $\zStr\,\mathbbX_2\not= 0$, as discussed in Refs.~\cite{Abel:2021tyt,Abel:2023hkk}, we shall assume that $\zStr\,\mathbbX_2=0$ throughout this discusion.

Let us begin by 
studying the behavior of
$\langle \calX\rangle_{  \calG}^{(4)}$ for extreme values of $\mu$, \ie, for 
$\mu=0$ as well as $\mu=M_s$.   Having fixed the behavior of 
$\langle \calX\rangle_{  \calG}^{(4)}$
at these extreme endpoints, we shall then investigate the implications of our theorem for the intermediate values $0<\mu<M_s$.

We first discuss the IR behavior of $\langle \calX\rangle^{(4)}_\calG(\mu)$, \ie, the behavior of this amplitude as $\mu\to 0$.
This behavior can be extracted~\cite{Abel:2023hkk} from the asymptotic properties of the Bessel function combinations $\calK$, yielding
\beqn
&&\langle \calX\rangle^{(4)}_\calG(\mu)
 ~\approx ~
\langle \calX\rangle^{(4)}
 \nonumber \\
&& ~~~~~~~+
  \sum_{0< M\leq \mu}\, b_M \, \log \left\lbrack \frac{1}{\sqrt{2}} e^{-(\eulerg+1)} \frac{\mu}{M} \right\rbrack ~,\nonumber\\
\label{approxDelta}
\eeqn
where 
\beq
b_M ~\equiv~ 
- 2\mStr \mathbbX_2 ~,
\label{eq:betaM}
\eeq
and where $\eulerg\approx 0.57721$ is the Euler-Mascheroni constant.
Note that the supertrace in Eq.~(\ref{eq:betaM}) is over only those string states whose masses are equal to $M$.  

Next, we discuss the behavior which arises in the extreme UV limit as $\mu\approx M_s$.
As discussed in Refs.~\cite{Abel:2021tyt,Abel:2023hkk}, our result for $\langle \calX \rangle_\calG^{(4)}(\mu)$ generally exhibits a {\it scale duality}\/ symmetry under which 
$\langle \calX\rangle_\calG^{(4)}(\mu)$
is invariant under 
\beq
 \mu ~\rightarrow~ \frac{M_s^2}{\mu}~.
 \label{eq:Mountbatten}
\eeq
Indeed, this symmetry applies to all such amplitudes regardless of the particular operator insertion $\calX$.  However, due to this symmetry and the fact that 
$\langle \calX\rangle^{(4)}_\calG(\mu)$
is a smooth function of $\mu$, we learn that 
\beq
  \frac{d\langle \calX\rangle^{(4)}_\calG(\mu)}{d\mu} \biggl|_{\mu=M_s}  ~=~ 0~.
\eeq
In other words, the corresponding {\it $\beta$-function}\/ $\beta_\calX(\mu)$ actually vanishes at the string scale, implying that there is an apparent {\it UV fixed-point}\/ regime around $\mu=M_s$. 
  As might be expected, this is a purely stringy effect which cannot be captured through an EFT-based analysis.

\label{subsubsec:power}
\label{sec:generalRG}

Having discussed the behavior of $\langle \calX\rangle^{(4)}_\calG(\mu)$ for $\mu\approx 0$ and $\mu\approx M_s$, we 
now investigate the behavior of the {\it running}\/ between these two endpoints.   It is here that our theorem leads to some additional surprising effects, connecting the two extremes $\mu \to  0$ and $\mu\approx M_s$ in sometimes unexpected ways.

In general, for large-volume compactifications, it is a natural (ultimately field-theoretic) expectation that at energy scales exceeding the compactification scale,
the accumulation of contributions from increasing numbers of Kaluza-Klein states running in loops will slowly deform an expected four-dimensional logarithmic running for a given amplitude into a {\it power-law}\/ running, as consistent with the emergence of extra spacetime dimensions (or equivalently an increase in the effective dimensionality of the theory) in this limit~\cite{Dienes:1998vh,Dienes:1998vg}.
More specifically, in the case of the one-loop inverse gauge couplings $\Delta_G(\mu)$,  and at scales $\mu\gsim R^{-1}$
where $R$ is a large compactification radius, 
one generally expects the logarithmic running we observe in Eq.~(\ref{approxDelta}) to follow a power law instead, with $\Delta_G(\mu)\sim (\mu R)^\delta$.
Indeed, in such situations the volume of the compactification manifold can be taken as $V_\delta\sim (2\pi R)^\delta$.  Note that we are here referring to $V_\delta$ as our compactification volume rather than $\widetilde V_T$ because we are discussing our field-theoretic expectations.

It is easy to see how such a result might arise from the logarithmic term within our result in Eq.~(\ref{approxDelta}), specifically the term 
\beq
 \Delta_G(\mu) \,\supset\,
 4 \,\effStr\!\! \left( \overline Q_H^2 - \frac{1}{12}\right) \! Q_G^2\, \log \left\lbrack \frac{1}{\sqrt{2}} e^{-(\eulerg+1)} \frac{\mu}{M} \right\rbrack ~.
\label{logterm}
\eeq
To keep the discussion simple, we shall focus on the most straightforward case in which our compactification is {\it untwisted}\/, so that each state in the theory with a given value of $\overline Q_H^2$ and $Q_G^2$ has an infinite spectrum of KK copies with higher masses but the same values of $\overline Q_H^2$ and $Q_G^2$. Of course, at any energy scale $\mu$, the supertrace is over all states in the theory with masses $0<M\lsim \mu$.     Thus, given this KK structure,
our supertrace {\it factorizes}, \ie, 
\beq
\effStr ~=~ 
     \effStrprime
   ~\cdot~
     \effTr^{\!\!\!\!\!\rm (KK)}~,
\label{sutracefactorization}
\eeq
where $\effStrprime$ is a supertrace over the different states in the theory {\it excluding}\/ the excited KK modes associated with the large dimensions --- indeed, the states in this sum may be regarded as the corresponding KK zero-modes --- and $\effTr^{\!\!\!\!\!\rm (KK)}$ is a trace over the excited KK modes associated with the large compactified dimensions.   Since these excited KK states necessarily have the same spins as their corresponding zero modes, the latter is a trace rather than a supertrace.  In other words, the $(-1)^F$ factor has already been absorbed into the primed supertrace rather than the KK trace.   We further note that we do not consider the effects of any winding modes, since this is meant to be a purely field-theoretic analysis.

Given the factorization in Eq.~(\ref{sutracefactorization}), 
we can rewrite Eq.~(\ref{logterm}) as 
\beqn
&& \Delta_G(\mu) ~\supset~ 
 4 \,\effStrprime \! \left( \overline Q_H^2 - \frac{1}{12}\right) \! Q_G^2\, \nonumber\\
&& ~~~~~~~~~~~~~~\times~
\effTr^{\!\!\!\!\! (\rm KK)} \,
 \log \left\lbrack \frac{1}{\sqrt{2}} e^{-(\eulerg+1)} \frac{\mu}{M} \right\rbrack ~.~~~~~~~~
\label{logtermfactorized}
\eeqn
Let us now focus for simplicity on the $\delta=1$ case in which our theory is compactified  on an untwisted circle of radius $R$.    In such a case, our excited KK spectrum consists of states 
with masses $M_{\widetilde m} \equiv \widetilde m/R$, with $\widetilde m = \pm 1,\pm 2, \pm 3, ...$.
For $\mu R\gg 1$, we then find that the final line of Eq.~(\ref{logtermfactorized}) becomes~\cite{Dienes:1998vh,Dienes:1998vg}
\beqn
&& 2 \sum_{\widetilde m=1}^{\mu R}
 \log \left\lbrack \frac{1}{\sqrt{2}} \,e^{-(\eulerg+1)} \,\frac{\mu R}{\widetilde m} \right\rbrack\nonumber\\
 && ~~~~=~ 2 \,\mu R \, 
 \log \left\lbrack \frac{1}{\sqrt{2}} \,e^{-(\eulerg+1)} \mu R \right\rbrack 
 - 2 \, \log (\mu R)!  ~~~~\nonumber\\
 &&  ~~~~\approx~ 2 \,
\left\lbrace 1+ \log \left\lbrack \frac{1}{\sqrt{2}} \,e^{-(\eulerg+1)} \right\rbrack\right\rbrace  \,\mu R 
\label{FTrunning}
\eeqn
where we used Stirling's approximation $\log N!\approx N\log N - N$ for $N\gg 1$ in passing to the final line.
We thus see that the sum over the KK modes associated with a single compactified extra dimension has changed our logarithmic running to a running which is {\it linear}\/ in
$\mu$.  This process easily generalizes to the KK excitations associated with $\delta$-dimensional compactification manifolds with $\delta>1$, yielding power-law running with correspondingly higher powers $ \Delta_G(\mu)\sim (\mu R)^\delta$.

This phenomenon whereby a sum over 
KK states deforms a running from logarithmic to power-law is well
known from phenomenological studies of theories with
large extra dimensions~\cite{Dienes:1998vh,Dienes:1998vg}.
Indeed, it played  a crucial
role in realizing low-scale gauge-coupling unification~\cite{Dienes:1998vh,Dienes:1998vg}.  As mentioned above, this result can ultimately be understood from the observation that a large
compactification radius $R$ effectively increases the overall
spacetime dimensionality of the theory for all energy scales $\mu\gsim R^{-1}$, thereby shifting
the mass dimensions of the gauge couplings and consequently shifting their corresponding runnings.

Before going further, several remarks are in order.  First, the above discussion has assumed an {\it untwisted}\/ compactification --- this is what enabled the complete factorization of the supertraces in Eq.~(\ref{sutracefactorization}).   However, even for twisted compactifications in which such a complete factorization does not apply, the theory can be separated into individual sectors, and such factorizations are valid within each sector.   One then obtains the same power-law results sector by sector.
Second, the discussion above has neglected the contributions of winding modes.   However, at first glance this would appear to be justified because we are performing a field-theoretic analysis, and also because we are  restricting our attention to cases with $\mu \ll M_s$ and $R^{-1} \ll \mu$.  In general, for large string compactifications with $R^{-1}\ll \mu$, the corresponding winding modes will have masses $\gg \mu$.  Thus --- from a field-theoretic perspective ---  such states will not directly affect the running at the scale $\mu$. Finally, we have taken all of our KK masses as $M= \widetilde m/R$.   This is correct for massless states within the primed supertrace. However, for $\mu < M_s$, this is a valid assumption since we are assuming that the KK modes associated with any large dimensions are already part of the KK trace, the winding states are above $M_s$, and the only other states --- the string excitations --- have masses which are at (or heavier than) the string scale.  Thus the only states which contribute to the primed supertrace are indeed massless.

There is, however, one important shortcoming to the above treatment: 
we did not handle the spacetime compactification in a fully modular-invariant way.   Indeed, we simply split our supertrace into separate pieces without considering the deeper aspects of the effects of the compactification on the underlying {\it partition function}.   However, it is at the level of the partition function that modular invariance must be maintained.
 Although the above treatment captures the expected higher-dimensional power-law running --- and would thus be sufficient for a field-theoretic analysis, as in Refs.~\cite{Dienes:1998vh,Dienes:1998vg} --- it misses the critical fact that 
{\it a fully modular-invariant theory in higher dimensions $D>4$ has more internal cancellations within its spectrum than does a four-dimensional theory}\/.   As we have seen, these extra internal cancellations are required by modular invariance, and in particular can be attributed to various supertrace identities,
such as $\Str\, {\bf 1}=0$,
that result from UV/IR mixing and misaligned supersymmetry.
Indeed the higher the spacetime dimensionality of a modular-invariant theory, the more internal cancellations of this sort exist within the spectrum.   

As we shall now demonstrate, our theorem implies that these cancellations ultimately have the effect of {\it eliminating} the above power-law running that results from these very same extra dimensions!

To understand this, let us begin by recalling that our running four-dimensional amplitude $\langle \calX\rangle^{(4)}$ is given by
\beq
 \langle \calX\rangle^{(4)}(\mu)~=~  \int _\calF
\frac{d^2\tau}{\tau_2^2} \, 
 Z_\calX^{(\rm base)} \, Z_{\rm KK/winding}\, \calG(\mu,\tau) ~.
\label{eq:Lilly}
\eeq
Here $\calG(\mu,\tau)$ is the ``regulator'' function whose purpose is not to regulate this amplitude (since the amplitude is already presumed finite), but rather to introduce the running scale $\mu$.  For example, as we have seen, we may identify $\calG(\mu,\tau)= \calG_\rho(a,\tau)$ where 
$\calG_\rho(a,\tau)$ is defined in
Eq.~(\ref{operator}) and where the scale $\mu$ is defined in terms of $\rho$ and $a$ in Eq.~(\ref{mudef}).
Ultimately, our claim is that $\langle \calX\rangle^{(4)}$ does not run as a power-law function of $\mu$.   However, it turns out that this absence of power-law running is wholly independent of the specific form of $\calG(\mu,\tau)$, and instead relies on deeper, more universal features associated with the modular invariance of the partition function $Z_\calX^{(\rm base)}$.

This can most easily be understood by recasting the above string-theoretic expression into its field-theory analogue.   This will ultimately enable us to make a direct comparison between the two, and thereby uncover the reason for the absence of power-law running in field-theory language.   The field-theory analogue of the expression in Eq.~(\ref{eq:Lilly}) is given by the Schwinger representation of the analogous field-theory amplitude, and takes the form
\beq
 \langle \calX\rangle_{\rm FT}^{(4)}(\mu)~=~  \int_{\Lambda^{-2} }^\infty \frac{dt}{t} \, 
 Z_\calX^{(\rm base)} \, Z_{\rm KK}\, \calG (\mu , t) ~.
\label{eq:Harry}
\eeq
Here $t$ is the Schwinger proper time (with a field-theoretic measure $dt/t$), while $\Lambda$ is an arbitrary cutoff on the Schwinger integral and $Z_{\rm KK}$ denotes a trace over {\it only}\/ the KK modes associated with our $\delta$ extra dimensions (since the winding modes are intrinsically stringy).  Likewise, $Z_\calX^{(\rm base)}$ as before tallies the contributions of the physical states in our theory  with the above KK states excluded.
Of course, in field theory we also now have a UV divergence.  We have therefore introduced a UV cutoff $\Lambda$ in Eq.~(\ref{eq:Harry}).
Finally, in this field-theory expression we are also introducing the running scale $\mu$ just as we would in string theory, through the introduction of a $\calG$-function $\calG(\mu,\tau)$
which suppresses the contributions to the integral from the region of integration with $t\gtrsim \mu^{-2}$.   Once again, as for the string-theory amplitude, the absence of running for this field-theory amplitude will be insensitive to the details of this regulator function.  We can therefore choose to model this function in the most simple way possible, namely as providing a hard step-function cutoff:
\beq 
    \calG(\mu,t) ~=~ \Theta(\mu^{-2}-t)~.
    \label{eq:fieldG}
\eeq
We then have
\beq
 \langle \calX\rangle_{\rm FT}^{(4)}(\mu)~=~  \int_{\Lambda^{-2} }^{\mu^{-2} } \frac{dt}{t} \, 
 Z_\calX^{(\rm base)} \, Z_{\rm KK}\, .
\label{eq:Harry2}
\eeq

The next step, as in string theory, will be to expand both $Z_\calX^{(\rm base)}$ and the product $Z_\calX^{(\rm base)}\cdot Z_{\rm KK}$ in powers of $t$ as $t\to 0$.  In analogy with our string-theory results, we have
\beqn
Z_\calX^{(\rm base)}~ &=&~ \frac{1}{t^2 } (C'_0+C'_1t + C'_2t^2 +\ldots) \nonumber\\
Z_\calX^{(\rm base)}\cdot Z_{\rm KK}
~&=&~ \frac{1}{t^2 } (C_0+C_1t + C_2t^2 +\ldots) ~~~~~~
\label{CCprime}
\eeqn
as $t\to 0$.   Note that in the string-theory calculation we would be instead  $\tau_2$-expanding the  $g(\tau_2)$ functions associated with our partition functions, not the partition functions directly.  However, in this analogous field-theory system our Schwinger time $t$ is the analogue of $\tau_2$;  in particular, we lack an analogue of $\tau_1$.   Thus, whereas in string theory we had both a partition function $Z$ and a corresponding $g$-function, the former involving both physical and unphysical states and the latter involving only physical states,
in our field-theory calculation our partition function is already restricted to physical states.  Thus both $Z$ and $g(\tau_2)$ have the same field-theory analogue, which we may simply regard as $Z$ itself.

It is important to properly interpret the $C$- and $C'$-coefficients in Eq.~(\ref{CCprime}).  If we were to take $\calX={\bf 1}$ as an example,  $C_0$ 
becomes the coefficient of the leading divergence, corresponding to
the power of $1/t^3$ that would appear  in a calculation of the standard Coleman-Weinberg (CW) potential.
Indeed, this would correspond to the quartic $\Lambda^4$ divergence of the CW potential.  Likewise, $C_1$ would correspond to the quadratic divergence of the CW potential, and so forth.

For this discussion, however, we are interested in situations in which we have $\delta$ large extra dimensions of radius $R$.  Our field-theory expectation is that $\langle \calX\rangle_{\rm FT}^{(4)}(\mu)$
in Eq.~\eqref{eq:Harry2}
will experience 
$\log \mu$ running for $\mu \lsim R^{-1}$, but that for $\mu \gg R^{-1}$ this logarithmic running is promoted to power-law running $\sim (\mu R)^\delta$ due to the presence of $Z_{\rm KK}$. 

It is easy to see how these expectations arise.   For $\mu \lsim R^{-1}$, we know that this integral must have at most a logarithmic dependence on $\mu$.
We also know that for $\mu\lsim R^{-1}$ only the zero-modes of $Z_{\rm KK}$ contribute, so that we can approximate $Z_{\rm KK}=1$.
It then follows that $C'_0= C'_1= 0$,
and hence  
\beq 
Z_\calX^{(\rm base)}~ = ~  \frac{1}{t^2} \left( C'_2 t^2 +\ldots \right)  ~~~{\rm as}~t\to 0~.
\eeq
At such energy scales $\mu\lsim R^{-1}$ we thus find 
\beq
 \langle \calX\rangle_{\rm FT}^{(4)}(\mu)~=~   C'_2 \log (\Lambda^{2} / \mu^{2} ) + {\rm const~.}
\label{eq:Harry3}
\eeq
This confirms that for $\mu\lsim R^{-1}$ we indeed obtain the expected logarithmic running.
{\it Moreover, we also see that $C'_2$ can be identified as the beta-function coefficient for this logarithmic running.}

By contrast, let us now consider scales far {\it above}\/ the KK scale, \ie, $\mu \gg R^{-1}$.
In this case we may sum over all of the KK states which are lighter than $\mu$.
This is the same calculation as in Eq.~\eqref{KKlimit}, whereupon we find  
\beq
 \langle \calX\rangle_{\rm FT}^{(4)}(\mu)~=~  \pi^{\delta/2 }R^\delta \int_{\Lambda^{-2} }^{\mu^{-2} } \frac{dt}{t^{1+\delta/2}} \, 
 Z_\calX^{(\rm base)} \, .
\label{eq:Harry4}
\eeq
Thus we identify $C_{j-\delta/2} = \pi^{\delta/2} R^\delta C'_j $, just as we found for string theory
in Eq.~(\ref{C-to-C_relation2}), modulo constants pertaining to the different definitions of the compactification volume.
Moreover,
evaluating our field-theory amplitude, we then find
\beq
 \langle \calX\rangle_{\rm FT}^{(4)}(\mu)~=~  \frac{2}{\delta} \,\pi^{\delta/2}\, R^\delta \,C'_2 \, (\Lambda^\delta -\mu^\delta) ~,
\label{eq:Harry5}
\eeq
as expected.  
This then reproduces the expected power-law running, with $C'_2$ now serving as the beta-function coefficient for this running.  Indeed, $C'_2$ plays this role regardless of $\delta$ --- \ie, regardless of the number of compactified extra dimensions which decompactify in the large-volume limit.

This much is field-theoretic.   However, for $\delta>2$, we have learned from our 
theorem [see Eq.~\eqref{firstCprimeconstraint}]
that modular invariance and misaligned supersymmetry actually force $C'_2=0$. Indeed, this result applies for all $\delta>2$ (although not $\delta =2$ itself).
This, then, kills not only the power-law running in Eq.~(\ref{eq:Harry5}) in cases with $\delta>2$,
but also even the logarithmic running in Eq.~(\ref{eq:Harry3}) coming from those sectors.
{\it We thus conclude that 
in any modular-invariant theory which has a decompactification limit in which $\delta>2$ extra spacetime dimensions appear, there is no running at all from those sectors which are involved in the decompactification process.}
The contributions from such sectors are therefore completely scale-invariant.

The fact that this running is eliminated ultimately rests on the UV/IR mixing inherent in modular invariance and misaligned supersymmetry.
Essentially, for theories with $\delta>2$ decompactification limits, our ``base'' theory (\ie, our theory without the KK/winding excitations) has a non-trivial cancellation 
purely amongst the zero-mode fields at all mass levels, and it is this cancellation that eliminates the power-law running that would have arisen from the KK excitations of such fields.
By the same token, this cancellation also eliminates the running contributions that would have arisen from the winding modes as well.

The case $\delta=2$ is somewhat special
because we can no longer claim that $C'_2=0$.  Thus, for $\delta=2$ we can indeed have logarithmic running for $\mu<R^{-1}$.
{\it However, as we shall now demonstrate,  all power-law running 
for $\mu> R^{-1}$ is eliminated even in this case.}

To see this, let us begin by noting 
that for $\delta=2$ we have
$C_1 = \pi R^2 C'_2$. Thus, just as for Eq.~(\ref{eq:Harry5}), in field theory we would expect a term of the form
\beq
\langle \calX\rangle_{\rm FT}^{(4)}(\mu)~=~   \pi R^2 \,C'_2 \, (\Lambda^2 -\mu^2) ~
\label{wouldbeterm}
\eeq
where $\Lambda$ is, as above, a fixed arbitrary cutoff on the Schwinger integral.
Unfortunately, such a result is inconsistent with the scale-duality symmetry in Eq.~(\ref{eq:Mountbatten}).   Thus, in string theory, the result in Eq.~(\ref{wouldbeterm}) must somehow be ``completed'' to form a scale-duality invariant quantity.  However, as we shall now demonstrate, any such completion will leave us with a fully scale-invariant quantity --- \ie, one which has no dependence on $\mu$ at all.
Thus, in this instance, scale duality actually requires scale invariance above the compactification scale!

It is relatively straightforward to construct a scale-duality invariant ``completion'' of the expression in Eq.~(\ref{wouldbeterm}).  
Bearing in mind that $\Lambda$ is an arbitrary but unknown UV cutoff which we expect to naturally be replaced by $M_s$ within our analysis, we can immediately write a duality-invariant completion of Eq.~(\ref{wouldbeterm}) in the form
\beqn
&& \langle \calX\rangle^{(4)}(\mu)~=~  -\pi R^2 C'_2 M_s^2 \, \times~~~~~~\nonumber\\
&& ~~~~~~~~~~~ \left( 
h({\mu}^2/{M_s^2}) \, \frac{\mu^2}{M_s^2}
 +  h( M_s^2/\mu^2) \, \frac{M_s^2}{\mu^2} 
 \right)~~~~~~~~~~
\label{uniqueform}
\eeqn
where $h(x)$ is an unknown function. 
However, we can immediately list a number of conditions that this $h$-function must satisfy:
\begin{itemize}
    \item First, we must demand that the $\mu\to 0$ limit of Eq.~(\ref{uniqueform}) reproduce
    Eq.~(\ref{wouldbeterm}).   Specifically, as $x\to 0$, this requires that
\beq
  xh(x) + x^{-1} h(1/x) ~\approx ~ 
   a_0 + a_1 x~,
   \label{eq:ansatz}
\eeq
 where $a_0$ and $a_1$ are arbitrary constants and where we are identifying $x\equiv \mu^2/M_s^2$.
\item  Second, because we are restricting our attention to the situation in which only two extra dimensions open up in the decompactification limit, 
the running of $\langle \calX\rangle^{(4)}(\mu)$ cannot involve terms $\mu^p$ for any $p>2$.
\end{itemize}
Our claim, then, is that
any $h$-function satisfying these two conditions must actually have $a_1=0$.
For a function that simply terminates at $x$, this indeed follows trivially from the two assumptions above. The fact that the left side of  Eq.~\eqref{eq:ansatz} is invariant under $x\to 1/x$ in turn implies that 
$a_0 +a_1 x=a_0 +a_1/x$,  which in turn implies $a_1=0$.

Of course one may attempt to  propose functions for the right side of Eq.~\eqref{eq:ansatz} that behave correctly as $x\to 0$ but which would appear to allow a non-trivial dependence on $x$ in the $x\to 0$ limit. For example, one can consider the choice
\beq
h(x) ~=~ \frac{a_0}{x} + a_1 \, \Theta(1-x)~
\eeq
which, as $x\to 0$, yields
\beqn
&&  xh(x) + x^{-1} h(1/x)  \nonumber\\
&& ~~~=~a_0 + a_1 x\,\Theta(1-x) + a_1 x^{-1} \Theta(1-x^{-1}) ~.~~~~~~~~~~
\eeqn
If we further assume that the function $h(x)$ should be differentiable, we can alternatively model the Heaviside $\Theta$-function as a sigmoid: 
\beq
\Theta(1-x)~\to ~ \frac{1+e^{-1}}{1+e^{x^n-1}} ~
\label{eq:sig}
\eeq
for any given $n>0$. We then find that  expanding around $x=0$ yields $x^{n+1}$ terms as well as $x$ terms. However, because we are considering the $\delta=2$ case (with only two extra dimensions opening up), we do not expect any physical power-law running beyond quadratic. If we wish to restrict the running to quadratic this in turn restricts us to $n=0$, which renders the expression in  Eq.~\eqref{eq:sig} equal to a constant. 

One can investigate alternative functions but one always runs into similar problems. Essentially, the underlying issue is that the scale-duality symmetry requires that any quadratic running automatically come along with running that has even higher powers.
Indeed, if we have any power-law running at all, then the required ``turn-over'' of this running near the self-dual point $\mu\approx M_s$ (as required by scale duality)
will itself require power-law running involving even higher powers.
However, we know from our analysis of the complete amplitude
and our result that $C_2'=0$ for $\delta>0$ 
that such higher powers are unphysical.
Thus, the theory ensures its own self-consistency by avoiding all power-law running altogether.
 This argument is, of course, ultimately a consequence of the UV/IR mixing inherent in the scale-duality symmetry.   As such, this absence of running --- like the scale-duality symmetry itself --- is intrinsically a string-theoretic phenomenon.

\begin{figure}
\centering
\includegraphics[keepaspectratio, width=0.45\textwidth]{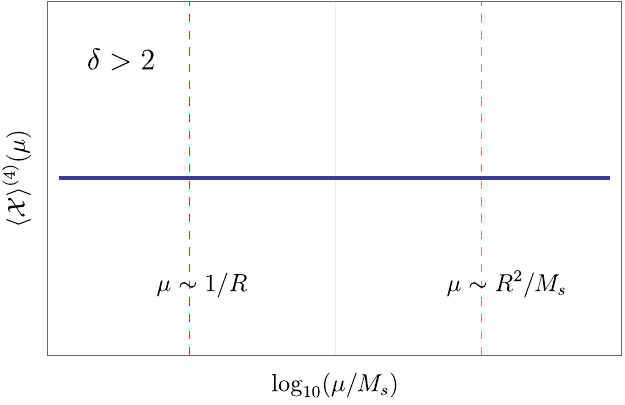}
\vskip 0.2 truein
\includegraphics[keepaspectratio, width=0.45\textwidth]{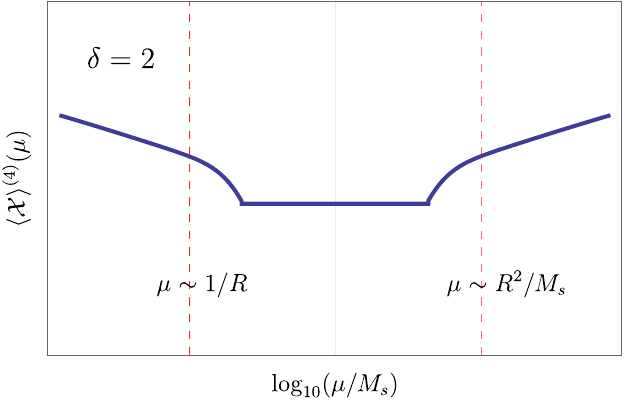}
\caption{Sketches of the generic one-loop running behavior for physical quantities that would otherwise run logarithmically in four-dimensional field theory, but now considered within the context of UV/IR-mixed closed string theories in which $\delta=D-4$ extra spacetime dimensions open up at a common scale $R^{-1}$.   For $\delta=2$ (top panel), the expected field-theoretic logarithmic running exists only at scales $\mu\gsim R^{-1}$ but (after a string-theoretic transient ``pulse'' around the scale $\mu\sim 1/R$) is killed beyond this scale and re-emerges only for $\mu \gsim M_s^2 R$, as mandated by the string-theoretic scale-duality symmetry under $\mu\to M_s^2/\mu$.
dsBy contrast, for $\delta>2$ (bottom panel), the UV/IR-mixing is sufficient to kill the running at {\it all}\/ scales, including the logarithmic running that would have existed at scales $\mu\lsim R^{-1}$.   It may seem strange that the emergence of extra spacetime dimensions at a given scale $\mu_\ast$ can kill the running at scales below $\mu_\ast$, but this is the direct consequence of the UV/IR mixing and misaligned supersymmetry which connects physics at all scales simultaneously.}
\label{fig:stringbeta_generic}
\end{figure}

The end result, then, is that the one-loop running of physical quantities 
$\langle \calX \rangle^{(4)}(\mu)$
in such theories can at most exhibit the behaviors shown in Fig.~\ref{fig:stringbeta_generic},  where we assume for simplicity that all directions of our compactification tori have equal radii $R$.
In particular, for $\delta=2$ (top panel), the physical quantities in our theory can exhibit at most logarithmic running, and this is limited to the $\mu\lsim R^{-1}$ regime.   By contrast,  for $\delta>2$ (bottom panel), these quantities cannot even exhibit logarithmic running at any scale, and our quantities remain scale-invariant at all scales.

Thus, to summarize the results of this section, we arrive at the following corollary of our theorem:

\begin{quote}
    {\it \underbar{Non-renormalization corollary}:~ Within any modular-invariant theory which has $\delta\equiv D-4$ large extra dimensions opening up at a scale $1/R$, misaligned supersymmetry and UV/IR mixing eliminate all running for $\mu \gtrsim R^{-1}$ regardless of the value of $\delta$. For $\mu < R^{-1}$, these same phenomena eliminate all running for $\delta>2$, and leave at most logarithmic running for $\delta =2$.  }
\end{quote}

It is important to understand how this running is eliminated at the level of the actual string spectrum.  Precisely what states are cancelling against what other states, as far as their contributions to the overall running are concerned?

The answer depends crucially on the value of $\delta$.   For all $\delta>2$, as we have seen, the power-law running is cancelled as a result of the vanishing of  $C'_{2}$, or equivalently as the result of a vanishing supertrace over the states in the corresponding base theory.  This means that there is a cancellation between different species $A$ and $B$ of particles in the base theory (or between different {\it collections}\/ $A$ and $B$ of species in the base theory).  Of course, viewed from the perspective of our compactification,  the base theory contains only the KK and winding zero modes.
However, any cancellation between the zero modes of $A$ and $B$ particles also naturally extends to the KK/winding excitations associated with these particles as well.  Thus, for $\delta>2$, we find that the KK modes associated with species $A$ cancel against those associated with species $B$, while the winding modes associated with species $A$ cancel against those associated with species $B$. 

This situation is quite different when $\delta=2$.   In this case, we do {\it not}\/ find that $C'_2=0$.   However, our arguments based on scale-duality invariance tell us that the $\mu$-dependence of the power-law running 
must nevertheless cancel, leaving behind at most a constant.  Because $C'_2\not=0$, this is {\it not}\/ a cancellation {\it within}\/ the base theory, but rather a cancellation within its spectrum of KK and winding excitations.   In fact, we can be more precise:
since the $A$ and $B$ zero modes are not cancelling against each other, the KK modes associated with $A$ are not cancelling with the KK modes associated with $B$, nor are the winding modes associated with $A$ cancelling against the winding modes associated with $B$.
 {\it Instead, what we have is a cancellation of the KK modes of $A$ against the winding modes of $A$, and a cancellation of the KK modes of $B$ against the winding modes of $B$.}

It may seem strange that a KK contribution can ever cancel against a winding contribution, specially since these contributions will generally have very different scales.   However, in a UV/IR-mixed context such as we have here, {\it all}\/ of our states --- both KK and winding --- give rise to effects that are simultaneously felt at {\it all}\/ energy scales in the theory.  

We also hasten to add that the cancellations we have been discussing here concern the contributions to the overall {\it running}\/ of $\Delta_G(\mu)$.
In particular, these are {\it not}\/ cancellations of the total contributions to $\Delta_G(\mu)$ from these states.  For example, if the KK states make a contribution of the schematic form $f(\mu^2/M_s^2)$ to $\Delta_G(\mu)$ within a certain range of energy scales $\mu$, then our assertion is that the corresponding winding states will make a contribution of the schematic form $C-f(\mu^2/M_s^2)$ to $\Delta_G(\mu)$ within that same range, where $C> f(\mu^2/M_s^2)$.    Both contributions have the same overall sign (since the KK and winding states necessarily have the same spin-statistics), and thus together they produce a total contribution which is non-zero.   However, it is the {\it running}\/ that cancels within this total contribution, leaving $\Delta_G(\mu)$ entirely $\mu$-independent.

This can also be understood at an algebraic level from the results we have outlined above.   Recall that our general contributions for $\delta=2$ take the form in Eq.~(\ref{wouldbeterm}).   Within this expression, we may identify the $\mu^2$ contribution as arising from KK states and the $\Lambda^2$ contribution (which we may associate with a fundamental high scale such as $M_s$) as arising from winding states. However what the field-theoretic expression in Eq.~\eqref{wouldbeterm} neglects is the fact that this term can itself carry $\mu$-dependence. Our argument concerning scale duality then guarantees the cancellation of the $\mu^2$ contributions coming from these two terms.  In other words, a dominant (leading) piece coming from the KK contribution is cancelling against a subdominant piece coming from the winding contribution. 
This argument is based on the assumption that $\mu< M_s$.   For $\mu> M_s$, by contrast, the roles of KK and winding states are reversed:  it is the KK states which give rise to the $\Lambda^2$ contribution, and it is therefore the subleading KK contributions which cancel against the leading winding contributions.

Thus far, our discussion has focused on factorizable compactifications in which each of the $\delta$ different compactifications share a common compactification scale $\mu\sim R^{-1}$.    However, just as with our previous results, these results can be extended to more general situations.
These include situations in which the compactification does not have degenerate radii (or more generally compactifications that do not involve a square torus), and also compactifications that are not factorizable. 

In the first of these cases the compactification introduces more than one KK scale into the theory. A typical example is a $\delta$-dimensional compactification on a $\delta$-torus with radii $R_i$, $i=1,2,...,\delta$.
Each of these radii is associated with a different KK scale $M_i \sim R^{-1}_i$.
If the compactification is still factorizable, then our previous discussion still applies. Specifically, at low energies, a state can contribute logarithmic running as usual in four dimensions. However as soon as $\mu$ reaches the lowest KK scale, all contributions to the running must cancel as embodied in our supertrace identities. Therefore there is no running beyond this point, and indeed all running ceases within the window
\begin{equation}
\min_i\left( M_i \right) ~\ll~
 \mu~\ll~ \max_i \left( 
 M_s^2 /M_i \right) ~.
\label{eq:differentKK}
\end{equation}
We will see a simple example of this situation in Sec.~\ref{KDL}, where we consider the case of compactification on a non-degenerate (and non-square) two-torus.   
Of course, for tori with non-trivial shape (complex-structure) moduli as well as non-trivial K\"ahler moduli,
the relevant scales within Eq.~\eqref{eq:differentKK} are
simply those associated with 
the lightest KK mode and the heaviest winding mode, respectively.  In general these can be complicated functions of the radii $R_i$ due to the non-trivial K\"ahler moduli.

As indicated above, our primary assertion is that there is no running within the window in  Eq.~(\ref{eq:differentKK}).   That said, there can (and usually will) be string-theoretic transient effects which do not represent true running but which nevertheless lead to localized changes in the values of the relevant amplitudes  as we transition {\it between}\/ different physical regions.  This is the ``pulse'' phenomenon shown in Fig.~\ref{fig:stringbeta_generic}, and we will see explicit examples of such pulses in Sect.~\ref{KDL}.~   However, our primary assertion stands:  beyond the existence of such pulses, all running ceases above the lightest KK scale.

In the second of these cases, that of non-factorizable compactifications, we have seen that our theorem applies individually for each of the independent base theories $Z^{\rm (base)}_s$.    Thus the states in each base theory will either contribute or not contribute to the overall running depending on the physics associated with that sector alone.

We close with a final comment concerning the implications of these results for cases in which the radii $R^{-1}$ are close to (or even equal to) the string scale.    Thus far in this paper, we have been implicitly assuming $R^{-1}\ll M_s$ --- in other words, that we are dealing with large-radius compactifications.  However, the results we have obtained are independent of the radii $R$, and thus our results will continue to hold even as the KK scales $R^{-1}$ approach $M_s$.  This then has some very important implications. 

For example, let us consider purely toroidal (untwisted) compactifications of ten-dimensional strings down to four dimensions.  The resulting four-dimensional strings clearly have a ``decompactification'' limit with $\delta=6$.  For such an untwisted compactification, this implies that our four-dimensional theory cannot exhibit {\it any}\/ running --- neither logarithmic nor power-law --- for $\mu < R^{-1}$, or equivalently for the entire range $0\leq \mu\leq M_s$.  Of course, if our ten-dimensional theory is supersymmetric, then our resulting four-dimensional theory has ${\cal N}=4$ supersymmetry.
In such a case, our result that this four-dimensional theory exhibits no running is not a surprise and is consistent with existing {\it supersymmetric}\/ non-renormalization theorems.
However, let us now instead imagine that we are toroidally compactifying the ten-dimensional {\it non}\/-supersymmetric tachyon-free $SO(16)\times SO(16)$ string.   In this case, the resulting four-dimensional theory will also be non-supersymmetric.   However, our theorem tells us that this theory will also fail to exhibit any running!    Indeed, in this case the UV/IR mixing inherent in misaligned supersymmetry and modular invariance have accomplished what supersymmetry is no longer available to accomplish, namely to suppress the running in the same way as in the supersymmetric case, only now through non-trivial conspiracies across {\it all}\/ of the states in the string spectrum rather than through boson/fermion pairings.    

Of course, this result applies only for untwisted compactifications.  For twisted compactifications, the situation described in Sect.~\ref{twisted} implies that certain contributions to the running will also be eliminated.  These are the contributions from the sectors that are involved in decompactification limits with $\delta>2$.   However, according to our theorem, logarithmic running may yet emerge from the sectors with $\delta=2$.   This feature is also reminiscent of what occurs in supersymmetric contexts.  However, once again, we are finding that this also holds in {\it non}\/-supersymmetric contexts, purely as a result of UV/IR mixing.

We consider this observation to be quite general.
It is well known that various supersymmetric theories are governed by non-renormalization theorems.  What we are finding, however, is that analogous non-supersymmetric theories {\it also}\/ seem to be governed by the same sorts of non-renormalization theorems.  Thus UV/IR mixing appears to play many of the same roles as supersymmetry.

\section{
 Explicit example:  Gauge couplings in ${\mathbbm T}^2$ string compactifications}

\label{KDL}

The discussion in the previous sections was completely general, and is thus applicable to all tachyon-free closed string theories.   However, it is instructive to revisit a well-known example and perform an explicit calculation within the framework of that example in order to see how the results of our theorem are realized in practice,  and how the cancellation of power-law running actually occurs.   Along these lines, this will also allow us to understand the critical role played by UV/IR mixing, and in particular by our insistence on maintaining worldsheet modular invariance throughout our calculation.

\subsection{Modular-invariant calculation versus traditional calculation}

For this purpose, we shall focus on a calculation which was historically
 at the center of an early triumph in string phenomenology, namely the first calculations of gauge threshold contributions 
 within an $\calN=1$ supersymmetric heterotic string model which is compactified from ten dimensions to four dimensions on an orbifolded six-dimensional torus. 
 This calculation was first performed in Ref.~\cite{Dixon:1990pc} using the formalism established in Ref.~\cite{Kaplunovsky:1987rp}.  The results of this analysis were subsequently developed in many later works and proved highly influential.
However, as we shall discuss, the analysis in Refs.~\cite{Kaplunovsky:1987rp, Dixon:1990pc} did not respect the full (worldsheet) modular invariance of the theory.
As a result, this calculation did not capture all of the features which we are analyzing in this paper.   By redoing this calculation within our fully modular-invariant framework, our goal is to see how the new features we have been discussing arise.   This includes the running of the gauge couplings, and the absence thereof beyond the appropriate KK scales.
For this reason, this calculation will prove to be a useful testing ground for our work.  

As in Ref.~\cite{Dixon:1990pc}, we shall consider the special case in which the six-dimensional torus can be factorized into a four-dimensional torus $\mathbbT^4$ and a two-dimensional torus $\mathbbT^2$.  With certain further assumptions outlined in Ref.~\cite{Dixon:1990pc}, we may disregard the physics associated with the $\mathbbT^4$ compactification.  
 This problem then reduces to a study of the two-dimensional compactification of an $\calN=1$ six-dimensional closed heterotic string to four dimensions on an orbifolded two-torus $\mathbbT^2$.

Finally, since our goal will be to compare our results with those of Ref.~\cite{Dixon:1990pc}, we stress that we will actually be calculating a quantity which is distinct from that calculated in Ref.~\cite{Dixon:1990pc}.
It will therefore be important to understand the difference between these quantities when attempting to make comparisons.

In general, our goal is to calculate $\Delta_G$, \ie, the one-loop contribution to $16\pi^2/g_G^2$.  Here $G$ is any of the unbroken gauge-group factors within the string model in question, 
$g_G$ is the corresponding gauge coupling, and 
 $\Delta_G$ is defined in 
Eq.~(\ref{oneloopcontribution}).
Moreover, we recall that prior to the introduction of a regulator 
we can identify $\Delta_G= \langle \calX \rangle$ where $\calX$ is given in Eq.~(\ref{X1X2presplit}) 
with the operator insertions $\mathbbX_\ell$
given in 
Eq.~(\ref{eq:Xs}).
The quantity $\Delta_G$ will be finite if there are no massless $\mathbbX_2$-charged states in the string spectrum. 

In general, however, such states {\it do}\/ appear, and therefore a regulator will be needed.   As discussed in Sect.~\ref{sec:implications_running},
we shall adopt the modular-invariant regulator function $\calG_\rho(a,\tau)$ described in Sect.~\ref{sec:regulator}.~
We therefore seek to calculate
\beq 
   \Delta_G~\equiv~
   \int_\calF \frac{d^2\tau}{\tau_2^2}
   \, Z_\calX\, \calG_\rho(a,\tau)
\label{ourexpression}
\eeq
where $\calX= \tau_2\mathbbX_1+\tau_2^2 \mathbbX_2$ [with the $\mathbbX_\ell$ defined in Eq.~(\ref{eq:Xs})].  Moreover, through the identification in Eq.~(\ref{mudef}), we may also view this as a scale-dependent quantity $\Delta_G(\mu)$.

In Ref.~\cite{Dixon:1990pc}, by contrast,
a somewhat different quantity $\Delta_{G}^{(\rm DKL)}$ is calculated.
For each gauge group $G$, the quantity $\Delta_{G}^{(\rm DKL)}$ represents a {\it threshold correction}\/ for the running of the corresponding gauge coupling $16\pi^2/g_G^2$ ---  a correction which arises due to the infinite towers of massive states in the string spectrum.   As such, $\Delta_G^{\rm (DKL)}$ tallies
the contributions to the running of $16\pi^2/g_G^2$ from only the {\it massive}\/ states in the theory.  As explained in Ref.~\cite{Kaplunovsky:1987rp}, this quantity is defined as
\beq
   \Delta_G^{(\rm DKL)}~\equiv~
   \int_\calF \frac{d^2\tau}{\tau_2^2}
   \, \left( Z_{\calX'} -\tau_2 \,b_G  \right) 
\label{DKLexpression}
\eeq
where $\calX'\equiv \tau_2^2 \mathbbX_2$ 
and where the beta-function coefficient $b_G$ is given by
\beq
     b_G ~\equiv~ \zStr \, \mathbbX_2~.
\eeq
Indeed, the subtraction of $\tau_2 \,b_G$ within the integrand of Eq.~(\ref{DKLexpression}) reflects the idea that $\Delta_G^{(\rm DKL)}$ includes the contributions of only massive states.  This subtraction also has the further benefit of rendering $\Delta_G^{(\rm DKL)}$ finite. 

We thus see that there are three important differences between our expression for $\Delta_G$ and the corresponding expression $\Delta_G^{(\rm DKL)}$ that is calculated in Ref.~\cite{Dixon:1990pc}.  First, we see that the operator insertion within $\Delta_G^{(\rm DKL)}$ is truncated, including only the $\mathbbX_2$ term but disregarding the $\mathbbX_1$ term.   Second, we see that these expressions utilize different methods of ensuring a finite result:  our expression in Eq.~(\ref{ourexpression}) utilizes a smooth, modular-invariant regulator function $G_\rho(a,\tau)$ that suppresses the potentially divergent contributions from the large-$\tau_2$ region of integration, while the expression in Eq.~(\ref{DKLexpression}) utilizes a sharp, brute-force subtraction of the otherwise divergent contribution from the massive $\mathbbX_2$-charged states. Finally, $\Delta_G^{(\rm DKL)}$ has no energy scale associated with it, but is simply a threshold that must be matched to an effective theory. By contrast, our $\Delta_G$ is calculated within the complete theory, with the regulator itself defining the energy scale.

As a result of these differences, $\Delta_G$ is modular invariant while $\Delta_G^{\rm (DKL)}$ is not.    Indeed, $\Delta_G$ includes the operator insertion $\mathbbX_1$, and
$\mathbbX_1$ is the modular completion of $\mathbbX_2$,  
as discussed in Ref.~\cite{Abel:2023hkk}.
Likewise, the regulator function $\calG_\rho(a,\tau)$ within $\Delta_G$ eliminates the divergences that would have arisen if $b_G\not= 0$, but does so in a fully modular-invariant way.   By contrast, the brute-force subtraction of massless contributions within $\Delta_G^{\rm (DKL)}$ breaks modular invariance, since the UV/IR mixing within modular invariance would have otherwise caused the massless states to mix non-trivially with all of the other states in the theory, thereby rendering such a targeted subtraction impossible.   

At first glance, it may seem surprising that $\Delta_G^{\rm (DKL)}$ 
is not modular invariant.  After all, the well-known expression for this quantity that is ultimately derived in Ref.~\cite{Dixon:1990pc} 
turns out to be a modular function of the compactification moduli.
However, as described above, we are calculating these quantities within the framework of a six-dimensional theory which is toroidally compactified to four dimensions.
There are therefore {\it two}\/ distinct modular symmetries that we expect to play a role in this calculation  --- not only the usual {\it worldsheet}\/ modular invariance that must be unbroken for all closed string theories, but also the {\it spacetime}\/ modular invariance associated with such a two-dimensional $\mathbbT^2$ toroidal compactification.
Indeed, while $\Delta_G^{\rm (DKL)}$ respects the spacetime modular invariance, it lacks the worldsheet modular invariance that is our main interest in this paper.   As such, it does not respect the stringy UV/IR mixing that drives our theorem and its consequences.   By contrast, $\Delta_G$ preserves both modular symmetries simultaneously, as it must.
It is therefore only within the calculation of the properly defined $\Delta_G$ that we expect the consequences of our theorem to become manifest.

\subsection{General setup}

We begin our evaluation of Eq.~(\ref{ourexpression}) by recalling the physical set-up. As discussed above, we assume a six-dimensional heterotic string with $\calN=1$ spacetime supersymmetry compactified on a two-torus $\mathbbm{T}^2$. Our goal is to calculate the amplitude $\Delta_G\equiv \langle \calX\rangle^{(4)}$.
To do this, we shall work within the approximation that our compactification volume is large.   This will not affect our final results but will allow us to make contact with the calculation in Ref.~\cite{Dixon:1990pc}.
Within this approximation, we can then utilize the result for $\langle\calX\rangle^{(4)}$ within the $\delta=2$ portion of Eq.~(\ref{eq:DeltaGdelta=2}):
\beq
\Delta_G ~=~ \frac{\pi}{3} \, \widetilde V_T\,
        C'_2
\label{regV}
\eeq
where
\beqn
C'_2 \,&=&\, 
-2\, \Str' \,(Q_G^2  \overline{Q}_H^2) + \frac{1}{6}\Str_E' \,Q_G^2~~~~~~      \nonumber \\
            && ~~-\frac{\xi}{2\pi }   \Str' \left( \overline{Q}_H^2 \mfrac\right)  + \frac{\xi }{24\pi }  \Str_E' \,\mfrac   ~.~~
\label{hereitisagain}
\eeqn

Before proceeding further, several comments are in order.
First, in stating that Eq.~(\ref{ourexpression}) takes the form in Eq.~(\ref{regV}) we are implicitly utilizing the integral definition for $\widetilde V_T$ in Eq.~(\ref{eq:lamd}), along with an extra factor of the regulator function $\calG_\rho(a,\tau)$ included in the integrand.
Indeed, it is the presence of the regulator which renders this definition for $\widetilde V_T$ valid and equivalent to that in Eq.~(\ref{eq:newdef}). 

Second, we note that the terms in the second line of Eq.~(\ref{hereitisagain}) correspond to the so-called ``$\calY$-terms'' in Ref.~\cite{Kaplunovsky:1987rp,Dixon:1990pc}.
Historically these terms tended to be disregarded because they are universal (\ie, independent of the gauge group in any given string model) and therefore play no role in gauge coupling unification --- a property which cares only about {\it differences}\/ of gauge couplings.  
Moreover, these terms receive contributions from only the massive string states in the base theory and are thus insensitive to any potential running of the gauge couplings at scales below $M_S$.
However, as we have seen, these terms are part of the full modular-invariant calculation and we shall therefore implicitly retain these terms within $C'_2$.

Finally, we also note that the contributions from the massless states within the top line of Eq.~(\ref{hereitisagain}) are nothing but the corresponding beta-function coefficient:
\beq 
C'_2\bigl|_{M=0} ~=~ b_G ~\equiv~ -2\, \zStr  \left( \overline{Q}_H^2  -\frac{1}{12} \right) Q_G^2 ~.
\label{eq:btwiddle}
\eeq
Indeed, these terms are nothing other than the divergent contribution $\zStr\,\mathbbX_2$.

Given these results, we therefore have 
\beq
 \Delta_G(T,U,a)~=~ 
C'_2
\int_{\mathcal{F}}\frac{d^2\tau}{\tau_2^2}  \, \tau_2 \, Z_\twotwo(\tau)  \, \calG_\rho(a,\tau) ~.
\label{eq:genexp}
\eeq
In writing this expression we have reintroduced the full expression for $\widetilde V_T$.   We therefore now seek to evaluate the integral in Eq.~(\ref{eq:genexp})) for a compactification on a two-torus.

To do this, we must know the physical spectrum of our theory.
We will therefore follow standard conventions by taking the general moduli for the two-torus to be  written as 
$T=T_1+iT_2$ and $U=U_1+iU_2$, where for reference the metric and $B$-field are defined in terms of the moduli as 
\beq
G_{ij} = \frac{T_{2}}{U_{2}}\left(\begin{array}{cc}
1 & U_{1}\\
U_{1} & |U|^{2}
\end{array}\right)\,,~~~~
B_{ij} = T_{1}\left(\begin{array}{cc}
0 & 1\\
-1 & 0
\end{array}\right) ~.
\eeq
Note that for $T_1=U_1=0$ (corresponding to rectangular tori without a Kalb-Ramond $B$-field), we can identify
\beq
   T_2 ~=~ M_s^2 R_1 R_2
 ~~~~{\rm and}~~~~
    U_2 ~=~ \frac{R_2}{R_1}~.
\label{T2U2R1R2}
\eeq
Although it will sometimes be useful for orientational purposes to think in terms of such rectangular tori with radii $R_1$ and $R_2$,
in the following we shall nevertheless let all four moduli be non-zero.
However, without loss of generality, we shall adopt the convention that $U_2>1$.

Given this, the $\delta =2$ KK/winding factor $Z_{\rm KK/winding}$ in this case is simply given by $Z_\twotwo$, where
\begin{align}
{Z}_\twotwo (\tau)~&=~\sum_{\vec k, \vec \ell \in \mathbb{Z}^2} e^{-\pi \tau_2 \alpha' M_\twotwo^2} \,
e^{2 \pi i \tau(k_2 \ell_1 - k_1 \ell_2 )} ~
\label{aim}
\end{align}
and where the squared-mass contribution coming from KK/winding states with KK/winding numbers $k_1,k_2,\ell_1,\ell_2 \in \IZ$ 
is given by 
\begin{equation}
\alpha'M_\twotwo^2 ~=~  \frac{|k_1 + Uk_2 + T\ell_1 + TU\ell_2|^2}{U_2 T_2}~.
\label{eq:masssquared}
\end{equation}

Before we present the fully modular-invariant expression for the gauge-coupling contribution,
it is useful to recall for comparison the result of Ref.~\cite{Dixon:1990pc} which was derived by computing the  integral in Eq.~\eqref{DKLexpression}. This gives 
\begin{align}
\Delta^{\rm (DKL)}_G\, &=\,\phantom{-} b_G\int_{\mathcal{F}}\frac{d^2\tau}{\tau_2^2}\big(\tau_2\,Z_\twotwo(\tau)-\tau_2\big) 
\nonumber\\
&=\, -b_G\log\left(
\frac{8\pi e^{1-\eulerg }}{3\sqrt{3}}T_2U_2|\eta(T)|^4|\eta(U)|^4\right),~~
\nonumber\\
\label{eq:KDLint} 
\end{align}
where $b_G$ is given by the coefficient in Eq.~\eqref{eq:btwiddle} and where $\gamma_E$ is the Euler-Mascheroni constant. 
We emphasize again that this $\tau_2$ subtraction, which is done to regulate the integral by removing the logarithmic divergence coming from the massless states, breaks worldsheet modular invariance. Therefore this result does not reflect the worldsheet modular invariance of the full theory from which it came.
However, as this $\tau_2$-subtraction is independent of $T$ and $U$, this integral correctly reflects the {\it spacetime}\/ modular symmetry associated with the spacetime toroidal compactification.  This spacetime modular symmetry is manifest in the $U$- and $T$-dependence of the one-loop gauge-coupling correction, and is inherited from the worldsheet modular symmetry. 

Our goal, of course, is to obtain an expression for $\Delta_G$ which preserves not only spacetime modular invariance but also worldsheet modular invariance.
It is to this task that we now turn.

\subsection{Summary of results}

Before plunging into our analysis, it may help to provide a bird's-eye view  by summarizing our results. 

Our main result, of course, is the evaluation of the full modular-invariant integral in Eq.~\eqref{eq:genexp}.  We shall find that this is highly non-trivial but eventually yields the expression
\begin{align}
\Delta_G & ~=~ \frac{-\,C'_2}{1+a^2 \rho}
\Bigg\{\log(c T_2 U_2  |\eta(T)\eta(U)|^4) \,+\, 2\log \sqrt{\rho} a \nonumber  \\ 
& +\frac{8}{\rho-1}\sum_{\gamma, \gamma' \in \Gamma_\infty \backslash \Gamma} \bigg[\mathcal{\tilde K}^{(0,1)}_0  \bigg(\frac{2 \pi }{a \sqrt{(\gamma \cdot T_2)( \gamma' \cdot U_2)}}   \bigg) \nonumber \\ 
 &\qquad\qquad ~~~~~-\frac{1}{\rho}\mathcal{\tilde K}^{(1,2)}_1  \bigg(  \frac{2 \pi}{a \sqrt{(\gamma \cdot T_2)( \gamma' \cdot U_2)}}\bigg) \bigg]  \Bigg\}~ 
 \label{eq:KDLresult}
\end{align}
where we have adopted the  short-hand notation $\gamma\cdot T_2 \equiv {\rm Im} (\gamma\cdot T)$, where we have defined
\beq 
c ~\equiv ~ 16 \pi^2 \,\rho^{-\frac{\rho+1}{\rho-1}}\,  e^{-2(\eulerg+1)} ~,
\eeq
and where we have defined the Bessel-function combinations~\cite{Abel:2021tyt}
\begin{equation}
\mathcal{\tilde K}_\nu^{(n,p)}(z,\rho)~\equiv~ \sum_{k,r=1}^{\infty} (krz)^n\Big[ K_\nu(krz/\rho)-\rho^p K_\nu(krz)\Big]~.
\label{eq:defK}
\end{equation}
Here $K_\nu(z)$ are modified Bessel functions of the second kind, and $\gamma$ and $\gamma'$ are the {\it spacetime}\/ modular transformations acting in the respective spacetime modular groups associated with  $T$ and $U$.  
 Likewise, $\Gamma_\infty \backslash \Gamma$ is the set of modular transformations which collectively ``unfold''  the fundamental domain of the full modular group $\Gamma$ into the 
``strip'' with $|\tau_2|\leq 1/2$ and $\tau_1>0$.  Indeed, this strip
is nothing but the fundamental domain of the modular subgroup $\Gamma_\infty$ which is generated only by the transformation $\tau\to\tau+1$ and which therefore leaves the cusp at $\tau=i\infty$ invariant.

As we see, the first line of the expression for $\Delta_G$  in Eq.~(\ref{eq:KDLresult}) clearly contains the classic 
moduli-dependent  
pieces  of Eq.~\eqref{eq:KDLint}.
However, we now see that  {\it this result also comes with the additional terms on the second and third lines of Eq.~(\ref{eq:KDLresult})}\/.  
These extra terms not only maintain the spacetime modular invariance that already existed in the first line, but also restore worldsheet 
modular invariance for the entire amplitude, as required.   

It is not difficult to recover the result for $\Delta_G^{\rm (DKL)}$ in
Eq.~(\ref{eq:KDLint}) from the full expression in Eq.~(\ref{eq:KDLresult}).
To see this, we first note that the factor $C'_2$ yields $b_G$ via the identification in Eq.~\eqref{eq:btwiddle}. In addition, within the result of Eq.~\eqref{eq:KDLresult}  the quantities $(\rho,a)$ parametrize the regulator function $G_\rho(a,\tau)$ which appears in Eq.~(\ref{eq:genexp}). Therefore  $\Delta_G^{({\rm DKL})}$ is equivalent to $\Delta_G$  with the logarithmic piece $\log \sqrt{\rho} a$ subtracted by hand and with $a$ sent to zero. In this limit the Bessel functions all vanish, and the remaining difference between the threshold of Eq.~\eqref{eq:KDLint} and the $a\to 0$ limit of Eq.~(\ref{eq:KDLresult}) with subtracted $\log\sqrt{\rho }a $ amounts to a difference in renormalization scheme.
This scheme-dependence is encapsulated within the parameter $c$.  Indeed, 
 by numerically equating the remaining terms we find that equivalent schemes would correspond to $\rho\approx 22$.

Given the result in Eq.~(\ref{eq:KDLresult}), and following the procedures outlined at the beginning of Sect.~\ref{sec:implications_running}, we can also analyze how $\Delta_G$ runs in the full modular invariant theory.
Identifying 
 the physical mass scale $\mu$ as in Eq.~(\ref{mudef}), we find that the running 
behavior for $\Delta_G$ can most easily be described by partitioning the full range of $\mu$ into five different regimes.
For presentational purposes, we shall assume that our $\delta=2$ compactification geometry consists of a rectangular two-torus with radii $R_1$ and $R_2$ with $R_2\gg R_1$  and with no Kalb-Ramond field. We then find that our five separate regions of interest are given by
\beqn
{\rm Region~I}\/:&&~~~
\mu \ll 1/R_2 
\nonumber\\
{\rm Region~II}\/:&&~~~
1/R_2 \ll \mu \ll 1/R_1
\nonumber\\
{\rm Region~III}\/:&&~~~
1/R_1 \ll \mu \ll M_s^2 R_1
\nonumber\\
{\rm Region~IV}\/:&&~~~
M_s^2 R_1 \ll \mu \ll M_s^2 R_2
\nonumber\\
{\rm Region~V}\/:&&~~~
\mu \gg M_s^2 R_2~.
\label{eq:mu-ranges}
\eeqn 
Within each of these regions and far from the boundaries {\it between}\/ these regimes, we  can then evaluate the approximate leading behaviors for the amplitude $\Delta_G$.   As we shall see, this ultimately yields the results
\beqn
&& {\rm Region~I}\/:\nonumber\\
&&~~~~~~~~
\Delta_G~\approx~
     \frac{\pi}{3} \left( M_s^{2} R_1 R_2 + \frac{R_2}{R_1} \right) - 2\log \Big(\mu R_2 \Big)
       \nonumber\\
&& {\rm Region~II}\/:\nonumber\\
&&~~~~~~~~
\Delta_G~\approx~
     \frac{\pi}{3} \left( M_s^{2} R_1 R_2 + \frac{R_2}{R_1} 
     \right)\nonumber\\
&& {\rm Region~III}\/:\nonumber\\
&&~~~~~~~~
\Delta_G~\approx~
     \frac{\pi}{3}  M_s^{2} R_1 R_2  ~\nonumber\\
&& {\rm Region~IV}\/:\nonumber\\
&&~~~~~~~~
\Delta_G~\approx~
     \frac{\pi}{3} \left( M_s^{2} R_1 R_2 + \frac{R_2}{R_1} 
     \right)  \nonumber\\
&& {\rm Region~V}\/:\nonumber\\
&&~~~~~~~~
\Delta_G~\approx~
     \frac{\pi}{3} \left( M_s^{2} R_1 R_2 + \frac{R_2}{R_1}\right) - 2\log \Big(\frac{M_s^2 R_2}{\mu} \Big)~\nonumber\\
\label{eq:massive_foreshadowing_tis_true}
\eeqn

These results provide confirmation of many of our previous assertions.   For example, within Regions II through IV, we see that there is no logarithmic or power-law running at all!    By contrast, within Region~I we have at most a logarithmic running which essentially ceases as we cross from Region~I to Region~II and encounter the lightest KK states.  
Moreover, as expected, we observe that Regions~IV and V are directly related to Regions~II and I respectively under the 
scale-duality transformation in 
Eq.~(\ref{eq:Mountbatten}), while Region~III is self-dual.  Thus the absence of running that we observed in Regions~II and III extends into Region~IV, with logarithmic running re-appearing only in Region~V. 

Two further comments are in order.   First, 
while the above results indicate the leading behaviors for our overall amplitudes $\Delta_G(\mu)$, the results for Regions~II and IV actually contain an overall coefficient which is weakly $\mu$-dependent.  This is the  overall coefficient in Eq.~(\ref{operator}), or equivalently the coefficient $1/(1+a^2 \rho) = 1/(1+\mu^2/M_s^2)$ in Eq.~(\ref{eq:KDLresult}). However,  this dependence is not a true running, but simply represents a residual $\mu$-dependence which is subleading within these regions.    By contrast, for Region~III,  this residual $\mu$-dependence is exactly cancelled within our result for Region~III.~  Thus, within Region~III, our result is fully $\mu$-independent, exhibiting no running at all and remaining truly constant.

Second, we observe that we have limited our analysis to the behaviors of $\Delta_G(\mu)$ within the {\it interiors}\/ of each region in 
Eq.~(\ref{eq:mu-ranges}).  However, interesting behavior can also be found at the boundaries between these different regions.
In particular, we shall 
find that between Regions~II and III, and also between Regions~III and IV, 
we have a transient ``pulse''.  This is not a pure ``running'', but rather a completely stringy phenomenon which serves to connect the 
constant value of $\Delta_G$ which emerges on one side of each of these boundaries with the different constant value of $\Delta_G$ which emerges on the other.

For the record, we emphasize that the above behavior is fairly generic for $\delta=2$, which has the potential for two different compactification radii $R_1$ and $R_2$.   Of course, for $\delta>2$, we have the potential for additional distinct regions opening up due to the possible appearance of additional KK scales.   Indeed, the behavior within all of these regions would continue to be arranged in a manner consistent with the scale-duality symmetry.   However, as we have explained in Sect.~\ref{sec:implications_running},  
we would find that $\Delta_G$ is a constant  in each region.

This concludes the summary of our main results.
Much of the rest of this section is devoted to explaining how these results can be extracted from our general result in Eq.~(\ref{eq:KDLresult}).
However, it turns out that this extraction will entail a number of interesting subtleties and technical maneuvers, some of which revolve around writing our general result in Eq.~(\ref{eq:KDLresult}) in a series of different but equivalent ways.
For this reason, the reader who does not wish to delve into those details and 
who is  willing to accept the above results can skip directly to Sect.~\ref{subsec:lessons}.


\subsection{Evaluating the one-loop contribution to the modular-invariant gauge coupling}

Let us start by explicitly deriving the fully modular-invariant result of Eq.~\eqref{eq:KDLresult}. Our derivation proceeds by evaluating the integral in Eq.~\eqref{eq:genexp}, where for convenience we will henceforth set $C'_2=1$, since $C'_2$ merely provides an overall factor.

In order to perform this calculation we first need to fully specify the physical spectrum over which we will eventually be taking supertraces.  In general, the states appearing in Eq.~\eqref{eq:masssquared} are not level-matched. Therefore we must first determine the spectrum of physical states. 
Given the expression in Eq.~(\ref{aim}), 
these are the states which satisfy the level-matching constraint 
\beq 
k_2 \ell_1 - k_1 \ell_2 ~=~ 0~.
\eeq
This condition straightforwardly yields two sets of solutions for the physical spectrum:
\begin{equation}
(k_1,k_2,\ell_1,\ell_2)
~\in ~  A \cup B
\end{equation}
where
\begin{align}
A &~\equiv~ \{(k_1,k_2,0,0) ~|~  (k_1,k_2) \in \mathbbZ^2 \}~, \nonumber \\
B &~\equiv ~ \{(c \tilde k_1, c \tilde k_2, d \tilde k_1, d \tilde k_2)~| ~(\tilde k_1, \tilde k_2) \in \mathbbZ^2, \nonumber \\ 
   &    \qquad \qquad\text{gcd}(\tilde k_1,\tilde k_2) = 1, ~ (c,d) \in \mathbbZ^2, ~ d \geq 1\}~.
\label{eq:physstates}
\end{align}

As described in Sect.~\ref{sec:implications_running}, knowledge of this spectrum then allows us to determine the fully modular-invariant amplitude  ${\Delta}_G(\mu)$.
As shown in Eq.~\eqref{operator2}, this amplitude can naturally be expressed in terms of a ``reduced'' amplitude $P(a)$, which from Eq.~(\ref{mainint}) now takes the form
\begin{equation}
    P(a) ~=~
     \int_{\mathcal{F}}\frac{d^2\tau}{\tau_2^2}
\,\tau_2\,Z_\twotwo(\tau)\, 
Z_{\rm circ}(a ,\tau)~
\label{eq:regint2}
\end{equation}
where $Z_{\rm circ}$ is defined in Eq.~(\ref{Zcircdef}).
Following the methods outlined
in Ref.~\cite{Abel:2021tyt}, we shall evaluate $P(a)$ by writing
\beq
         P(a) ~=~ P'_1(a) + P'_2(a) ~
\eeq
where
\beqn
        P'_1(a) ~& = &~  \frac{1}{a} \int_{\mathcal{F}}\frac{d^2\tau}{\tau_2^2}
\, \tau_2\,Z_\twotwo(\tau)~-~ \frac{1}{a} \int_t^\infty  \frac{d\tau_2}{\tau_2} ~\nonumber\\
        P'_2(a) ~& = &~  P_2(a) ~+~ \frac{1}{a} \int_t^\infty  \frac{d\tau_2}{\tau_2} ~
\label{Pprimes}
\eeqn
with
\beq
P_2(a) 
       ~=~ \frac{2}{a} \int_0^\infty \frac{d\tau_2}{\tau_2^2} \, \gx(\tau_2) \sum_{\ell =1}^\infty  e^{-\pi \ell^2/(a^2 \tau_2)}~.
\eeq
Note that in Eq.~(\ref{Pprimes}), the final terms for $P'_1(a)$ and $P'_2(a)$ cancel in the sum $P(a)$.  However, the reason for introducing these terms is that they allow us to shuffle
logarithmic divergences between $P'_1(a)$ and $P'_2(a)$, with the arbitrary finite parameter $t$ allowing these two quantities to be independently convergent.      It will be an important self-consistency check on our calculation that all dependence on $t$ will naturally drop out of the sum $P(a)$. 
As discussed in Ref.~\cite{Abel:2021tyt}, this reshuffling does not have to preserve modular invariance in the 
individual terms because modular invariance is  ultimately restored in the sum. 

Note that we can identify $P_1'(a)$ as the ``traditional'' {\it non}\/-modular-invariant minimally-subtracted integral in Eq.~\eqref{eq:KDLint}.  The remaining quantity $P_2'(a)$ 
thus contributes all the ``extra'' terms in Eq.~\eqref{eq:KDLresult} which render our effective cutoffs smooth and moreover restores 
worldsheet modular invariance  to our calculation.

As discussed in Ref.~\cite{Abel:2021tyt}, the integral for $P_1'$ can be performed using the Rankin-Selberg-Zagier techniques~\cite{rankin1a,rankin1b,selberg1} which, upon adapting the results of  Ref.~\cite{Angelantonj:2011br} and performing the sum over the physical spectrum in Eq.~\eqref{eq:physstates}, gives   
\begin{equation}
 P_1'(a)\, =\,  -\frac{1}{a} \log\left( 4 \pi T_2 U_2 |\eta(T)\eta(U)|^4\right) + \frac{1}{a}\log (e^\eulerg t )~.
\label{eq:p1prime}
\end{equation}

To complete the calculation for $P(a)$ --- and indeed to restore the worldsheet modular invariance --- we must add the $P'_2(a)$ integral to this result. Following Ref.~\cite{Abel:2021tyt} and utilizing the same ``unfolding'' techniques as above, this integral can also be expressed as a sum over the physical spectrum in Eqs.~\eqref{eq:masssquared} and \eqref{eq:physstates}. 
This then yields a total expression for $P(a)$:
\begin{align}
    &  P(a) ~=~ -\frac{1}{a} \log(4 \pi T_2 U_2 |\eta(T)\eta(U)|^4) \nonumber\\
    &~~~~~+\frac{8}{a} \sum_{k,r=1}^{\infty} \sum_{\gamma, \gamma' \in \Gamma_\infty \backslash \Gamma} K_0 \bigg(\frac{2 \pi}{a}  \frac{rk}{\sqrt{(\gamma \cdot T_2)( \gamma' \cdot U_2) }} \bigg) \nonumber\\
    &~~~~~\qquad\qquad\qquad\qquad -\frac{1}{a}\log(4\pi a^2 e^{-2\eulerg})~
    \label{eq:PKDL}
\end{align}
where we have adopted the same conventions and notations as described below Eq.~(\ref{eq:defK}).
As required, this result is now invariant under worldsheet modular symmetries.   It is also finite for every $a>0$;  it is independent of $t$, as required; 
and it also exhibits spacetime modular invariance for the compactification moduli $T$ and $U$.

 Note that for practical purposes the sums over $r,k$ can be combined with the sums over the coset to yield the result
\begin{align}
    &  P(a) ~=~ -\frac{1}{a} \log(4 \pi T_2 U_2 |\eta(T)\eta(U)|^4) \nonumber\\[3pt]
    &~~~~~~~\qquad +\frac{8}{a} \sum_{c,d,c',d'}  K_0 \bigg(\frac{2 \pi}{a}  \frac{\left| c T+d \right|\left| c' U+d' \right| }{\sqrt{T_2 U_2}} \bigg) \nonumber\\
    &~~~~~\qquad\qquad\qquad\qquad -\frac{1}{a}\log(4\pi a^2 e^{-2\eulerg})~
    \label{eq:PKDLprimed}
\end{align}
where $(c,d)\in {\mathbbm Z}^2\backslash \{( 0,0)\}$ and $(c',d')\in {\mathbbm Z}^2\backslash \{( 0,0)\}$.

Having determined the reduced amplitude $P(a)$, all that remains is to insert this into
Eq.~\eqref{operator2}, whereupon the full amplitude $\Delta_G$ given in Eq.~\eqref{eq:KDLresult} is obtained. 

\subsection{Running gauge couplings}

Thus far we have calculated the one-loop contribution to the gauge coupling as a fixed quantity, much as was originally done in Ref.~\cite{Dixon:1990pc} except that we have done this in a manner that fully respects worldsheet modular invariance.
However, as we have discussed above and in 
Refs.~\cite{Abel:2021tyt, Abel:2023hkk}, we may now go one step further and proceed to interpret our regulator variables $\rho$ and $a$ as defining a running scale $\mu$ through the identification
$\mu\equiv \sqrt{\rho} a M_s$.
In this way we can 
then interpret our results as yielding a {\it running}\/ coupling in the low-energy effective field theory derived from the string. To map out the behavior of this running, in this section we will  analyze the result in Eq.~\eqref{eq:KDLresult} in various limits and in various energy windows.

In order to derive the behavior of $\Delta_G$ in various limits and regions, as in 
Eq.~(\ref{eq:massive_foreshadowing_tis_true}), 
it will prove convenient to  work with the reduced amplitude $P(a)$ of Eq.~\eqref{eq:PKDL} rather than the full amplitude $\Delta_G$ of Eq.~\eqref{eq:KDLresult}, as the former is far simpler to analyze. However it is important to appreciate that the identification made in Eq.~\eqref{mudef} of the energy scale $\mu$ with $\sqrt{\rho}a M_s$ is a crucial physical step in going from $P(a)$ to $\Delta_G(\mu)$, as discussed at length in Ref.~\cite{Abel:2021tyt}.   Indeed due to the scale-duality symmetry in Eq.~(\ref{eq:Mountbatten}), the alternative identification $\mu\equiv M_s/(\sqrt{\rho}a)$ would be an equally valid choice.   Making this choice is tantamount to choosing which direction of our worldsheet theory should be identified as UV versus IR physics in spacetime.   As discussed in Ref.~\cite{Abel:2021tyt}, it is inevitable that such a choice must be made in order to extract an EFT from our underlying UV/IR-mixed string theory.

We shall now proceed to evaluate $P(a)$ --- and ultimately $\Delta_G(\mu)$ --- within the different regions 
outlined in Eq.~(\ref{eq:mu-ranges}).
To do this, in each case we shall find an approximation for $P(a)$ that is valid within the appropriate region, and thereby deduce the leading running that emerges within that region.  This requires some care, as the approximations that are appropriate in each case are very different from each other.  We shall therefore be relatively explicit in how these approximations are made in each case.

\subsubsection{Region I:  Field-theory limit}

Having chosen to identify $\mu= \sqrt{\rho}a M_s$, let us now as a first step examine the low-energy behavior of Region~I which extends to $\mu\to 0$, or equivalently to $\sqrt{\rho}a\ll 1$.
 Here we expect that Eq.~\eqref{eq:KDLresult} yields the effective four-dimensional field-theoretic behavior together with finite gauge thresholds. Indeed, we know that in this limit the regulator should ``turn off'' entirely, \ie, ${\cal{G}}_\rho(a,\tau) \rightarrow 1$, whereupon the original unregulated integral should be restored. Of course the unregulated integral had a  logarithmic divergence  that emerges because $\zStr\,\mathbbX_2\not= 0$.  Thus we expect that in this limit Eq.~\eqref{eq:KDLresult} gives the threshold correction result in Eq.~\eqref{eq:KDLint} plus a term that diverges logarithmically as $\mu\to 0$.  

This is indeed what happens.
Recognizing the Bessel-function asymptotic behavior
$K_\nu(z) \sim \sqrt{\pi/(2z)}\,e^{-z}$ as $z\rightarrow\infty$,
we see that  the terms involving Bessel functions do not survive in this limit.
We then find 
by  direct analysis of Eq.~(\ref{eq:KDLresult}) that the only remaining terms are 
\beq
\Delta_G (\mu) ~\approx~ -\log(c\, T_2 U_2 |\eta(T)\eta(U)|^4) -2\log \Big(\frac{\mu}{M_s}\Big)~
\label{eq:stringthresh}
\eeq
as $\mu\to 0$.
Assuming $T_2\gg1$ and recalling that $\eta{(iz)} \sim e^{-\pi z/12} $ for $z\gg 1$,  we can extract the leading volume dependence of $\Delta_G(\mu)$. Up to additional terms of order unity, we obtain 
\beq 
 \Delta_G(\mu) ~\approx~ \frac{\pi}{3}(T_2+U_2) -\log \Big(\frac{\mu^2}{M_s^2}T_2U_2 \Big)~
 \label{eq:d1}
\eeq
as $\mu\to 0$.  
This approximation is in accord with the $T$-volume scaling relation in Eq.~\eqref{eq:genexp}, and is valid in the region $a\ll 1/\sqrt{T_2 U_2}$ or equivalently $\mu\ll 1/R_2$ for the rectangular torus, yielding the Region~I~behavior described in Eq.~\eqref{eq:massive_foreshadowing_tis_true}.  
Note that this is the energy scale below which Kaluza-Klein modes are effectively inactive and may be considered to have been integrated out.

It is important to understand the  differences between our modular-invariant result $\Delta_G(\mu)$ in Eq.~\eqref{eq:stringthresh} and the traditional  result $\Delta^{\rm (DKL)}$ of Eq.~\eqref{eq:KDLint}. Indeed, explicitly restoring the factors of $b_G$, we see that the two are related for $\mu\ll M_s$ via 
\beq 
\Delta_G(\mu) ~=~  -b_G\log 
\frac {\mu^2}{M_s^2} - b_G 
\log  
\left( 
\frac{3\sqrt{3}c }
{8\pi e^{1-\eulerg}}
\right) +
 \Delta^{\rm (DKL)}~.
\label{eq:comparison}
\eeq
Thus, 
while $\Delta_G(\mu)$ is an amplitude which is a function of a running mass scale $\mu$, we see that $\Delta^{\rm (DKL)}$ is a scale-independent {\it threshold correction}.  Of course, as we have repeatedly stressed, 
$\Delta_G^{\rm (DKL)}$ is not a modular-invariant quantity.  It is only the full expression $\Delta_G(\mu)$ which respects the full modular invariance of the theory.

We thus see that our modular-invariant calculation not only keeps track of the running, but also keeps track of the natural degrees of freedom that are dynamical in the EFT associated with the scale $\mu$.  
As explained in more detail in Refs.~\cite{Abel:2021tyt} and \cite{Abel:2023hkk}, for any scale $\mu$ our regulator implicitly keeps track of which states can be classified as either ``light'' (with masses $M\lsim \mu$) or ``heavy'' (with masses $M\gsim \mu$):
the contributions from the heavy states are suppressed by our regulator and have thus effectively been integrated out, while the contributions from the light states are retained.   Indeed, this establishes $\mu$ as a floating mass scale, which in turn enables us to take  $\mu$ to $M_s$ and beyond. 

\subsubsection{Region~V:  Ultra-high energies}

Let us now consider what happens in the $\mu\rightarrow\infty$ limit. In order to do this we note that, as discussed in Ref.~\cite{Abel:2021tyt}, the scale-duality symmetry $\mu\to M_s^2/\mu$ requires that the  reduced amplitude $P(a)$ have an $a\to 1/a$ symmetry, and this is in turn ensured by the  explicit $a\rightarrow 1/a$ symmetry of $Z_{\rm circ}(a)$. Thus in order to study the $a\to\infty$  limit 
we can simply replace $a\to 1/a$ in Eq.~\eqref{eq:PKDL}, yielding a dual $P(a)$ of the form
\beq
    P(a) \,=\, -a \log(4 \pi T_2 U_2 |\eta(T)\eta(U)|^4) 
   ~-~a\log\bigg(\frac{4\pi e^{-2\eulerg}}{a^2} \bigg)~~~
\label{eq:alternativePKDL}
\eeq
up to Bessel-function terms which are exponentially suppressed as $a\to\infty$.   This then yields 
\beq
\Delta_G ~\approx~ -\log\Big({c}\, T_2 U_2 |\eta(T)\eta(U)|^4\Big) + 2\log \Big(\frac{\mu}{M_s} \Big)~
\label{eq:inflimit}
\eeq
as $\mu\to \infty$.
Thus for $T_2\gg1$ the leading volume dependence up to additional terms of order unity becomes 
\beq 
 \Delta_G(\mu)  ~\approx~\frac{\pi}{3}(T_2+U_2) + \log \Big(\frac{\mu^2}{M_s^2}T_2U_2 \Big)~~~
 {\rm as}~\mu\to\infty~.
 \label{eq:d2}
\eeq
Indeed, for the rectangular torus  this approximation is valid for $a\gtrsim\sqrt{T_2 U_2}$, or 
equivalently for
\begin{equation}
 \mu  \, \gtrsim \, R_2 M_s^2 ~,
\end{equation}
yielding the Region~V~behavior of Eq.~\eqref{eq:massive_foreshadowing_tis_true}.

\subsubsection{Regions~II, III, and IV:  Stringy energies}

Since both the $\mu\rightarrow 0$ and $\mu \rightarrow\infty$ limits  display only logarithmic behavior, one might then suspect that power-law running behavior is present near the self-dual point $a=1$. Indeed, motivated by the field-theoretic result in Eq.~\eqref{wouldbeterm}, one may even suspect that in this region the contribution from $P_2'$ cancels that from $P_1'$ so as to give $\Delta_G\to 0$ at the self-dual point. 

To investigate this, let us focus more closely on  the  $a\approx 1$ region by writing $P_2'$ in the alternative form
\begin{align}
& P_2'(a) ~=~ 
\frac{1}{a} \log(4 \pi T_2 U_2 |\eta(T)|^4 |\eta(U)|^4) -\frac{1}{a} \log (e^\eulerg t) \nonumber \\
&
-~\sqrt{T_2 U_2} \log 
\left|
\eta \left( i a \sqrt{T_2/U_2} \right) \eta 
\left( i a^{-1} \sqrt{T_2/U_2} \right) 
\right|^4 \nonumber \\
& -~4  \sum_{
\stackrel{
\gamma \in
\Gamma_\infty \backslash \Gamma}
{
\gamma \neq\mathbbm{1} } }\,
\sum_{k=1}^{\infty}
\Bigg\{ 
\sqrt{T_2 (\gamma \cdot U_2)}
\log\left(1-e^{-2 \pi ak \sqrt{\frac{T_2}{\gamma \cdot U_2}}}\right) \nonumber \\
 &\quad \quad \quad \quad \quad \quad +\sqrt{U_2 (\gamma \cdot T_2)} \log\bigg(1-e^{-2 \pi a k\sqrt{\frac{U_2}{\gamma \cdot T_2}}}\bigg) \Bigg\} \nonumber\\
&+\,\frac{8}{a} \sum_{
\stackrel{
\gamma \in
\Gamma_\infty \backslash \Gamma}
{
\gamma \neq\mathbbm{1} } }
\,\sum_{k,r=1}^{\infty} 
K_0 \bigg(\frac{2 \pi}{a}  \frac{rk}{\sqrt{(\gamma \cdot T_2 )(\gamma' \cdot U_2)}}\bigg)\, . \label{eq:alternativeP2} 
\end{align}
Note that in writing Eq.~(\ref{eq:alternativeP2}) we have neglected terms that vanish in the derivative definition of $\Delta_G(\mu)$ in Eq.~\eqref{operator2}.  
Rendering the sums explicitly, we then find that this becomes 
\begin{align}
& P_2'(a) ~=~ 
\frac{1}{a} \log(4 \pi T_2 U_2 |\eta(T)|^4 |\eta(U)|^4) -\frac{1}{a} \log (e^\eulerg t) \nonumber \\
&
\qquad~~ - ~\sqrt{T_2 U_2} \log 
\left|
\eta \left( i a \sqrt{T_2/U_2} \right) \eta 
\left( i a^{-1} \sqrt{T_2/U_2} \right) 
\right|^4 \nonumber \\
& \qquad~~- ~4  \, \sum_{\stackrel{ {\mbox{\scriptsize $c,d$} }}{c\neq 0}  }
\,\Bigg\{ 
\frac{\sqrt{T_2U_2}}{|cU+d|} 
\log\left(1-e^{-2 \pi a |cU+d| \sqrt{T_2/U_2}} \right) \nonumber \\
 & \quad \quad \quad \quad \quad ~~~ +
 \frac{\sqrt{T_2U_2}}{|cT+d|} 
\log\left(1-e^{-2 \pi a |cT+d| \sqrt{U_2/T_2}} \right)  \Bigg\} \nonumber\\
&\qquad ~~~ +\,\frac{8}{a} 
\sum_{\stackrel{ {\mbox{\scriptsize $c,d,c',d'$} }}{c,c'\neq 0}  }
K_0 \bigg(\frac{2 \pi}{a}  \frac{\left| c T+d \right|\left| c' U+d' \right| }{\sqrt{T_2 U_2}} \bigg) ~ ,
\label{eq:alternativeP2prime} 
\end{align}
where as before $c,d, c',d'\in {\mathbbm Z}$.

Before proceeding further, we observe that the calculation that leads to this result requires considerable care. 
 In particular, we must regulate all of the terms using methods similar to those in Ref.~\cite{Dixon:1990pc}. 
The steps in this calculation are as follows.  First one splits the spectrum of $M_\twotwo^2$ in Eq.~\eqref{eq:masssquared} into three sets: the first has $l_1 = l_2 = 0$ and coincides with the $A$~set in Eq.~\eqref{eq:physstates}; the second has $k_2 = l_2 = 0$;  and the third has $l_2 \neq 0$.  In performing the calculation for the first two sets we Poisson-resum $(k_1, k_2)$ and $(l_1,k_1)$ respectively. Then, performing a partial sum over the greatest common divisors of $(k_1,k_2)$ and $(l_1,k_1)$ and subtracting the double-counted contribution from the states with $l_1 = l_2 = k_2 =0$ gives the first four lines of Eq.~\eqref{eq:alternativeP2}.  Finally, the contribution from  the third set of states is computed directly without any Poisson resummation, 
and yields the Bessel functions on the final line.  These simply resemble the equivalent terms appearing in the previous incarnation of $P_2'$ in Eq.~\eqref{eq:PKDL}. 

The expression in Eq.~\eqref{eq:alternativeP2prime} is most useful for considering $P_2'$ near
$\mu\approx M_s$.  Moreover, the first line of Eq.~\eqref{eq:alternativeP2prime} precisely cancels $P_1'$ in Eq.~\eqref{eq:p1prime}.  Thus,  at first sight 
it seems that the one-loop contribution to the gauge-coupling could indeed cancel in this region in a manner that mimics Eq.~\eqref{wouldbeterm}.

However, this is not the case.
To see this, let us consider the window 
\begin{equation}
\frac{1}{\sqrt{T_2U_2}} ~ \ll ~ a ~ \ll ~ \sqrt{T_2 U_2} ~.
\label{eq:window0}
\end{equation}
For rectangular tori this corresponds to 
\begin{equation}
1/R_2 ~ \ll ~ \mu  ~ \ll ~ {M_s^2}R_2 ~,
\label{eq:window}
\end{equation}
which may be taken as defining
an ``extended stringy regime'' that encapsulates Regions~II, III, and IV.~  By inspection we find that within this window the terms in the third through fifth lines of Eq.~\eqref{eq:alternativeP2} are all exponentially suppressed. 
Therefore, neglecting these exponentially suppressed terms and adding $P_2'$ to  
$P_1'$ in Eq.~\eqref{eq:p1prime} we obtain a very compact approximation for $P(a)$ in this region:
\beq
 P(a)  ~\approx~
-\sqrt{T_2 U_2} \log 
\left|
\eta \left( i a \sqrt{\frac{T_2}{U_2}} \right) \eta 
\left( i a^{-1} \sqrt{\frac{T_2}{U_2}} \right) 
\right|^4 ~.
\label{eq:alternativeP}
\eeq
Defining for convenience a dimensionless energy scale
$\muresc\equiv \mu/M_s$,
we find that Eq.~\eqref{operator2}  near $\hat\mu\approx 1$ 
gives 
\beqn
    \Delta_G &\approx & \frac{\pi}{3}\, T_2 \, \frac{E_2\left(\frac{i}{\muresc}\sqrt{\frac{T_2}{\rho U_2}}\right)}{(1-\rho)(1+\muresc^2)} ~ + ~\rho \rightarrow \frac{1}{\rho}   \, ~ \nonumber \\
    ~~ && ~~~~~+~\,  \muresc\rightarrow \frac{1}{\muresc}  \,~+~\,  (\rho,\muresc) \rightarrow  \left(\frac{1}{\rho},\frac{1}{\muresc}\right)  ~~~~~~~~~
\label{eq:DeltaGPlanck}
\eeqn
where $E_2$ is the Eisenstein function defined in Eq.~\eqref{defG2here} and where each substitution within the right side of  Eq.~(\ref{eq:DeltaGPlanck}) is meant to operate purely on the first term (thereby leaving us with a total of only four terms on the right side of this equation).  
Given the definition of the Eisenstein function, we see that for large $T_2$ all the $E_2$ factors are unity up to terms that are exponentially suppressed as long as $M_s\sqrt{U_2/T_2}\ll \mu \ll M_s \sqrt{T_2/U_2}$.  Of course, this is nothing but Region~III.~ Thus, within Region~III we find that Eq.~\eqref{eq:DeltaGPlanck} is exponentially well-approximated by 
\beq 
 \Delta_G(\mu) ~\approx~
\frac{\pi}{3} \,T_2 ~~~~{\rm for}~~\mu\approx M_s~.
\label{eq:d3}
\eeq
Recalling that $T_2$ is the $T$-volume $\widetilde V_T$, we thus
verify, as anticipated,
that there is no quadratic $\mu$-dependence --- or indeed running of any kind --- within this energy-scale window. 
Indeed, within this region, $\Delta_G(\mu)$ is exponentially well approximated by a constant.

Finally, going beyond this region, we see that Eqs.~\eqref{eq:d1}, \eqref{eq:d2}, and \eqref{eq:d3} together imply that 
$\Delta_G(a) \approx
(\pi/3) T_2$ everywhere.
Indeed, this holds regardless of the energy scale $\mu$ so long as $U_2\approx 1$.

\begin{figure}
\begin{center}
\includegraphics[width=0.495\textwidth]{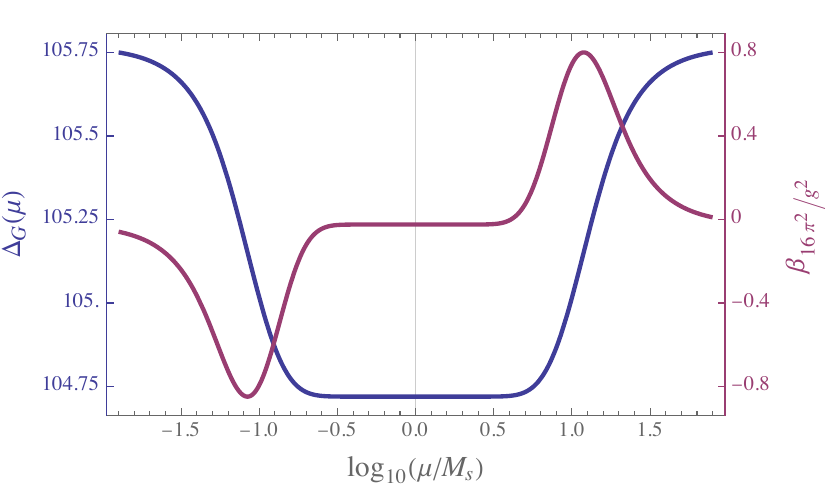}
\caption{The precise one-loop contribution to $16\pi^2/g^2$ (blue) in Eq.~(\ref{eq:DeltaGPlanck}), 
along with the corresponding beta function (red) in Eq.~(\ref{eq:beta_def}), plotted as functions  of the energy scale $\mu$ within the extended stringy regime of Eq.~(\ref{eq:window}) for a toroidal $\mathbbT^2$ compactification with compactification moduli 
$T_2=100$ and  $U_2=1$. 
These curves are evaluated using the full expression in Eq.~\eqref{eq:alternativeP},
and provide an explicit example of the behavior anticipated in Fig.~\ref{fig:stringbeta_generic}.
\label{fig:stringbeta_oo} }
\end{center}
\end{figure}

The full analytical expression within Eq.~\eqref{eq:DeltaGPlanck}
can also be evaluated numerically.
In Fig.~\ref{fig:stringbeta_oo} 
we plot an example of the resulting running behavior for $\Delta_G(\mu)$
that emerges in the case of a square torus with $T_2=100$ and $U_2=1$ within the extended stringy regime of Eq.~(\ref{eq:window}) (blue curve).  The fact that this torus is square implies that our two KK scales $1/R_1$ and $1/R_2$ collapse to become the same scale, thereby rendering Regions~II and IV non-existent.
We also plot the corresponding beta function (red curve). 
We observe, as promised, that there is essentially no running within this regime;  for example, within the range 
$|\log (\mu/M_s)|\lsim 0.7$ our blue curve is completely flat.  However, for 
$|\log (\mu/M_s)|\gsim 0.7$ we
see that there exists a small
transient ``pulse'' as the theory crosses the KK thresholds on either side of this regime, in accord with the behavior anticipated in Fig.~\ref{fig:stringbeta_generic}.
We also observe that the string-scale value of $\Delta_G$ differs from the low-energy value near $\mu\approx 0$ by a term that grows with volume only as $\log T_2$ rather than as the total number of Kaluza-Klein modes with masses below the string scale.  Indeed, the latter behavior would have been expected  if there had been a dominant region of power-law running between the low scale and $M_s$.

This behavior changes if our compactification torus is not square.   As an example, let us consider the case in which our compactification torus is {\it rectangular}\/, with $U_2\gg 1$
but a purely imaginary shape modulus.  In this case Regions~II and IV open up and become part of our extended stringy regime,
potentially inducing an arbitrarily large 
splitting of the KK scales associated with each of the two dimensions of the torus.
In such cases, Regions~II, III, and IV become separated within the extended stringy regime.  Within this extended regime there is suppressed running of 
$\Delta_G(\mu)$ with transient behavior localized near the boundaries.
Once again, field-theoretic running ceases as soon as the first KK states are encountered, as described in Sect.~\ref{sec:implications}.

We can learn more about the nature of this transient ``pulse'' by concentrating on Regions~II and IV.~
Recall that our expressions for
$P(a)$ and $\Delta_G(\mu)$ in Eqs.~(\ref{eq:alternativeP})
and (\ref{eq:DeltaGPlanck}) respectively
were valid across the entire extended stringy regime in Eq.~(\ref{eq:window}) consisting of Regions~II, III, and IV. 
By contrast, the approximate result in Eq.~(\ref{eq:d3}) was valid only within Region~III.~
Indeed, within Region~II the argument appearing in  $\eta \left( i a \sqrt{{T_2}/{U_2} } \right)$ is no longer large. 
 However, within this window we may modular-transform the first Dedekind eta function of Eq.~\eqref{eq:alternativeP} to find yet another form for $P(a)$:
\beqn
  P(a) & \approx&
-\sqrt{T_2 U_2} \log 
\left|
\eta \left( i a^{-1} \sqrt{\frac{U_2}{T_2}} \right) \eta 
\left( i a^{-1} \sqrt{\frac{T_2}{U_2}} \right) 
\right|^4 \nonumber \\
&&
\qquad \qquad 
-\sqrt{T_2U_2} \log \left( a^2\frac{U_2}{T_2} \right)     
~.
\label{eq:yetanotherP}
\eeqn
Neglecting constant terms that vanish under the derivative definition of $\Delta_G$ in Eq.~\eqref{operator2}, this then becomes 
\beq
P(a) ~\approx~  \frac{\pi }{3}\,\frac{1}{a} \, ( T_2 + U_2) - 2\sqrt{T_2U_2} \log  a~.
\label{eq:surely_the_last_P}
\eeq 

Given the $\log a$ term within 
Eq.~(\ref{eq:surely_the_last_P}), it may at first sight seem that we have discovered  linear power-law running
within Region~II.~  Indeed the individual terms in Eq.~\eqref{operator2} are proportional to $a^2 \partial_a P(a) = a$, which, when multiplied by the prefactor $\sqrt{T_2U_2}\equiv M_s R_2$, would imply linear contributions to $\Delta_G$ that are proportional to $\mu R_2$.  
As we have discussed above, this would appear to make sense because Region~II is an approximately five-dimensional regime, with $1/R_2 \ll \mu \ll 1/R_1$.   However, what actually appears in the definition of $\Delta_G$ in Eq.~\eqref{operator2} is not a single $P(a)$ but rather the combination $P(\rho a)- P(a)$, and within this difference these logarithm terms cancel!  Indeed, simple $\log(a)$ terms within $P(a)$ do not make contributions to $\Delta_G$. This is not a coincidence, since 
the appearance of the combination
$P(\rho a)- P(a)$ is ultimately dictated by the scale-duality symmetry.  We therefore conclude that we do not have any running in Region~II.~ Needless to say, the same conclusion also applies to the dual Region~IV.~

\FloatBarrier 
\subsection{Putting it all together\label{alltogethernow}}

\begin{figure*}
\includegraphics[width=0.8\textwidth]{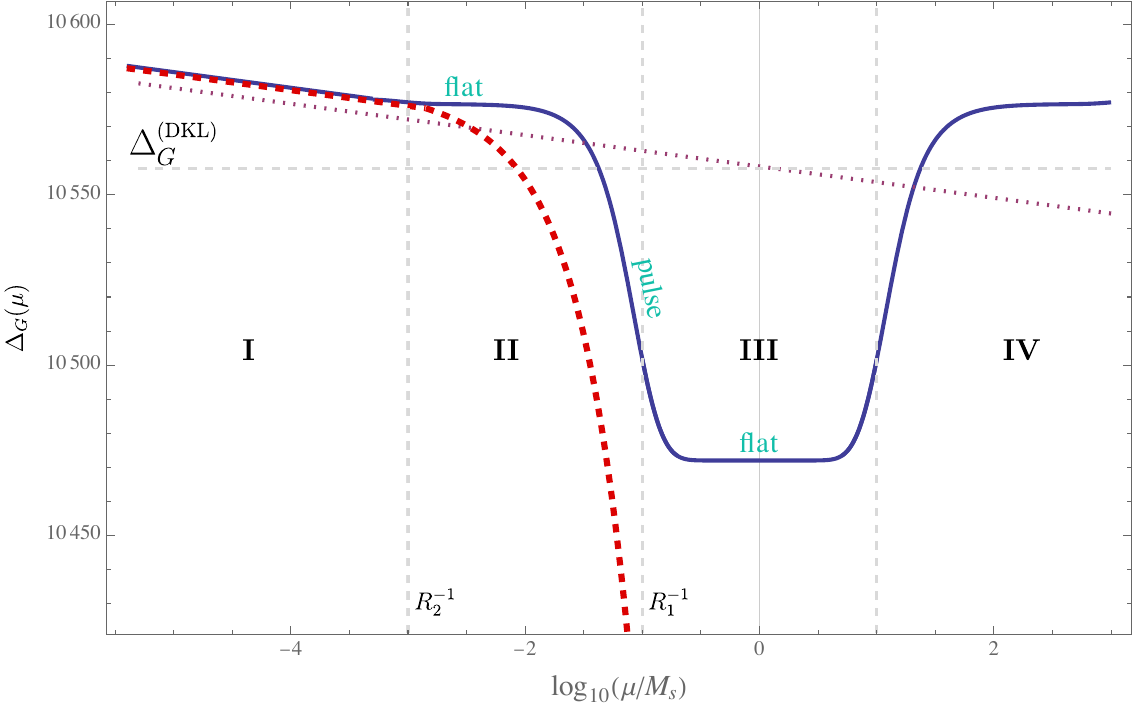}
\caption{Comparison of the results of three different running calculations.  The first (faint dotted  red line) is the minimally subtracted  EFT result in which the
four-dimensional coupling runs purely logarithmically across all energy scales and is 
simply matched to $\Delta_G^{\rm (DKL)}$ at the scale $\mu=M_s$.
The second (dotted red curve)
indicates
 the traditional field-theoretic power-law running that is expected from the accumulated effects of the KK modes beyond the first KK threshold at $\mu=R_2^{-1}$, with our theory becoming effectively five-dimensional in Region~II and eventually six-dimensional in Region~III.~
By contrast, the third (solid blue curve) 
shows what actually 
 emerges when the full worldsheet modular invariance of our theory is taken into account, including the use of a fully modular-invariant regulator (in this case, $\calG_\rho$ with $\rho=2$).
As indicated, all running ceases beyond the first KK threshold, except for the appearance of a transient ``pulse'' localized between Regions~II and III.~  All three plots correspond to choosing a rectangular compactification torus with   
$T_2=10000$ and $U_2=100$.
Despite the fact that all three runnings exhibit logarithmic behavior at low scales $\mu\ll 1/R_2$, we see that they diverge significantly at higher energies, with the fully modular-invariant result ultimately leading to a fixed-point regime at $\mu\approx M_s$ and a dual regime beyond $M_s$. This result also automatically avoids the Landau pole that might otherwise appear below $M_s$. 
\label{fig:dkl_comp}} 
\end{figure*}

Pulling together all of our previous results,  
we may  summarize our expressions for 
$P(a)$ within each of the five regions corresponding to Eq.~\eqref{eq:mu-ranges} as follows:
 \beqn
&& {\rm Region~I }\/: ~~~a~\ll~ 1/\sqrt{T_2 U_2} \nonumber\\
&&~~~
P(a) ~\approx~ \frac{\pi}{3}  \frac{1}{a} \,(T_2+U_2) -\frac{1}{a}\log ( 16\pi^2 e^{-2\eulerg} a^2 T_2 U_2 )~, 
       \nonumber\\
&& {\rm Region~II }\/: ~~~1/\sqrt{T_2 U_2}~\ll~ a ~\ll~ 1/\sqrt{T_2/ U_2}\nonumber\\
&&~~~
P(a) ~\approx ~ \frac{\pi }{3}\frac{1}{a} ( T_2 + U_2) ~,
      \nonumber\\
&& {\rm Region~III}\/: ~~~1/\sqrt{T_2/ U_2}~\ll~ a~\ll ~ \sqrt{T_2/ U_2}\nonumber\\
&&~~~
P(a) ~\approx~ \frac{\pi}{3} \left( a + \frac{1}{a}\right) \,T_2  ~ ,  ~\nonumber\\
&& {\rm Region~IV}\/: ~~~\sqrt{T_2/U_2} ~\ll~ a ~\ll~ \sqrt{T_2 U_2} \nonumber\\
&&~~~
P(a) ~\approx ~ \frac{\pi }{3} a ( T_2 + U_2) ~, \nonumber\\
&& {\rm Region~V}\/:~~~ a ~\gg~ \sqrt{T_2U_2}\nonumber\\
&&~~~
P(a) ~ \approx~
        \frac{\pi}{3}  a \,(T_2+U_2) -a
        \log ( 16\pi^2 e^{-2\eulerg} a^{-2} T_2 U_2 ) ~.\nonumber\\
\label{eq:allthePs}
\eeqn 
Note that within these expressions we have omitted terms which do not yield any contributions to the corresponding amplitudes $\Delta_G$.   These include not only constant terms but also terms scaling as $\log\,a$, as discussed above.

These expressions generate the suite of behaviors that were summarised in Eq.~\eqref{eq:massive_foreshadowing_tis_true}. 
Specifically, proceeding upwards in energy scale, we find
\begin{itemize} 
\item  At the lowest energy scales (Region~I), we find that $\Delta_G$ evolves as $\log\mu$.
This behavior assumes that our underlying theory has $\zStr\mathbbX_2\not=0$;  otherwise this logarithmic running is absent.
\item  Next, after crossing the lowest KK threshold (Region~II), we find that $\Delta_G(\mu)$ flattens out and becomes $\mu$-independent.
\item  Next, as we approach the second KK threshold, $\Delta_G(\mu)$ experiences a ``pulse'' and then enters Region~III.~   Within Region~III, $\Delta_G(\mu)$ is again flat, but with a different constant value.   In general, this pulse is a
genuinely transient effect that appears when we flow {\it between}\/ (rather than within) our asymptotic regions.
The size of this pulse depends on the shape modulus associated with  this toroidal compactification;   for a square (or rectangular) torus this magnitude is approximately $\pi U_2/3$.
\item  The above behavior persists all the way to the string scale $\mu=M_s$.   Beyond this scale, the behavior of $\Delta_G(\mu)$ is fixed by scale-inversion duality, with 
$\Delta_G(\mu)= \Delta_G(M_s^2/\mu)$.
\end{itemize}

This running is shown in Fig.~\ref{fig:dkl_comp}, along with two other runnings:
 the minimally
subtracted EFT result in which the four-dimensional coupling runs purely logarithmically across all energy scales and is simply
matched to $\Delta_G^{\rm (DKL)}$ at the scale $\mu=M_s$, and
the traditional field-theoretic power-law running that is expected from the accumulated effects of the KK modes beyond the first KK threshold at $\mu=R_2^{-1}$.
For this example we have chosen values of $T_2=10^4$ and $U_2=100$ (on a rectangular torus) in order to make the different behaviors in Regions~I, II, and III all evident.

As evident from Fig.~\ref{fig:dkl_comp}, 
all of these different runnings share a common feature, specifically a logarithmic running at extremely low energies $\mu \ll M_s$ within Region~I.~  
However, once we cross the first KK threshold  at $\mu=1/R_2$, the behaviors diverge. 
The four-dimensional EFT approach is of course completely insensitive to these thresholds that correspond to heavy modes because this approach is tantamout to a simple matching procedure. 
 As such, it is incapable of capturing any physics at energy scales that exceed the lowest KK scale. Meanwhile the traditional approach which involves summing the contribution of the KK states experiences a strong power-law running above the lowest KK scale.  Indeed, in the example shown in Fig.~\ref{fig:dkl_comp}, this running even leads to a Landau pole below the string scale!  The existence of such a Landau pole would signal a fundamental inconsistency, indicating that our theory becomes non-perturbative at such energy scales. If this were true, a one-loop perturbative approach would no longer be consistent.

 Fortunately, our fully modular-invariant calculation leads to an entirely different behavior.  Indeed, we see from Fig.~\ref{fig:dkl_comp} that all running ceases above $\mu=1/R_2$.  As a result, $\Delta_G(\mu)$ becomes flat.   In Fig.~\ref{fig:dkl_comp} this flat behavior is evident within the lower portion of Region~II.~ However, we see from 
 Fig.~\ref{fig:dkl_comp} 
 that this flat behavior soon gives way to our transient ``pulse''  near the second threshold at $\mu=1/R_1$.  This pulse is an entirely ``stringy'' phenomenon which smoothly connects the flat behavior in Region~II to the flat  behavior in Region~III, causing a sudden drop in the value of $\Delta_G(\mu)$.  Despite appearances, we emphasize that this drop in the value of $\Delta_G(\mu)$ is relatively small (a mere 1\% effect) in the example shown.  Moreover, this pulse is only a transient effect in the sense that its duration does not grow  into the asymptotic regions, but instead is localized to the boundary between Regions~II and III.~   Finally, upon entering Region~III,
the running then remains flat all the way to the string scale $M_s$. 
 Beyond this, the  running is governed by the entrance into a dual phase of the theory beyond $M_s$. 

The cessation of running beyond the lightest KK scale serves to protect our theory from the possible appearance of Landau poles.  This helps to render this approach self-consistent.   Moreover, as extensively discussed in Refs.~\cite{Abel:2021tyt,Abel:2023hkk}, the existence of the scale duality between the $\mu<M_s$ and $\mu> M_s$ regions indicates that there is a fundamental limit on the degree to which our theory can exhibit such field theoretic UV behavior.    

\subsection{The interplay of KK and winding states } 

\label{subsec:lessons}

We conclude this section with one final comment concerning the roles of Kaluza-Klein states and winding states in our analysis.

As we have seen, there is no logarithmic or power-law running for $\Delta_G(\mu)$ within Regions~II, III, and IV.~ 
Indeed, within Region~III, it is an exact result that $\Delta_G(\mu)$ contains no $\mu$-dependence at all. 
Moreover, as we explained 
below Fig.~\ref{fig:stringbeta_generic}
(at the very end of Sect.~\ref{sec:implications_running}), the cessation of running in this $\delta=2$ case arises because the contributions from Kaluza-Klein states are ultimately cancelling against those of winding states.
While Kaluza-Klein states begin to appear as we cross from Region~I to Region~II, at first glance it might appear that winding states
would not appear until the crossing from Region~III into Region~IV.~   However, in a UV/IR-mixed theory, this conclusion is too narrow --- indeed, the effects of both Kaluza-Klein and winding states are felt {\it throughout}\/ the string spectrum.   Indeed, it is precisely because of this mixing that the cancellation of running occurs at all, and that it persists throughout the entire extended stringy regime stretching from Region~II through Region~IV.~

Given our previous results, this can also be understood at an algebraic level.
Indeed, the two factors within the logarithm that appears in  Eq.~\eqref{eq:alternativeP} can be identified with Kaluza-Klein and winding modes respectively.
Accordingly, within Region~III, the KK modes make a contribution to the gauge coupling correction of 
\beq
 \Delta_G(\mu)\biggl|_{\rm KK} ~=~  \frac{\pi}{3}\,  \frac{\mu^2/M_s^2}{1+\mu^2/M_s^2} \, R_1 R_2 ~.
\label{KKcontribution}
\eeq
This quantity is proportional to the number of ``active'' KK modes at the scale $\mu$ (\ie, the number of KK modes with masses less than $\mu$), as one would expect from power-law running in a two-dimensional compactification.
However, the corresponding  winding-mode contribution is given by
\beq
 \Delta_G(\mu)\biggl|_{\rm winding} ~=~  \frac{\pi}{3}\,  \frac{1}{1+\mu^2/M_s^2} \, R_1 R_2 ~.
\label{windingcontribution}
\eeq
We thus see that the overall $\mu$-dependence cancels in the sum, 
leading to the $\mu$-independent quantity quoted for Region~III in
Eq.~(\ref{eq:massive_foreshadowing_tis_true}).

Interestingly, for $\mu<M_s $,
the power-law running contribution to $\Delta_G$ in Eq.~(\ref{KKcontribution}) does not dominate;  rather, it is the corresponding winding-mode contribution in Eq.~(\ref{windingcontribution}) which dominates. By contrast, for $\mu>M_s$, these roles are exchanged:  the Kaluza-Klein modes dominate, and it is the winding modes that contribute the subdominant piece. 

\section{Conclusions, discussion and future directions}

\label{sec:discussion}
\label{sec:conclusions}

In this paper we have derived a new  renormalization theorem for theories that exhibit UV/IR mixing.  The natural setting for our theorem, namely theories exhibiting modular invariance, includes all closed string theories.  It has been known for a long time that in four dimensions all such tachyon-free theories satisfy a remarkable supertrace constraint~\cite{Dienes:1995pm}
\beq 
\Str \,{\bf 1} ~=~ 0~
\label{boneconstraint}
\eeq
where the supertrace --- as suitable for theories with {\it infinite}\/ towers of states --- has the regulated definition given in Eq.~\eqref{supertrace_regulated}.  Of course, if the theory in question exhibits spacetime supersymmetry, this identity is satisfied level-by-level, with the contribution from each state cancelling in pairwise fashion against that of its superpartner.  However  --- even {\it without}\/ spacetime supersymmetry --- Eq.~\eqref{boneconstraint} continues to hold exactly as a consequence of a hidden {\it misaligned supersymmetry}\/~\cite{Dienes:1994np,Dienes:1995pm,Dienes:2001se} that exists within the spectrum of all tachyon-free modular-invariant string theories.  This misaligned supersymmetry is the manifestation of the underlying UV/IR mixing inherent in modular invariance.   Moreover, in the absence of spacetime supersymmetry,  the constraint in Eq.~(\ref{boneconstraint}) is not satisfied through pairwise cancellations;  indeed, no such pairwise cancellations are possible.    Instead, the result in Eq.~(\ref{boneconstraint}) holds as the result of a non-trivial cancellation between {\it all}\/ of the states across the {\it entire}\/ spectrum of the theory --- exactly as one would expect in a UV/IR-mixed theory.

In complete analogy with the constraint in Eq.~(\ref{boneconstraint}), another similar ``core'' constraint relates the 
one-loop cosmological constant 
(or more precisely, the zero-point one-loop amplitude) $\Lambda$ to the physical mass spectrum of the theory~\cite{Dienes:1995pm}:
\beq
    \Lambda ~=~ \frac{1}{24} \, \calM^2\,\Str\, M^2~
\label{boneconstraint2}
\eeq
where $\calM$ is the reduced string scale $M_s/(2\pi)$.  This constraint also applies to all tachyon-free modular-invariant theories, and is true in part because Eq.~(\ref{boneconstraint}) is true.  Indeed, both constraints ultimately emerge together within the same analysis~\cite{Dienes:1995pm}.

As we have seen, the constraints in Eqs.~(\ref{boneconstraint}) and (\ref{boneconstraint2}) have important physical ramifications.  It turns out that the first of these constraints 
can actually be regarded as a constraint on the zero-point one-loop amplitude $\Lambda$.
Indeed, within a purely field-theoretic context, this identity tells us that the one-loop quartic divergence that would ordinarily have  arisen for the one-loop cosmological constant is ``magically'' cancelled  --- indeed, cancelled {\it exactly}.
 However, within a string-theory context, Eq.~(\ref{boneconstraint}) is actually sufficient to kill {\it all}\/ of the divergences of $\Lambda$, not only the quartic divergence that might have been expected in field theory.  Because of this, the value of $\Lambda$ in string theory is actually {\it finite}\/, and this finite value is then given in Eq.~(\ref{boneconstraint2}).

This cancellation and the results in Eqs.~(\ref{boneconstraint}) 
and (\ref{boneconstraint2}) arise regardless of the particular phenomenology that the string might exhibit (such as its gauge symmetry or particle content)~\cite{Dienes:1995pm}.
These constraints are even preserved in the face of radiative corrections or if the theory passes through phase transitions in which the fundamental degrees of freedom change, so long as modular invariance is preserved.   Indeed, these results require only the preservation of modular invariance and the absence of physical tachyons, or equivalently on the existence a misaligned supersymmetry in the string spectrum.
As such, we regard these constraints as a core part of the UV-completeness of the theory. 

At first glance, it might appear that these constraints hold only at one-loop order.   However, this guess would not be correct.
To understand this, we must first realize that modular invariance itself is an {\it all-orders}\/ symmetry.   This point might seem somewhat counter-intuitive, since modular invariance is {\it motivated}\/ by the requirement that one-loop amplitudes be consistent with worldsheet reparametrization invariance.  Indeed, modular invariance is the symmetry that ensures that reparametrization invariance continue to hold for genus-one diagrams --- and not only genus-zero diagrams --- despite the extra ``large'' transformations that are possible around the non-contractible cycles of the torus.  For this reason this symmetry is sometimes called ``one-loop'' modular invariance.  However, the important point is that this symmetry --- regardless of its motivations --- is an exact symmetry which must be enforced exactly within any string theory.  Indeed, if it were not enforced, one-loop diagrams would not be consistent.   Of course, the requirement that the two-loop amplitudes also be consistent with reparametrization invariance might provide {\it additional}\/ constraints on our theory.
However, such constraints would  merely {\it augment}\/ the one-loop constraints.  They do not replace the constraints from modular invariance any more than the constraints of modular invariance replace those of reparametrization invariance at tree level.

Given this, one might wonder what masses appear within the supertrace in  Eq.~(\ref{boneconstraint2}). The answer is relatively simple.   To any order in perturbation theory, our theory will have a spectrum of (potentially corrected) masses and likewise will have a (potentially corrected) gravitational background.   Within this background, there will be a corresponding value of the zero-point one-loop amplitude $\Lambda$.   Our claim, then, is that these new masses and this new value for 
$\Lambda$ will continue to be exactly related by Eq.~(\ref{boneconstraint2}), and that Eq.~(\ref{boneconstraint}) will continue to hold as well.   Indeed, this is why we regard these two constraints as fundamental universal truths:  their validity holds regardless of the order at which we are performing our calculations, and stems directly from modular invariance in any tachyon-free theory.

In this paper, we have presented a new theorem which also governs the spectra of four-dimensional tachyon-free modular-invariant theories.  
Like the above constraints, this theorem also ultimately arises because of the non-trivial cancellations inherent in misaligned supersymmetry.   Moreover, as we have demonstrated, this theorem also yields many additional spectral constraints which are cousins of Eqs.~(\ref{boneconstraint}) and (\ref{boneconstraint2}).

This theorem holds for modular-invariant tachyon-free theories in which there is at least one decompactification limit, \ie,
a limit in which a geometric compactification volume can be taken to infinity, resulting in a higher-dimensional theory.
Such four-dimensional theories can therefore be viewed as geometric compactifications of higher-dimensional theories.
As we have described in Sect.~\ref{subsec:clash},
our theorem rests on the observation that there is a subtle mathematical clash that arises within the modular structure of such theories as the volume of compactification of the four-dimensional theory is taken to infinity.
Resolving this clash implies that the original four-dimensional theory must satisfy not only the universal constraint in Eqs.~\eqref{boneconstraint} --- a constraint which is appropriate for four dimensions --- {\it but also}\/ certain additional constraints that are appropriate for the higher-dimensional theory.
These additional constraints further restrict the properties of the theory, and likewise affect many more amplitudes than just the cosmological constant. Moreover, these additional conditions are independent of the radius of compactification. 

Even more interestingly, we found that these additional constraints immediately lead to a new {\it non-renormalization}\/ theorem for our four-dimensional theory.    This theorem also applies for all compactification radii, and holds even without spacetime supersymmetry. Like its antecedents, this theorem also holds for large classes of physical quantities, including (but not limited to) the case of gauge couplings.

The implications of this non-renormalization theorem are different for each possible decompactification limit experienced by our four-dimensional theory.  This is ultimately the case because (as we have discussed in Sect.~\ref{sec:implications_running})
each decompactification limit contributes its own constraints that must hold in the original four-dimensional theory.  

For limits involving decompactifications in which $\delta>2$ new dimensions open up, our theorem implies that there is {\it no running at all}\/ from the states involved in the sectors producing that limit.  In other words, the states which survive the limit do not collectively yield any running at all, regardless of the compactification radius associated with that limit.

By contrast, for limits involving decompactifications in which only $\delta=2$ new dimensions open up, the implications of our theorem are more complex.   In this case, the states from such sectors can give rise to at most a logarithmic running, but this can exist only at scales below the lightest of the KK thresholds associated with that limit.  Beyond this critical scale, we once again find that there are no collective contributions to the running.

For a given string model involving combinations of decompactification limits, these results generally imply that there is never any running at all beyond the lightest of the relevant KK scales.  Indeed, the  only changes in the value of the relevant amplitude beyond the lightest of the KK scales is that which arises as the result of our string-theoretic ``pulse''.
This effect is transient and highly localized to certain KK thresholds.  As such, this pulse is not a true running.

Along with our non-renormalization theorem, in this paper we have also proven a ``$T$-volume scaling rule''. This rule applies when 
the compactified volume is large but {\it not}\/ infinite, and asserts that the amplitudes in this regime are always proportional to the product of the decompactified $(4+\delta)$-dimensional amplitude and a so-called  ``$T$-duality-invariant compactification volume'' $\widetilde V_T$.  This result is given in 
Eq.~(\ref{QueenCamilla}), with $\widetilde V_T$ defined in 
Eq.~(\ref{eq:newdef}).
Although a similar phenomenon has previously been observed in the literature for certain calculations pertaining to certain string models, 
in this paper we have formulated an appropriate definition 
for $\widetilde V_T$ which respects the full symmetries of the theory (such as modular invariance and $T$-duality), and then proceeded to prove the resulting scaling rule in full generality.   Indeed, our proof holds 
even in cases without supersymmetry and regardless of the spacetime background.

Even beyond their phenomenological implications, the new supertrace constraints we have found could have far-reaching implications pertaining to the possible structure of string theory itself. Indeed one might reasonably ask if these constraints are so tight that the only solution is that all of the supertraces vanish level-by-level in a pairwise fashion.  This would then be a cancellation between a given state and a corresponding partner of opposite spin --- \ie, a cancellation between states and their superpartners. 
This would then imply that the only consistent four-dimensional theories are those that can decompactify exclusively to higher-dimensional theories which are supersymmetric.  Moreover, if the relevant compactification is untwisted, this would in turn imply that the original four-dimensional theory would have to be supersymmetric as well. 

However, it is not true that the only way to solve these constraints is through boson/fermion pairings.  Indeed, there exists an entire {\it landscape}\/ of tachyon-free non-supersymmetric heterotic string models~\cite{Dienes:2006ut,Dienes:2007ms}, and each of these furnishes us with an explicit example of a modular-invariant spectrum for which such constraints are satisfied without boson/fermion pairings.  It is not difficult to understand why such solutions exist.  Within any of these string models the spectrum exhibits
exponentially growing degeneracies of states.  Moreover, each of these states exhibits its own helicity and gauge charges.  As a result, even within the bounds of modular invariance,  the spectrum has far more ways of arranging itself and its charges than there will be constraints on it.  Indeed, the existence of an entire landscape of such strings provides direct verification of this fact.
Of course, a separate question concerns the {\it stability}\/ of such tachyon-free non-supersymmetric strings.  Addressing this question is beyond the scope of this work, although we note that there exist non-supersymmetric tachyon-free string models in four dimensions in which various instabilities can be exponentially suppressed~\cite{Abel:2015oxa}.

We close with four important remarks.
The first of these concerns the generality of our results with respect to our choice of regulator.
By implicitly distinguishing which states are to be considered dynamical and which are to be considered heavy and therefore ``integrated out'', the regulator helps to establish an appropriate energy scale $\mu$ for an effective field theory derived from the string. 
However, as we have stated throughout this paper, our conclusions are  independent of the specific form of the regulator $\calG_\rho(a)$.
This is ultimately due to the fact that our results stem directly from the modular invariance of the regulator, and any suitable regulator must be modular invariant.
Indeed, the precise definition of energy scale cannot affect the constraints that must be satisfied in order to cancel the accumulated contributions from the infinite towers of string states.  Likewise, our $T$-volume defined in Eq.~(\ref{eq:newdef}) and the scaling rule which results from this definition are
independent of the regulator as well.

Secondly, we note that the 
new constraints that we have found are also independent of the compactification radius. 
This in turn implies that the states which collectively correspond to a $\delta>2$ decompactification direction can \emph{never}\/ contribute to running, regardless of the size of the compactification volume. This remains the case even when the compactification volumes are of order the string scale, provided that the theory does not develop a tachyon at some radius.  Indeed, so long as the theory is merely \emph{capable}\/ having a large-volume decompactification, the constraints that correspond to every possible decompactification direction must all simultaneously apply in the four-dimensional theory.
As a result, our non-renormalization theorems hold even beyond the framework of large-volume compactifications,  and have a generality that matches that of their supersymmetric counterparts.
Indeed, as discussed at the end of 
Sect.~\ref{sec:implications_running}, our results suggest that non-supersymmetric theories seem to be governed by the same sorts of non-renormalization theorems as apply to supersymmetric theories, and that UV/IR mixing appears to play many of the same roles as supersymmetry.  This observation is worthy of further exploration.

Third, we note that our conclusion that the running of gauge couplings exhibits a fixed-point behavior at high scales bears a superficial similarity to the results of Ref.~\cite{Dienes:2002bg}.  
In Ref.~\cite{Dienes:2002bg}, a purely field-theoretic analysis was performed --- based on the results of Refs.~\cite{Dienes:1998vh,Dienes:1998vg} --- in which it was shown that although the accumulating effects of Kaluza-Klein states would generically convert a four-dimensional logarithmic gauge-coupling running into a higher-dimensional power-law running, this power-law growth could occasionally be cancelled in certain theories by the effects of switching from the true four-dimensional gauge coupling to an effective  loop expansion parameter which is sensitive to the number
of Kaluza-Klein levels that have been crossed at a given scale $\mu$.
This would then result in an apparent UV fixed point for the higher-dimensional theory.  
However, despite this similarity, there is ultimately no connection between these results:   
in this paper we are considering the actual gauge couplings, not the effective loop-expansion parameter;  we are not performing a field-theoretic analysis involving only Kaluza-Klein states but rather a fully modular-invariant analysis involving not only Kaluza-Klein states but also winding states;  and our non-renormalization theorem is completely general, serving as a core feature of {\it all}\/ tachyon-free modular-invariant theories and emerging directly from the UV/IR-mixed modular invariance of the theory.
Thus, despite these superficial similarities, the underlying physics is completely different.

Finally, we note that our work may have important connections to the  recent swampland program which seeks to ascertain the limits of viability of four-dimensional theories as low-energy approximations to UV-complete theories of quantum gravity~\cite{Vafa:2005ui,Ooguri:2006in}.
In particular, decompactification limits have played a central role in 
 swampland discussions of the so-called ``distance conjecture'' (see, {\it e.g.}\/, Refs.~\cite{Ooguri:2006in,
Baume:2016psm,Klaewer:2016kiy,Blumenhagen:2017cxt,Blumenhagen:2018nts,Grimm:2018ohb, Ooguri:2018wrx, Grimm:2018cpv,Corvilain:2018lgw,Lee:2019xtm,Lee:2019wij,Blumenhagen:2019qcg,Marchesano:2019ifh,Font:2019cxq,Grimm:2019wtx,
Lust:2019zwm,Baume:2019sry,Gendler:2020dfp,Baume:2020dqd,Perlmutter:2020buo,Bastian:2020egp,Klaewer:2020lfg,Calderon-Infante:2020dhm,Grana:2021zvf,Rudelius:2021oaz,
Montero:2022prj,Aoufia:2024awo}).
Our theorem clearly has relevance to this question. Indeed, while previous work in this  area has mostly developed along generic lines, utilizing traditional concepts of energy scales, mass splittings, and relations to the cosmological constant, 
it is clear from our discussion that 
these concepts and relations may be highly modified when UV/IR mixing is properly taken into account. Indeed, as we have seen, a generic theory will not be able to accommodate a decompactification limit unless it already possesses some very special and non-generic properties. The ramifications of these ideas for the distance conjecture and for the swampland program as a whole will be discussed in future work. 

\begin{acknowledgments}
The research activities of SAA were supported by the STFC grant ST/P001246/1. The research activities of KRD were supported in part by the U.S.\ Department of Energy
under Grant DE-FG02-13ER41976 / DE-SC0009913, and also 
by the U.S.\ National Science Foundation through its employee IR/D program.
The opinions and conclusions
expressed herein are those of the authors, and do not represent any funding agencies.
\end{acknowledgments}

\appendix 

\section{Amplitudes with $\overline E_2$ factors}

\label{ap:entwined}

In this Appendix we provide a prescription for evaluating the supertraces of operators whose definitions include factors of the Eisenstein function $\overline{E}_2$. 
The details behind this prescription are given in Ref.~\cite{Abel:2023hkk}.

Towards this end, let us assume that we have an operator of the form
\beq
\calX ~=~ \calA + \overline E_2 \calB~ 
\eeq
where $\calA$ and $\calB$ are functions of $\tau_2$.
Let us also assume that $\calA$ and $\calB$ have separate $\tau_2$-expansions of their own:
\begin{align}
{\calA} ~&=~ \mathbbA_0+\tau_2 \mathbbA_1+\tau_2^2 \mathbbA_2 ~\nonumber  \\
{\calB} ~&=~ \mathbbB_0+\tau_2 \mathbbB_1+\tau_2^2 \mathbbB_2~.
\end{align}
Our goal is to evaluate $\Str\,\calX$, or more general expressions such as $\Str\,[\calX f(M)]$ where $f(M)$ is some function of the mass $M$ of the states across the string spectrum.

To do this, let us recall that $E_2(\tau)$ has its own $q$-expansion 
\beq
E_2(\tau) ~=~ \sum_{r=0}^\infty \,\chi_r\, q^r~
\label{defG2here}
\eeq 
where the coefficients are given by 
\beq 
     \chi_r ~=~ \begin{cases}
        ~~~~~~ 1 &  r=0 \\
          -24 \sigma(r) & r>0 ~
     \end{cases}
\label{cr0}
\eeq
with $\sigma(r)$ denoting the sum-of-divisors function $\sigma(r) \equiv \sum_{d|r} d$.   For example, we find
$\sigma(r)= 1,3,4,7,\ldots$ for $r=1,2,3,4,\ldots$.

Given this definition for the coefficients $\chi_r$, we now define what we shall call an ``$E$-entwined'' supertrace, to be denoted $\Str_E$.
This is given in terms of the $\chi_r$-coefficients as
\beq
     {\rm Str}_E \,\calX ~\equiv ~
               \sum_{r=0}^\infty \,
               \chi_r \, \Str^{(r)} \calX~
\label{Esupertrace}
\eeq
where $\Str^{(r)} \calX $ denotes the {\it shifted}\/ supertrace over all the states in the theory that are ``level-mismatched'' by $r$ units:
\beq 
    {\rm Str}^{(r)} X ~\equiv~ 
    \lim_{\tau_2\to 0} \,\sum_{n}\, a_{n-r,n} \,X_{n-r,n} \, e^{- \pi  \alpha' M_L^2  \tau_2}~.
\label{shifted_supertrace}
\eeq
In Eq.~(\ref{shifted_supertrace}), as in Eq.~\eqref{eq:M2}, the quantity $M_L$ denotes the left-moving contribution to the total mass of the state, \ie, 
$\alpha' M_L^2=4n$, while $a_{n-r,n}$
is the net number of (bosonic minus fermionic) 
states which have 
\beq\alpha' (M_L^2-M_R^2) ~=~4r~.
\eeq
\mbox{\vspace{0.1cm} }

Our claim, then, is that $\Str\,[\calX f(M)]$ can be evaluated by formally replacing~\cite{Abel:2023hkk}
\beq
    \Str\, \Bigl[\calX \, f(M)\Bigr]~ \longrightarrow 
    ~
    \Str\,\Bigl[ \calA\, f(M)\Bigr] + \Str_E\, \Bigl[\calB \,f(M_L)\Bigr]~.~~
\label{replacement}
\eeq
Equivalently, for each component $\mathbbX_\ell$, we may replace 
\beq
\Str\, \Bigl[\mathbbX_\ell\, f(M)\Bigr] ~ \longrightarrow~
    \Str\,\Bigl[ \mathbbA_i\, f(M)\Bigr] + \Str_E\, \Bigl[\mathbbB_i \,f(M_L)\Bigr]~.~~~~~~~~
\label{replacement2}
\eeq
This procedure applies to all of our expressions in Eq.~\eqref{finalRSresult}. 

With these replacements, our theorem then continues to apply exactly as described in the main text.   

\bibliography{}
\end{document}